\documentclass[]{rmaa}
\pdfoutput=1

\title{The effects on a core collapse of changes in the number and size of
turbulent modes of velocity}

\altaffiltext{1}{Departamento de Investigaci\'on en F\'{\i}sica\\
Universidad de Sonora. Mexico.}

\author{Guillermo Arreaga-Garc\'{\i}a\altaffilmark{1}}

\fulladdresses{\item Guillermo Arreaga: Departamento de Investigaci\'on en F\'{\i}sica\\
Universidad de Sonora. \\
Apdo. Postal 14740, C.P. 83000, Hermosillo, Sonora, Mexico.\\
garreaga@cifus.uson.mx.}

\shortauthor{Arreaga}
\shorttitle{The effects of turbulent modes on a collapse}

\resumen{Consideramos 28 simulaciones de part\'{\i}culas dise\~nadas para comparar
el colapso gravitacional de un n\'ucleo de gas uniforme y esf\'ericamente
sim\'etrico, en el cual dos tipos extremos de espectros turbulentos de
velocidad han sido inicialmente inducidos, tales que $\nabla \cdot \vec{v}=0 $
(14 simulations) y $\nabla \times \vec{v}=0 $ (14 simulations). En todas las
simulaciones las razones de energ\'{\i}a cin\'etica y energ\'{\i}a t\'ermica
con respecto a la energ\'{\i}a gravitacional han sido fijadas en  $\beta$=0.21 y
$\alpha$=0.24, respectivamente. La mayor\'{\i}a de las simulaciones terminan
formando una sola protoestrella, excepto en dos simulaciones en las se forma un sistema binario
de protoestrellas. Con el prop\'osito de cuantificar las diferencias (o similitudes)
entre los dos tipos de simulaciones, calculamos algunas propiedades integrales de las
protoestrellas resultantes, tales como la masa M$_f$ y las razones $\alpha_f$ and $\beta_f$.}
\abstract{We consider 28 particle-based simulations aimed at comparing
the gravitational collapse of a spherically symmetric, uniform gas
core in which two extreme types of turbulent spectra of velocity have
been initially induced, so that $\nabla \cdot \vec{v}=0 $ (14 simulations)
and $\nabla \times \vec{v}=0 $ (14 simulations). For all the simulations, the ratios
of the kinetic energy and thermal energy to the gravitational energy were fixed
at $\beta$=0.21 and $\alpha$=0.24, respectively. Most of the
simulations finish by forming a single protostar, except for two simulations that form a
binary system of protostars. In order to quantify the differences (or similarities)
between the two types of simulations, we calculate some integral properties of the
resulting protostars, such as the mass M$_f$ and the ratios $\alpha_f$ and $\beta_f$.}

\addkeyword{--stars: formation}
\addkeyword{--physical processes: gravitational collapse}
\addkeyword{--physical processes:hydrodynamics}
\addkeyword{--methods: numerical}

\begin{document}
\maketitle
\newpage
\section{Introduction}
\label{intro}
\setcounter{footnote}{2}

Turbulence can provide support to gas cores against their
gravitational collapse. \citet{chandra} modeled turbulence support as
an additional pressure term in the equation of state in the
classical Jeans gravitational instability theory, so that the
turbulent velocity dispersion enlarged the Jeans length. Later,
\citet{bona} found that for large $n$ ($n > 3$) in the energy spectrum
$E(K) \approx K^{-n}$, the small scales can become unstable against
gravitational instability while the large scales can become stable;
this is just the opposite behavior to that expected from the
classical Jeans theory. \citet{leorat} performed simulations of
gravitational collapse in which a turbulent forcing was injected
at small scales, and found an effective support.

Gas cores are in general embedded in larger gas structures
called gas clouds. If turbulence is also present at the cloud scale,
then the large and intermediate scales of turbulence can favor the
fragmentation of the cloud, while the smaller scales still provide core
support, see \citet{sasao}, \citet{elmegreen} and
\citet{ballesteros1999}.

With regard to its numerical aspects, turbulence is usually generated
by a random driving force $\vec{f}$. According to the Helmholtz
decomposition theorem, the vector field $\vec{f}$ can be decomposed
into two extreme parts, a divergence-free (or solenoidal) part and a
curl-free (or compressive) part. \citet{fede} presented a very
detailed statistical comparison of the properties not only of the
two types of turbulence mentioned, but also of a mixed type of
turbulence that includes a desired ratio of both solenoidal and
compressive types. \citet{fede} considered driven turbulence, as the
forcing term is included in the right hand side of the Navier--Stokes
equations and it is kept during all the evolution time of typical
molecular clouds of the interstellar medium. Furthermore,
\citet{fede} also studied the properties and consequences of all
these mentioned types of turbulence in the star-formation theory,
although their simulations did not include self-gravity.

With regard to the observational side, it must be emphasized that
there is observational evidence suggesting the possibility that
turbulence on scales larger than the size of a molecular cloud can
significantly affect it; see \citet{brunt}. Besides, some starless
dense cores have been observed, which clearly show inward motions,
so that they are likely to evolve to the formation of one or more
low-mass stars; see \citet{tafalla1} and \citet{tafalla2}. However,
the inward motions are complex and do not correspond to a rotating
core under collapse, but to a core where turbulence provides
support against gravity; see \citet{caselli}. Besides, there are
observations suggesting that the mass structure of pre-stellar cores
is strongly centrally condensed, with a nearly uniform density in
their innermost region (ranging from a few to $10^3$ \ AU) and for
the outer region, a falling density profile that varies with the radius
as $r^{-\eta}$, with $\eta$ a constant; see \citet{myers} and
\citet{bou}.

In fact, the pre-stellar core L1544 has been well observed by
\citet{tafalla2}, and also modeled theoretically by, among others,
\citet{whithworth}, who computed analytically its evolution with the
simplifying assumptions of negligible pressure and rotation.
\citet{miAA} considered a rotating gas model to study the effects
of the extension of a gas envelope on a collapsing central core that
resembles the structure proposed for the L1544 pre-stellar core.
They found that a sufficient initial rotational energy must be
supplied initially  to favor a fragmentation of the core. It was also
modeled numerically by \citet{goodwin2004a} and \citet{goodwin2004b},
who considered the collapse and fragmentation of a core whose
collapse was triggered by using only a divergence-free turbulence
type. In two related papers, \citet{goodwin2006} and
\citet{goodwin2004a} again considered the influence of low levels of
solenoidal turbulence on the fragmentation and multiplicity of dense
star-forming cores. All these papers suggested that turbulent
fragmentation can be a natural and efficient mechanism for forming
binary systems. A step further along this direction was achieved by
\citet{attwood}, who introduced an energy equation that provided a
more realistic description of the core thermodynamics and compared
their results with the simulations made with the barotropic
equations of state reported by \citet{goodwin2004a},
\citet{goodwin2004b}, and \citet{goodwin2006}.

Shortly after, \citet{walch} implemented a mixed turbulence velocity
spectrum, which resulted in a ratio of solenoidal to compressive
modes of 2:1; a cubic mesh of 128$^3$ grid elements was then
populated with Fourier modes with wave-numbers $K$ between $K_{min}$
and $K_{max}$, so that the particles obtained a  velocity from a
linear interpolation within their corresponding grid element.
\citet{walch} then calculated the collapse of a gas core of radius
R$_0$ under the influence of modes with a peak wavelength
$\lambda_{max}$ that varied within the range R$_0$/2, R$_0$, 2 R$_0$
and 4 R$_0$. It must be emphasized that \citet{walch} observed core
fragmentation only for the models with R$_0$/2 $\leq \,
\lambda_{max} \, \leq$ 2 R$_0$.

In this paper we present a set of self-gravitating simulations to
follow the collapse of a core, in which the initial density
fluctuations come from random collisions of particles whose velocities
have been assigned according to a turbulent spectrum. We here only
focus on decaying turbulence (not driven). One-half of
the simulations are of the divergence-free turbulence type, while the
rest are of the curl-free turbulence type. We extend the
range of $\lambda_{max}$, so that here it goes from 1--4 R$_0$ and
6--10 R$_0$. All the models are calculated using a Fourier mesh that
changes not only in size, but also in the number of Fourier modes, so
that we consider a mesh of 64$^3$ or 128$^3$ grid elements per
model.

All the simulations have been calibrated so that the
values of the dimensionless ratios of thermal energy to
gravitational energy, $\alpha$, and kinetic energy to gravitational
energy, $\beta$, maintain values fixed at 0.24 and 0.21,
respectively. These dimensionless ratios are very important in
collapse simulations since two decades ago, since theorists proposed
collapse and fragmentation criteria of rigidly rotating cores by
constructing configuration diagrams in terms of $\alpha$ and
$\beta$; see for instance \citet{miyama}, \citet{hachisu1},
\citet{hachisu2} and \citet{tsuribe1}. From these diagrams, it has
also been possible to study the formation of binaries of low-mass stars.
Thus, the characterization of our simulations in terms of $\alpha$
and $\beta$ will allow comparison between the collapse of rotating
and turbulent cores.

For comparison with the turbulent models of \citet{goodwin2004a} and
\citet{goodwin2004b}, as well as with the rotating models of \citet{miAA},
\citet{RMAA2016}, and \citet{miAAS}, we implement in this paper initial
conditions to represent only the central core, so that it resembles
the dense core L1544.

Finally we mention that computer simulations of turbulence is a very active field of
research; see for instance the review of \citet{padoan}, in which the authors reported
recent advances of current simulations focusing on the connection of the physics of turbulence
with the star formation rate in molecular clouds.

\section{The physical system and the computational method}
\label{subsec:met}

\subsection{The core and the initial setup of the particles.}
\label{subsec:core}

We consider the gravitational collapse of a variant of the so-called
"standard isothermal test case'' which was
first calculated by \citet{boss1979} and later calculated by
\citet{burkertboden93} and \citet{bateburkert97}.

In this paper, the core radius is R$_0$=4.99 $\times 10^{16} \,$ cm
$\equiv$ 0.016 pc $\equiv$ 3335 AU and its mass is M$_0$=5
M$_{\odot}$. Thus, the average density and the corresponding free
fall time of this core are $\rho_0$=1.91 $\times 10^{-17}\,  g$
cm$^{-3}$ and $t_{ff} \approx $4.8 $\times 10^{11} \,$ s $\equiv$
15244 yr, respectively.

We set N$_p$ particles on a rectangular mesh of side length  2 \, R$_0$
(the core diameter), to represent the initial core configuration.
We then partition the simulation volume into small
elements, each with a side length given by $\Delta x = \Delta y =
\Delta z$ = R$_0$ /N$_{xyz}$ where N$_{xyz}$ is an integer given by
133, so that there are 133$^3$ grid elements in the entire volume;
at the center of each small grid  we place a particle, which is
next displaced from its initial location a distance of the order
$\Delta x/4.0$ in a random spatial direction. The boundary conditions
applied to the simulation box, both in the initial setup and during the time evolution
are named vacuum boundary conditions ( so that we have a constant volume
with non-periodic boundary conditions).

It should be noted that we here consider only a uniform density
core, so that its average density is  given by $\rho_0$ or a density
number of n$_0$=9.5 10$^6$ particles/cm$^{3}$. This is a very dense
core, whose density is 10 times higher than that of the ``standard
isothermal test case.''

Thus,  for all the simulations, the particles
have the same mass, according to
$m_i= \rho_0 \times \Delta x\, \Delta y\, \Delta z$ with
i=1,...,N$_p$, where N$_p$=996 040. Then the mass
is given by m$_p$=5.0 $\times \, 10^{-6} \, $ M$_{\odot}$.

\subsection{Evolution Code}
\label{subs:code}

The gravitational collapse of our models has been followed by using
the fully parallelized particle-based code Gadget2;
see~\citet{gadget2} and also \citet{serial}.
$Gadget2$ is based on the $tree-PM$ method for
computing the gravitational forces and on the standard smoothed
particle hydrodynamics (SPH)  method for
solving the Euler equations of hydrodynamics.
Gadget2 implements a
Monaghan--Balsara form for the artificial viscosity;
see~\citet{mona1983} and \citet{balsara1995}. The strength of the
viscosity is regulated by the parameter $\alpha_{\nu} = 0.75$ and
$\beta_{\nu}=\frac{1}{2}\, \times \alpha_v$; see eqs. 11 and 14
in~\citet{gadget2}. In our simulations we fixed the Courant factor to
be $0.1$.
\subsection{Resolution and thermodynamics considerations}
\label{subs:resol}

Following \citet{truelove} and \citet{bateburkert97}, in order to
avoid artificial fragmentation, any code for collapse calculation
must have a minimum resolution given in terms of the Jeans
wavelength $\lambda_J$:

\begin{equation}
\lambda_J=\sqrt{ \frac{\pi \, c^2}{G\, \rho}} \; , \label{ljeans}
\end{equation}
\noindent where $c$ is the instantaneous speed of sound and $\rho$ is the
local density. To obtain a more useful form for a particle based code, the Jeans
wavelength $\lambda_J$ can be transformed into a Jeans mass, given by

\begin{equation}
M_J \equiv \frac{4}{3}\pi \; \rho \left(\frac{ \lambda_J}{2}
\right)^3 = \frac{ \pi^\frac{5}{2} }{6} \frac{c^3}{ \sqrt{G^3 \,
\rho} } \;. \label{mjeans}
\end{equation}
\noindent The values of the density and speed of sound must be updated according to
the following equation of state:

\begin{equation}
p= c_0^2 \, \rho \left[ 1 + \left(
\frac{\rho}{\rho_{crit}}\right)^{\gamma -1 } \, \right] ,
\label{beos}
\end{equation}
\noindent which was proposed by \citet{boss2000}, where $\gamma\, \equiv
5/3$ and for the critical density we assume the value
$\rho_{crit}=5.0 \times 10^{-14} \, $ gr $\,$ cm$^{-3}$.

The smallest mass particle that a SPH calculation must resolve in
order to be reliable, is expressed in terms of the particle mass,
m$_r$, given by m$_r \approx M_J / (2 N_{neigh})$, where
$N_{neigh}$ is the number of neighboring particles included in the
SPH kernel; see \citet{bateburkert97}. Hence, a simulation
satisfying all these requirements must satisfy $\frac{m_p}{m_r}<1$.

Thus, for the turbulent core under consideration we have $M_j
\approx 1.1 \times 10^{-3}\,$ M$_{\odot}$ and m$_r \approx 1.4
\times 10^{-5}\,$ M$_{\odot}$, since we took $N_{neigh}=40$. The
ratio of masses is given by m$_p$/m$_r$=0.33 and then the desired
resolution is achieved in our simulations.
\subsection{The turbulent velocity of the particles.}
\label{subsec:turbspect}

To generate the turbulent velocity spectrum, we set a second
mesh, in which the partition is determined by the values of N$_x$,
N$_y$ and N$_z$, which will initially be given by 64 and later on
they will be increased to 128, so that the total number of grid
elements will change from 64$^3$ to 128$^3$. It should be noted
that the velocity vector of a particle changes with the number of
modes.

Let us denote the side length of this second mesh by L$_0$, so
that it is proportional to the core radius R$_0$

\begin{equation}
L_0=C_R \times R_0
\label{defCR}
\end{equation}
\noindent where $C_R$ is a constant, the value of which
will also determine the collapse model under consideration, see Table~\ref{tab:mod}.

Thus, the size of each grid element is given by $\delta x = L_0/N_x$, $\delta
y=L_0/N_y$ and  $\delta z=L_0/N_z$. In Fourier space the partition
is given by $\delta K_x=1.0/\left( N_x \times \delta x \right)$ , $\delta
K_y=1.0/\left( N_y \times \delta y \right)$ and $\delta
K_z=1.0/\left( N_z \times \delta z \right)$.

Each Fourier mode has the components $K_x=i_{K_x}  \delta K_x$,
$K_y=i_{K_y} \, \delta K_y $ and $ K_z=i_{K_z} \, \delta K_z $,
where the indices $i_{K_x}$, $i_{K_y}$ and $i_{K_z}$ take values in
$[-N_x/2,N_x/2]$, $[-N_y/2,N_y/2]$ and $[-N_z/2,N_z/2]$,
respectively. The magnitude of the wave number is
$K=\sqrt{K_x^2+K_y^2+K_z^2}$. Then $K_{max}$ is proportional to
$\frac{\sqrt{3}\, N_x}{2 \, L_0} $. A Fourier mode can equally be
described by a wave length $\lambda=2\, \pi/K$. Then we see that

\begin{equation}
\lambda_{max} \approx L_0 \hspace{1 cm} \lambda_{min} \approx \delta_x \,.
\label{kylvsR}
\end{equation}
\noindent

We will consider two types of turbulence, so that we will be able to compare how the nature of the
initial turbulent spectrum affects the core collapse. The initial power of the
velocity field will be given by

\begin{equation}
P(\vec{K})=\left< \left| \vec{v}(\vec{K}) \right|^2 \right> \approx \left| \vec{K}\right|^{-n}
\label{power}
\end{equation}
\noindent where the spectral index $n$ is a constant.

Let us first consider the Fourier transform ${\cal F}$ of the velocity field, so that
$\vec{v}(\vec{r})={\cal F} \, [\vec{v}(\vec{K})]$ or explicitly

\begin{equation}
\vec{v}(\vec{r})=\int \vec{v}(\vec{K}) \exp \left( i \vec{K}\cdot \vec{r} \right) \; d^3K.
\label{defF}
\end{equation}

\noindent It can then be shown that the velocity $\vec{v}(\vec{K})$ can be written as

\begin{equation}
\vec{v}(\vec{K}) = -i \vec{K} \Phi(\vec{K}) + i \vec{K} \times \vec{A}(\vec{K})
\end{equation}
\noindent where the scalar and vector potential functions are given by

\begin{equation}
\begin{array}{l}
\Phi(\vec{K}) = -i \vec{K} \cdot \vec{v}(\vec{K}) \vspace{0.25 cm} \\
\vec{A}(\vec{K})= i \vec{K} \times \vec{v}(\vec{K})
\end{array}
\end{equation} so that their Fourier transforms are $\Phi(\vec{r})={\cal F} \, [\Phi(\vec{K}) ]$ and
$\vec{A}(\vec{r})={\cal F} \, [\vec{A}(\vec{K})]$, respectively. According to the Helmholtz
decomposition theorem, the velocity field in physical space will then be determined in general by

\begin{equation}
\vec{v}(\vec{r}) = -\nabla \Phi(\vec{r}) + \nabla \times \vec{A}(\vec{r})
\label{decom}
\end{equation}
\subsubsection{Divergence-free turbulence.}
\label{subsubsec:divfreeturb}

Let us consider the case $\Phi(\vec{K}) \equiv 0$, which implies that $\vec{K} \cdot \vec{v}(\vec{K}) =0 $.
The velocity field can then be written in terms of a vector potential $\vec{A}$ alone.
According to Eq.~\ref{decom}, $\nabla \cdot \vec{v}(\vec{r})=0$; this is the divergence free turbulence
spectrum, whose Fourier transform can be approximated by the following discrete summation:

\begin{equation}
\vec{v}(\vec{r}) \approx  \Sigma \; i \vec{K} \times \vec{A}(\vec{K}) \,
\exp \left( i \vec{K}\cdot \vec{r} \right) \,.
\label{velcasea}
\end{equation}

In order to have the power described in Eq.~\ref{power}, we choose
the vector potential $\vec{A}(\vec{K})$ given by

\begin{equation}
\vec{A}(\vec{K})= \left| \vec{K} \right|^{-\frac{n+2}{2}} \; \vec{C}_K \; \exp \left( i \Phi_K \right)
\label{velA}
\end{equation}
\noindent where $\vec{C}$ is a vector whose components
are denoted by $\left( C_{K_x}, C_{K_y}, C_{K_z}\right)$, and take
values obeying a Rayleigh distribution. Hence, the magnitude of $\vec{C}$
is calculated by means of the formula $C=\sigma\times \sqrt{ -2.0
\times \log \left(1.0-u\right) }$, where $u$ is a random number in
$(0,1)$ and $\sigma=1.0$ is a fixed parameter. There must be one phase
$\Phi_K$ for each component of the vector $\vec{C}$, so that each of
$\left( \Phi_{K_x}, \Phi_{K_y}, \Phi_{K_z} \right)$ takes random
values in the interval $[0,2\,\pi]$.

Thus, following \citet{dobbs}, the components of the particle velocity are given by

\begin{equation}
\begin{array}{l}

\vec{v} \, \approx \Sigma_{-N/2}^{N/2} \left| \vec{K}\right|^{ \frac{-n-2}{2} }
\times

\left\{

\begin{array}{l}

\left[ K_z \, C_{K_y} \sin \left( \vec{K}\cdot \vec{r} + \Phi_{K_y}\right) -
K_y \, C_{K_z} \sin \left( \vec{K}\cdot \vec{r} + \Phi_{K_z}\right)\right] \; \mbox{for}\, v_x\\

\left[ - K_x \, C_{K_z} \sin \left( \vec{K}\cdot \vec{r} + \Phi_{K_z}\right) +
K_z \, C_{K_x} \sin \left( \vec{K}\cdot \vec{r} + \Phi_{K_x}\right)\right] \; \mbox{for} \, v_y\\

\left[ -K_x \, C_{K_y} \sin \left( \vec{K}\cdot \vec{r} + \Phi_{K_y}\right) +
K_y \, C_{K_x} \sin \left( \vec{K}\cdot \vec{r} + \Phi_{K_x}\right)\right] \; \mbox{for} \, v_z\\
\end{array}

\right.
\end{array}
\label{velturb}
\end{equation}
\subsubsection{Curl-free turbulence.}
\label{subsubsec:rotfreeturbspect}

On the other hand, let us consider the case in which $\vec{A}(\vec{K}) =0 $, so
that $\vec{K} \times \vec{v}(\vec{K}) =0 $. According to Eq.~\ref{decom}, this
velocity field satisfies $\nabla \times \vec{v}=0$ and for this reason we call
it the curl-free turbulence spectrum.

In order to have the power spectrum
shown in Eq.~\ref{power}, we choose the scalar potential function given by

\begin{equation}
\Phi(\vec{K})= \left|\vec{K}\right|^{-\frac{n+2}{2}} \; \exp \left( i \Phi_K \right)
\label{Phi}
\end{equation}
\noindent where there is now only one wave phase random function
$\Phi_{K}$ that takes random values in the interval $[0,2\,\pi]$. Therefore, the velocity
field will be determined by

\begin{equation}
\vec{v}(\vec{r})  \approx \Sigma_{-N_x/2,-N_y/2,-N_z/2}^{N_x/2,N_y/2,N_z/2} \left| \vec{K}\right|^{\frac{-n-2}{2}} \;
\vec{K} \sin \left( \vec{K}\cdot \vec{r} + \Phi_K \right)
\label{velPhi}
\end{equation}
\noindent where the spectral index $n$ will be fixed in our
simulations to be $n=-1$ and thus we will have $v^2\approx K^{-3}$; see
\citet{ArreagaKlapp}.

In the two types of turbulence spectra, the SPH particles have
initially a Gaussian distribution of velocity. Subsequently, by using all
the simulation particles, we obtain that the average Mach number is,
remarkably, almost the same for all our simulations, and is
${\cal M}$=1.5. The velocity dispersion is also very similar for
the two types of spectrum: $\sigma_v$= 0.21 km/s. Following the
definitions of skewness (or third moment) and kurtosis (or fourth
moment) of a given distribution, see \citet{numreci}, we observe
differences in the values of the skewness and kurtosis of the velocity
distribution spectra: 0.4 and 0.003 for the divergence-free type and
0.38 and 0.03 for the curl-free type, respectively.
\subsection{Initial energies}
\label{subs:energies}

It is well known that the global dynamical evolution of the core is
determined by the ratio of the thermal energy to the gravitational
energy, denoted by $\alpha$ and
the ratio of the kinetic energy
to the gravitational energy, denoted by $\beta$; see, for instance,
\citet{miyama}, \citet{hachisu1} and \citet{hachisu2}, who obtained a
criterion of the type $\alpha \, \beta <  0.2 $ to predict the
collapse and fragmentation  of a rotating isothermal core.

The gravitational and kinetic energies can
be approximated in terms of the physical parameters of the core
considered in this paper:

\begin{equation}
\begin{array}{l}
E_{grav} \approx - \frac{3}{5} \; \frac{G\, M_0^2}{R_0} \\
E_{kin} \approx  \frac{1}{2} \; M_0 <V>^2
\label{egrav}
\end{array}
\end{equation}
\noindent where $G$ is Newton's universal gravitation
constant, M$_0$ is the mass, R$_0$ is the radius, and $<V>$ the average
velocity. The thermal energy $E_{therm}$ (kinetic plus potential
interaction terms of the molecules) can be approximated by

\begin{equation}
E_{ther}\approx \frac{3}{2} {\cal N} \, k_B \, T = \frac{3}{2} M_0\, c_0^2
\label{etherm}
\end{equation}
\noindent where ${\cal N}$ is the total number of molecules in
the gas, $k_B$ is the Boltzmann constant, and $T$ is the
temperature of the core.

These energies can be calculated in terms of the SPH particles as follows:

\begin{equation}
\begin{array}{l}
E_{ther}=\frac{3}{2}\sum _{i}m_{i}\frac{p_{i}}{\rho _{i}}\\
E_{kin}=\frac{1}{2}\sum _{i} m_{i} v_i^{2},\\
E_{ grav}=\frac{1}{2}\sum _{i}m_{i}\Phi _{i}
\label{energiespart}
\end{array}
\end{equation}
\noindent where $p_i$ is the pressure and $\Phi _{i}$ are the values
of the pressure and gravitational potential at the location of particle
$i$, with velocity given by $v_i$ and mass $m_i$; the summations
include all the simulation particles.

Now, the values of the speed of sound $c_0$ and the level of
turbulence, which is adjusted by multiplying the velocity vector by
an appropriate constant, are chosen so that the velocity field
fulfills the following energy requirements:

\begin{equation}
\begin{array}{l}
\alpha \equiv \frac{E_{ther}}{\left|E_{grav}\right|}=0.24 \vspace{0.25 cm}\\
\beta \equiv \frac{E_{kin}}{\left|E_{grav}\right|}=0.21 \,.
\end{array}
\label{alphaybeta}
\end{equation}
\noindent where the energies entering in these ratios have been
calculated using the relations of Eq.~\ref{energiespart}.

The virial theorem states that  if a cloud is in thermodynamical equilibrium,
then the dimensionless energy ratios satisfy the following relation:

\begin{equation}
\alpha + \beta=\frac{1}{2}\;,
\label{abvirial}
\end{equation}
\noindent which will be used in a plot of the next section. It
should be noted that we do not include a turbulence term,
$\gamma_{turb}$, as all the turbulent energy  has already entered in
the kinetic ratio, $\beta$.

In order to compare our simulations with other papers, we consider
Eq. 1 of \citet{walch}, in which the turbulent energy of a core is
approximated by the sum of two terms: the first one contains the
FWHM velocity width reported by \citet{andre} for the Diazenylium
molecule N$_2$H$^+$, and the second one depends on the thermal
energy $k_B \, T$.

As we mentioned at the end of Section~\ref{subsec:turbspect}, our
simulations have an average Mach number of 1.5, so by using our
value of c$_0$=35 820 cm/s, we get that the average velocity
$<v>$= 0.53  km/s. However, an estimate of the velocity dispersion
used in observations, denoted here by $\sigma_v^e$, can be obtained
from the empirical relation $E_{kin}$=M$_0 \, (\sigma_v^e)^2$/2, and
using  Eq.~\ref{egrav} and the second relation of
Eq.~\ref{alphaybeta}, we thus obtain $\sigma_v^e=\sqrt{ \frac{ \beta
6 G M_0}{ 5 R_0} }$=0.57 km/s, which is of the same order as
$<v>$, as expected. By using  a mass of 29 uma for the Diazenylium
molecule and the value of $\sigma_v^e$ in eq.~1 of \citet{walch},
we obtain for our simulations a ratio of the turbulent energy
to the gravitational energy of $\gamma_{turb}=0.11$. In the
simulations reported by \citet{walch}, the ratio
$\gamma_{turb}=0.010$ was used.
\subsection{The models}
\label{subs:mod}

The models considered in this paper are summarized in
Table~\ref{tab:mod}, whose entries are as follows. Column 1 shows
the model number;  column 2 shows the value of the size constant
C$_R$ as defined in Eq.~\ref{defCR}; column 3 shows the partition
used in the Fourier mesh, so that the total number of grid elements
of the cubic mesh is the cube of that number; we mentioned in
Section~\ref{subs:energies} that the average Mach number of the
simulations are very similar, however, less than 1 percent of the
simulation particles can attain very high velocities: column 4
shows the peak Mach number of the simulations at the initial time (the
snapshot zero); column 5 gives the peak velocity found in the
collapsed region where the resulting protostar is formed; column 6 shows the
number of the figure in which the model configuration is shown;
column 7 shows the type of turbulence used to generate the initial
velocity spectrum where the label D-F means
divergence-free and C-F means curl-free; column 8 shows the
resulting configuration, where the label S means a single protostar and B means
a binary system of protostars.

\begin{figure}
\begin{center}
\includegraphics[width=5.2 in]{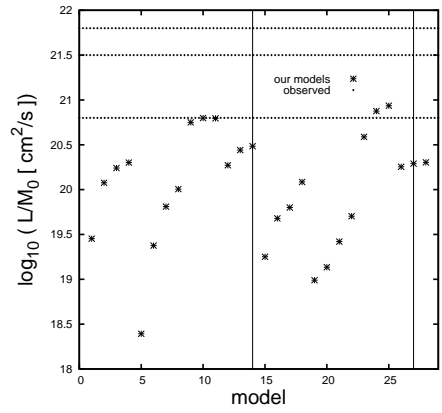}
\caption{The initial specific angular momentum $L/M_0$ against
the model number. The horizontal lines indicate the observed values of
$L/M_0$ for real cores; see ~\citet{goodman93,boden95}. }
\label{SpecAngMom}
\end{center}
\end{figure}

In order to further characterize our models, we calculate the total
angular momentum of each model by using all the simulation
particles, so that $\vec{L}= \sum_{i=1}^{N_p} \, \vec{r} \, \times
\, \vec{p}$, with the linear momentum $\vec{p}$ given by $\vec{p}=
m_p \, \vec{v}$, where $m_p$ is the mass of the particle and
$\vec{v}$ its velocity; see Section \ref{subsec:core}.
Fig.\ref{SpecAngMom} shows the specific angular momentum $L/M_0$
(the total angular momentum $|\vec{L}|$ of the initial distribution
of  particles divided by the core mass) against the model number. By
looking at Fig.\ref{SpecAngMom}, one can see that (i) when the
Fourier mesh size increases in terms of the core diameter, so does
the $|\vec{L}|$ and (ii) when the number of Fourier modes increases
from 64$^3$ to 128$^3$, $|\vec{L}|$ decreases. The three horizontal
lines in Fig.\ref{SpecAngMom} indicate the observed values for real
cores with radii varying within 1-5 $\times \, 10^{17}$ cm; see
~\citet{goodman93,boden95}. In fact, the observed values of $L/M_0$
are higher than those of our models. The two vertical lines in
Fig.\ref{SpecAngMom} indicate the models where we will observe
binary formation; see Section\ref{sec:results}.

\section{Results}
\label{sec:results}

\begin{figure}
\begin{center}
\begin{tabular}{cc}
\includegraphics[width=3.5 in]{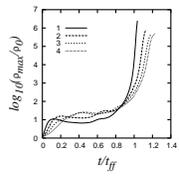} & \includegraphics[width=3.5 in]{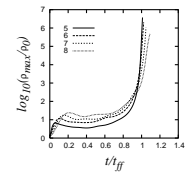} \\
\includegraphics[width=3.5 in]{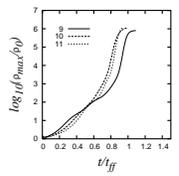} & \includegraphics[width=3.5 in]{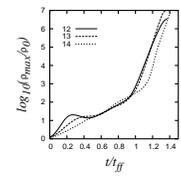} \\
\end{tabular}
\caption{\label{evoluciondensidadDF} Time evolution of the peak
density $\rho_{max}$ of the divergence-free turbulence models.}
\end{center}
\end{figure}

\begin{figure}
\begin{center}
\begin{tabular}{cc}
\includegraphics[width=3.5 in]{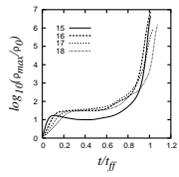} & \includegraphics[width=3.5 in]{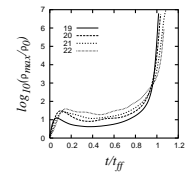} \\
\includegraphics[width=3.5 in]{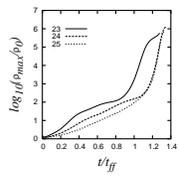} & \includegraphics[width=3.5 in]{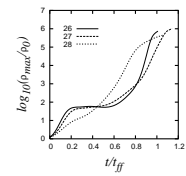} \\
\end{tabular}
\caption{\label{evoluciondensidadRF} Time evolution of the peak
density $\rho_{max}$ of the curl-free turbulence models.}
\end{center}
\end{figure}

We first note that all the models considered in this paper show a
clear tendency to collapse within a free fall time $t_{ff}$ (defined
in Section~\ref{subsec:core}) or a little after; see
Figs.~\ref{evoluciondensidadDF} and \ref{evoluciondensidadRF}. This
is to be expected because the values chosen for the initial energies
in Eq.~\ref{alphaybeta} favor a core collapse.

We observe many different transient effects in the first stage of
the dynamical evolution of the models. In spite of this,  the outcome of
most of the models is a single protostar. But two models form a
binary protostar system. In order to illustrate the results, the
outcome of each model is shown in a mosaic composed of three panels.

The first two panels are iso-density plots constructed using a
small set of particles (around ten thousand) located within a slice parallel to the
equatorial plane of the spherical core. In order to compare the
models, these panels are all taken at the same snapshot times: 0.06 $t_{ff}$
and 0.67 $t_{ff}$. A bar located at the
bottom of these panels shows the range of
values for the $log$ of the density $\rho(t)$ at time $t$ normalized to the
average initial density (  that is $log_{10}(\frac{\rho}{\rho_0})$, where $\rho_0$ was
defined in Section \ref{subsec:core}  ) and the color allocation set by the
program {\it pvwave} version $8$.

The third panel is also an iso-density plot, but in this case all
the particles are used  in order to make a 3D rendered image ( not
only a slice ) taken at the last snapshot available for each
selected simulation, so there is a variation in the snapshot time,
within 1.0--1.3 $t_{ff}$, depending on the model. A comparison of
these final outcome configurations is possible at slightly different
output times because the configurations have already entered (or are
about to enter) a stable stage. The bar is also located at the
bottom of these panels, but now the values are the $log$ of the
column density $\rho(t)$, calculated in code units  by the program
{\it splash} version $2.7.0$. The density unit is given by uden=1.6
$\times 10^{-17}$, so that the average density in code units is
$\rho_0$/uden=1.19. The color bar shows values typically in the
range 0-6, so that the peak column density is 10$^6 \, \times \, $
uden = 1.6 $\times 10^{-11}$ g cm$^{-3}$.

It must  be mentioned that the vertical and horizontal axes of all the panels
indicate the length in terms of the radius R$_0$ of the sphere (approximately
3335 AU). So, the Cartesian axes $X$ and $Y$ vary initially from -1 to 1. In order to
facilitate a comparison of the resulting last configurations, we use the same length
scale of 0.2 per side (approximately 667 AU) in all the third panels.

\subsection{Models 1--4}
\label{subsec:results1-4}

\begin{figure}
\begin{tabular}{ccc}
\includegraphics[width=2 in]{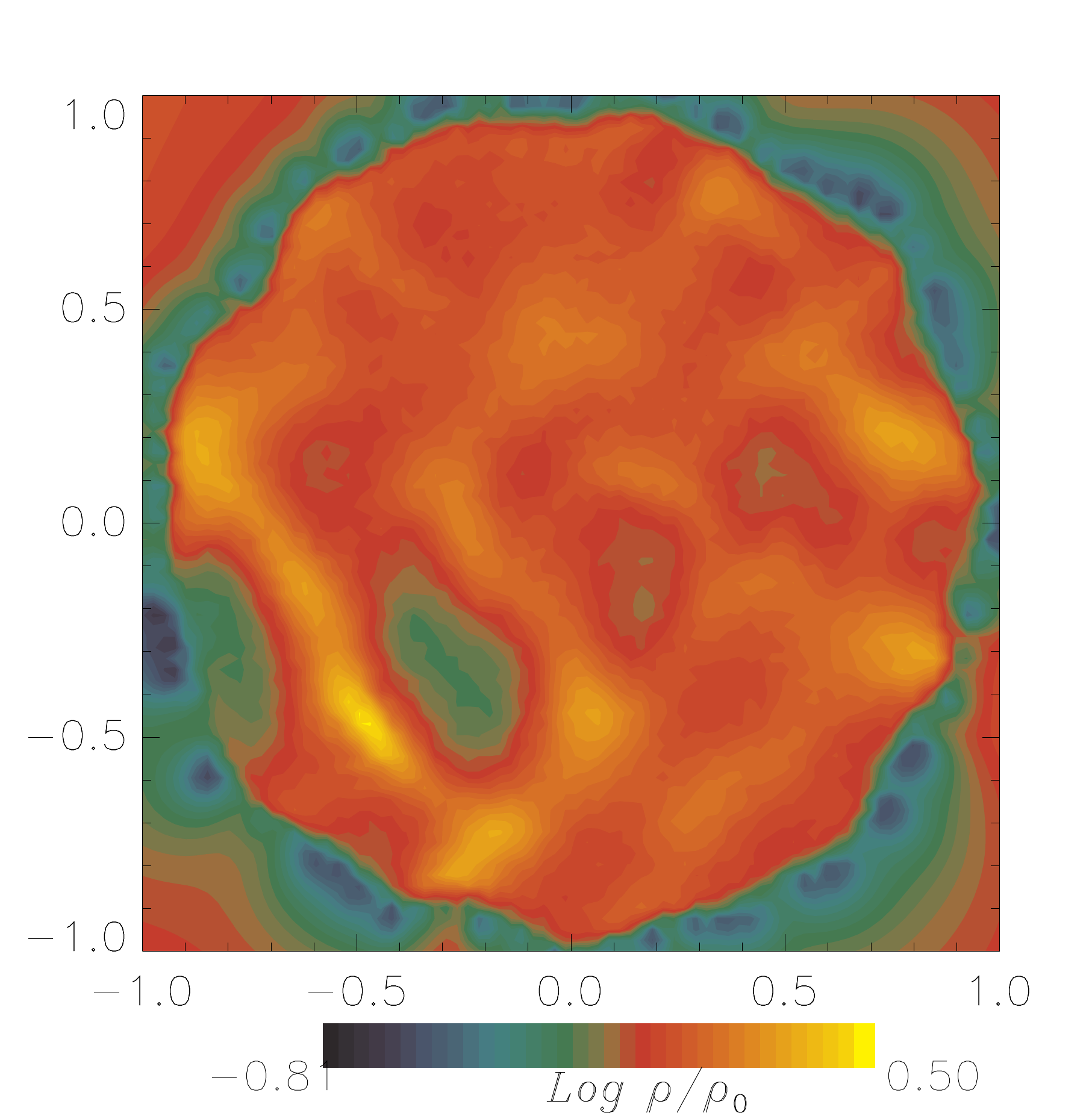} & \includegraphics[width=2 in]{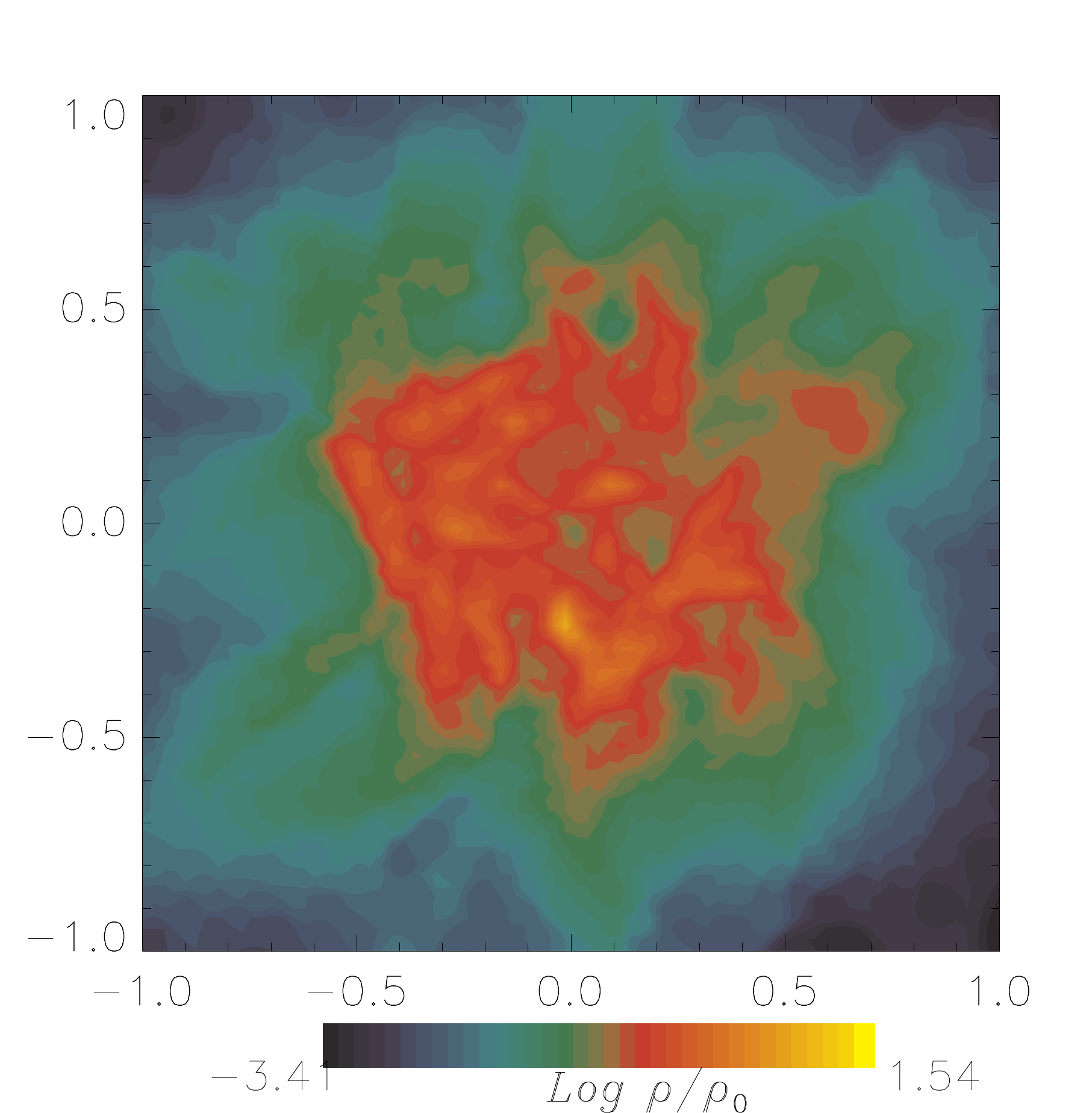} &
\includegraphics[width=2 in]{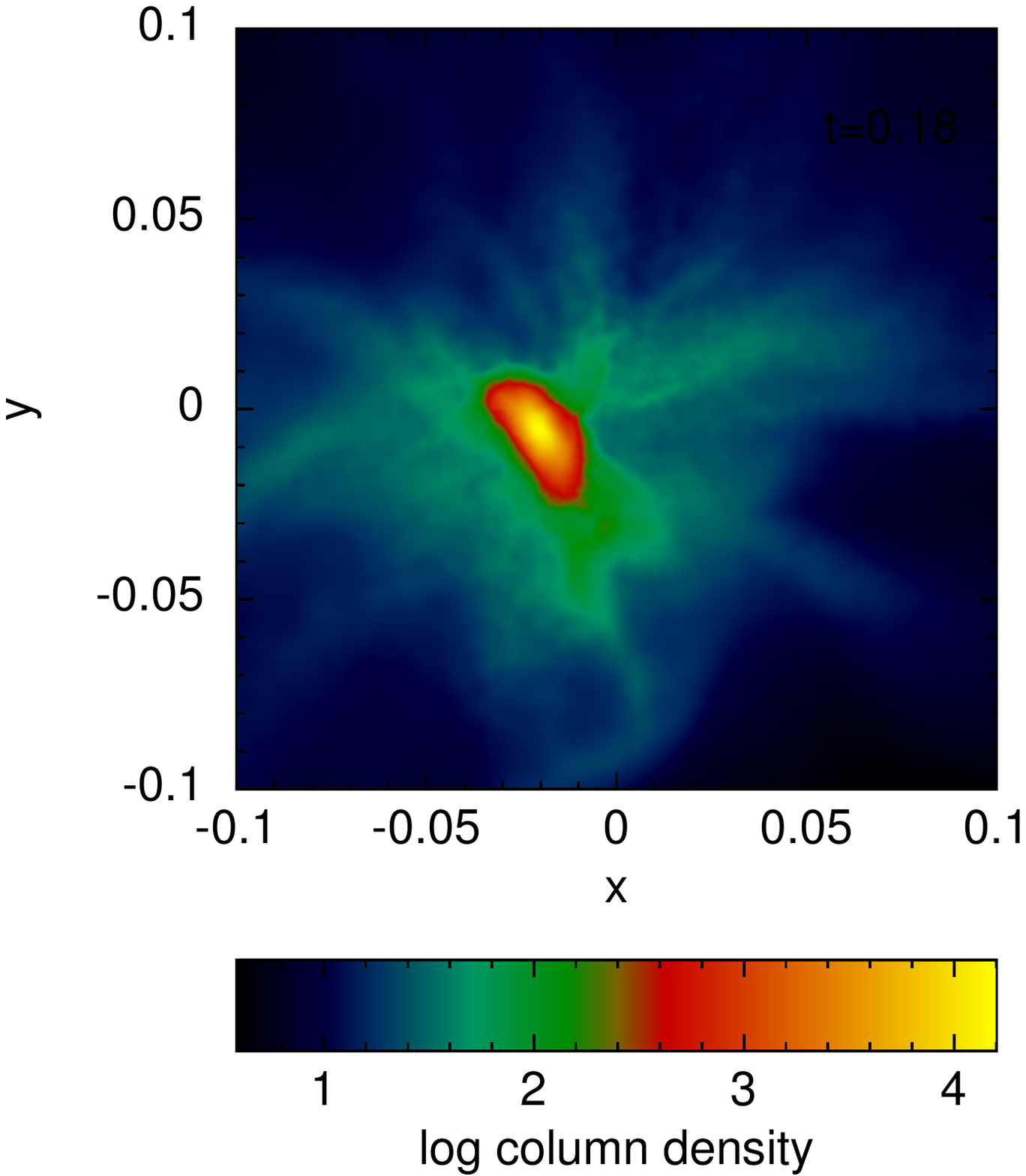}
\end{tabular}
\caption{\label{MosBs2rm} Iso-density plots for
model 2.}
\end{figure}

\begin{figure}
\begin{tabular}{ccc}
\includegraphics[width=2 in]{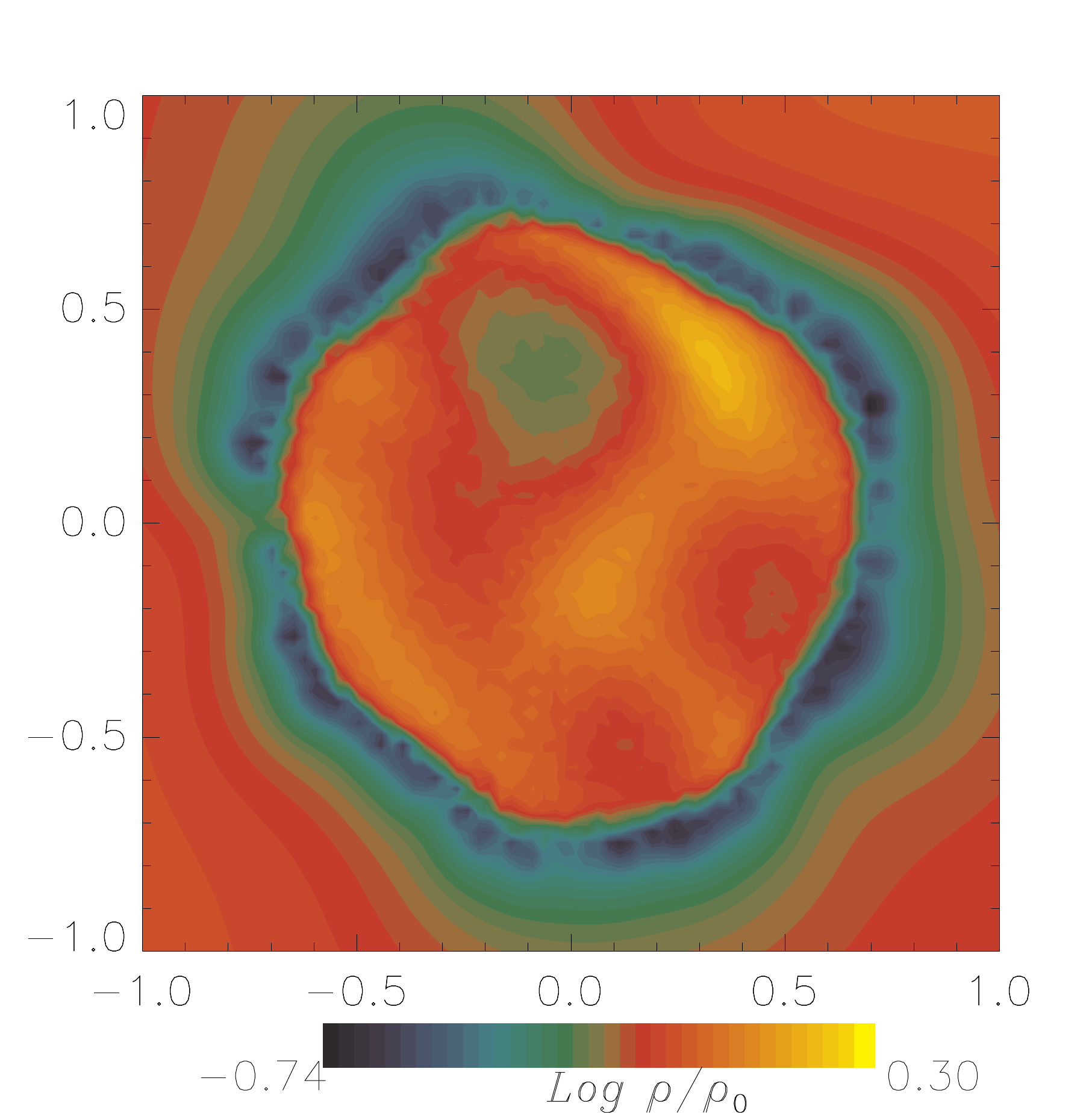} & \includegraphics[width=2 in]{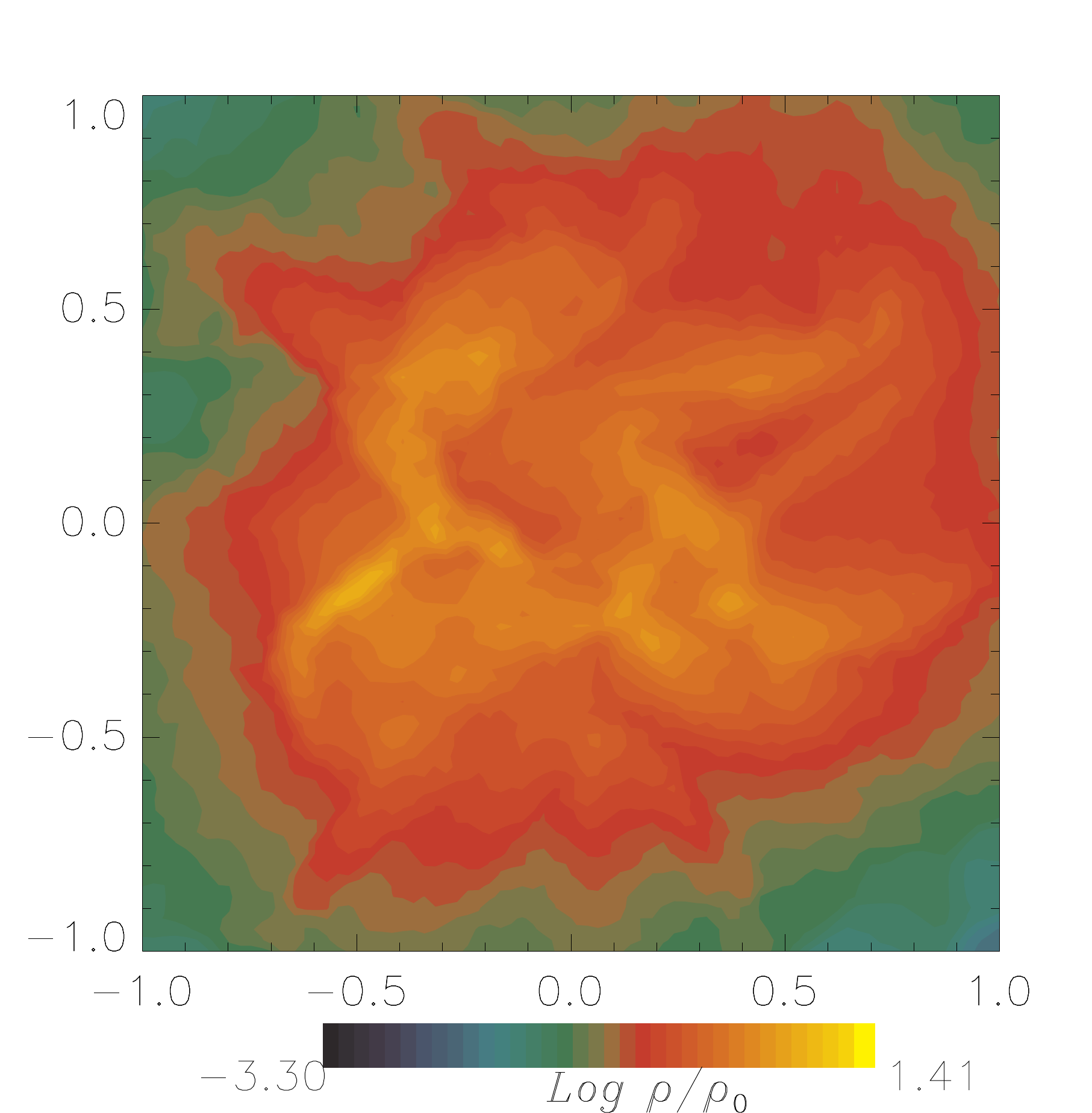} &
\includegraphics[width=2 in]{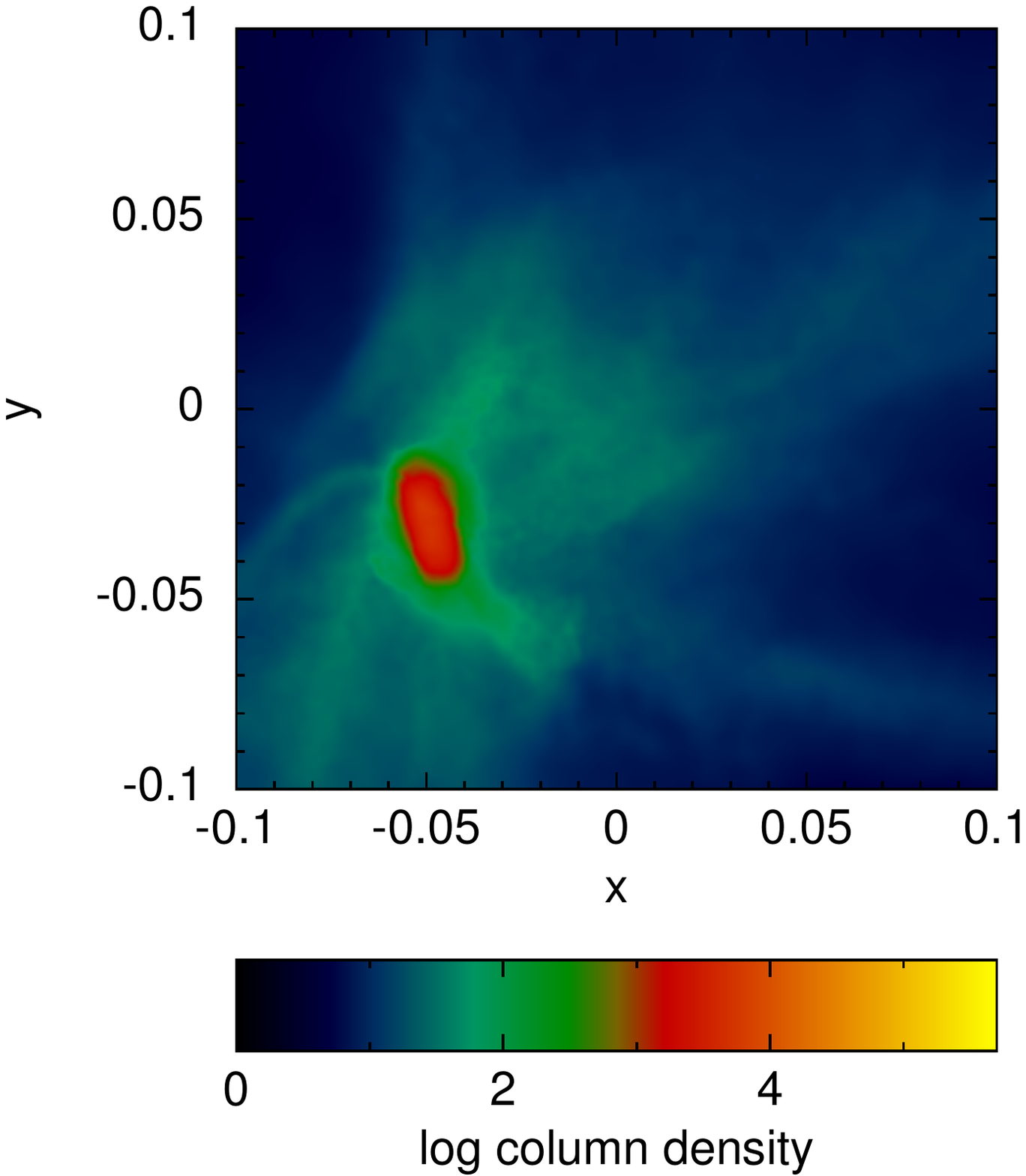}
\end{tabular}
\caption{\label{MosBs3rm} Iso-density plots for
model 3.}
\end{figure}

\begin{figure}
\begin{tabular}{ccc}
\includegraphics[width=2 in]{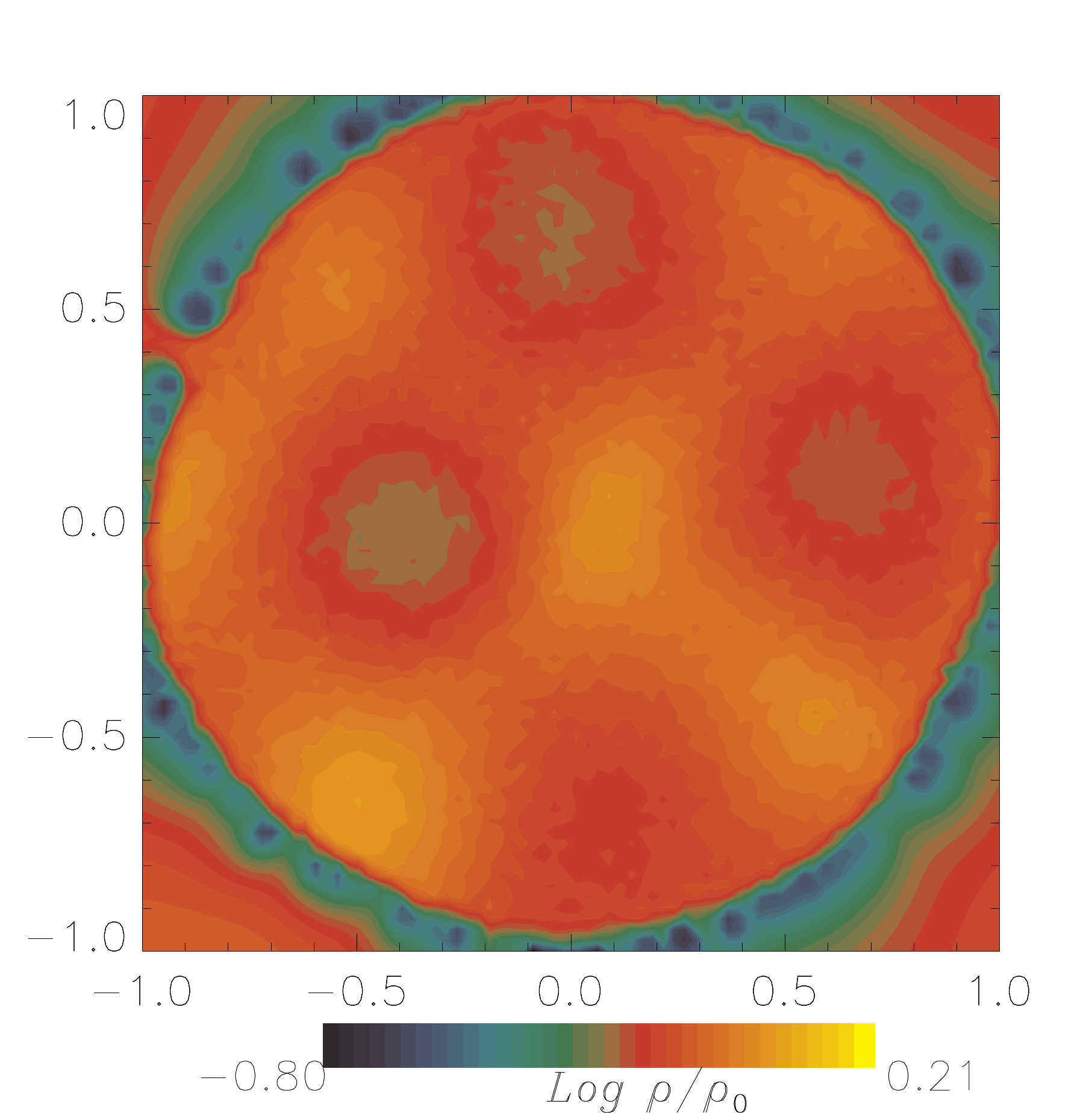} & \includegraphics[width=2 in]{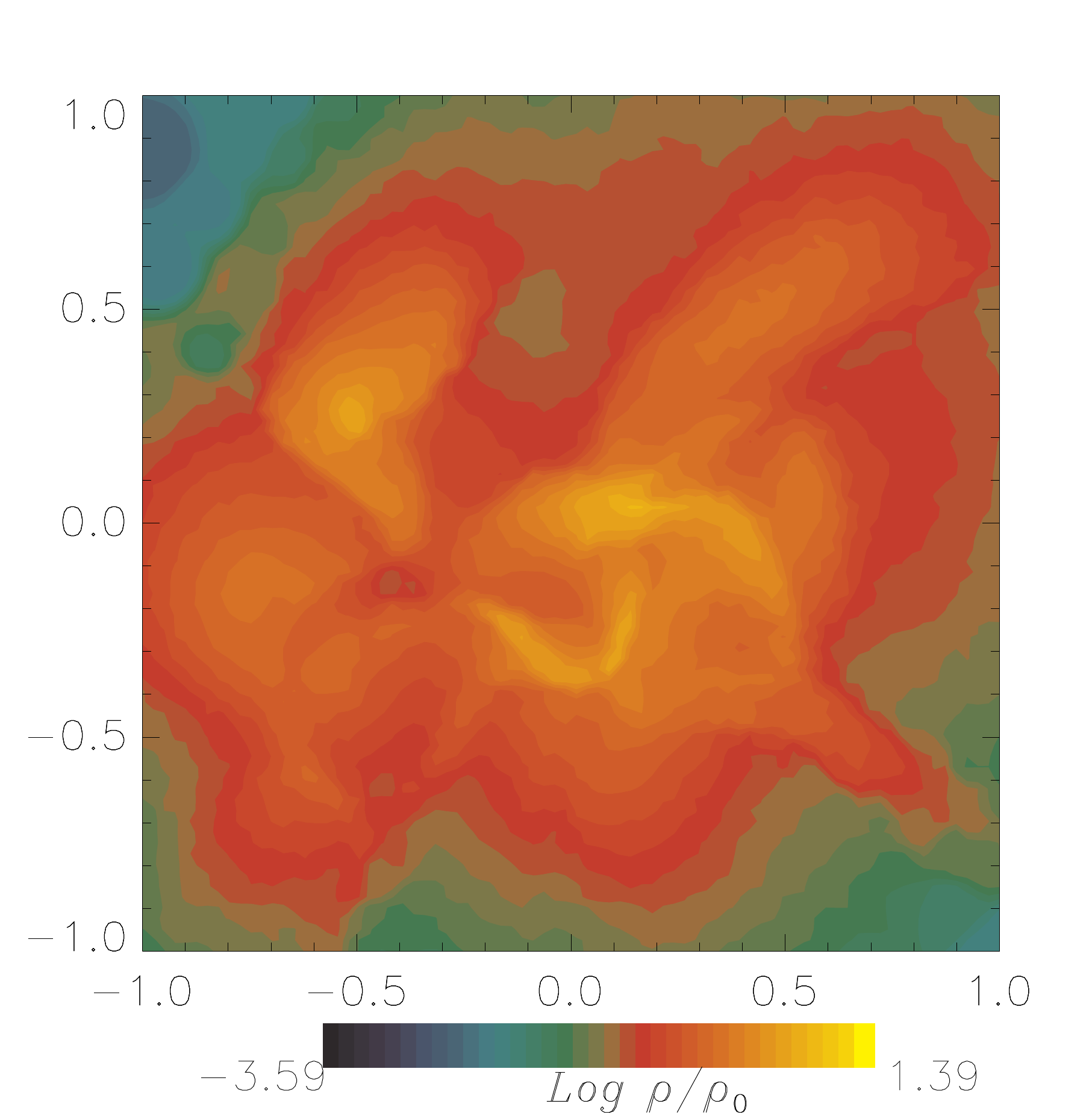} &
\includegraphics[width=2 in]{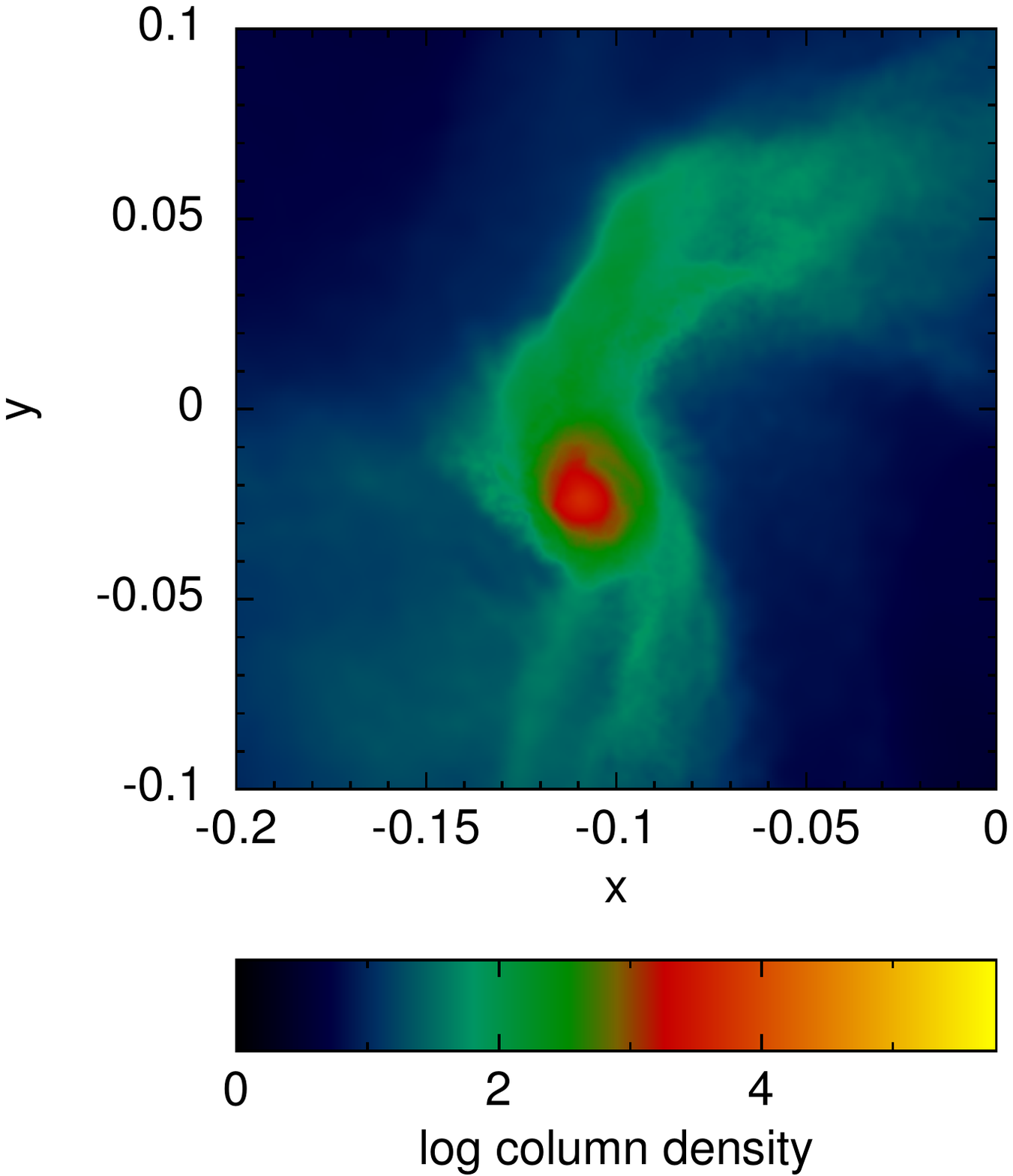}
\end{tabular}
\caption{\label{MosBs4rm} Iso-density plots for
model 4.}
\end{figure}

By looking at the left and middle panels of Figs.~\ref{MosBsrm},
\ref{MosBs2rm}, \ref{MosBs3rm}, and \ref{MosBs4rm}, which
correspond to models 1, 2, 3, and 4, we note that as the
wavelength of the initial perturbation increases, so do the sizes
of the initial over-density clumps, and the number of clumps
that form across the core decreases.

Due to the loss of homogeneity in the initial distribution of
the over-dense clumps, the protostar in formation accretes mass from the
surroundings in an anisotropic manner; as a consequence, the
protostar moves slightly from the center, in the south and southwest
directions, as can be seen in the right panels of
Figs.~\ref{MosBs3rm} and \ref{MosBs4rm}, which correspond to
models 3 and 4.

Fig.~\ref{evoluciondensidadDF} shows that for models 1--4, the larger
the initial perturbation size, the earlier the local peak in
the density curve occurs. Besides, we note that the peak velocity
in the collapsing region decreases as the size of the initial
perturbation increases; see column 5 of Table~\ref{tab:mod}.

\subsection{Models 5--8}
\label{subsec:results5-8}

\begin{figure}
\begin{tabular}{ccc}
\includegraphics[width=2 in]{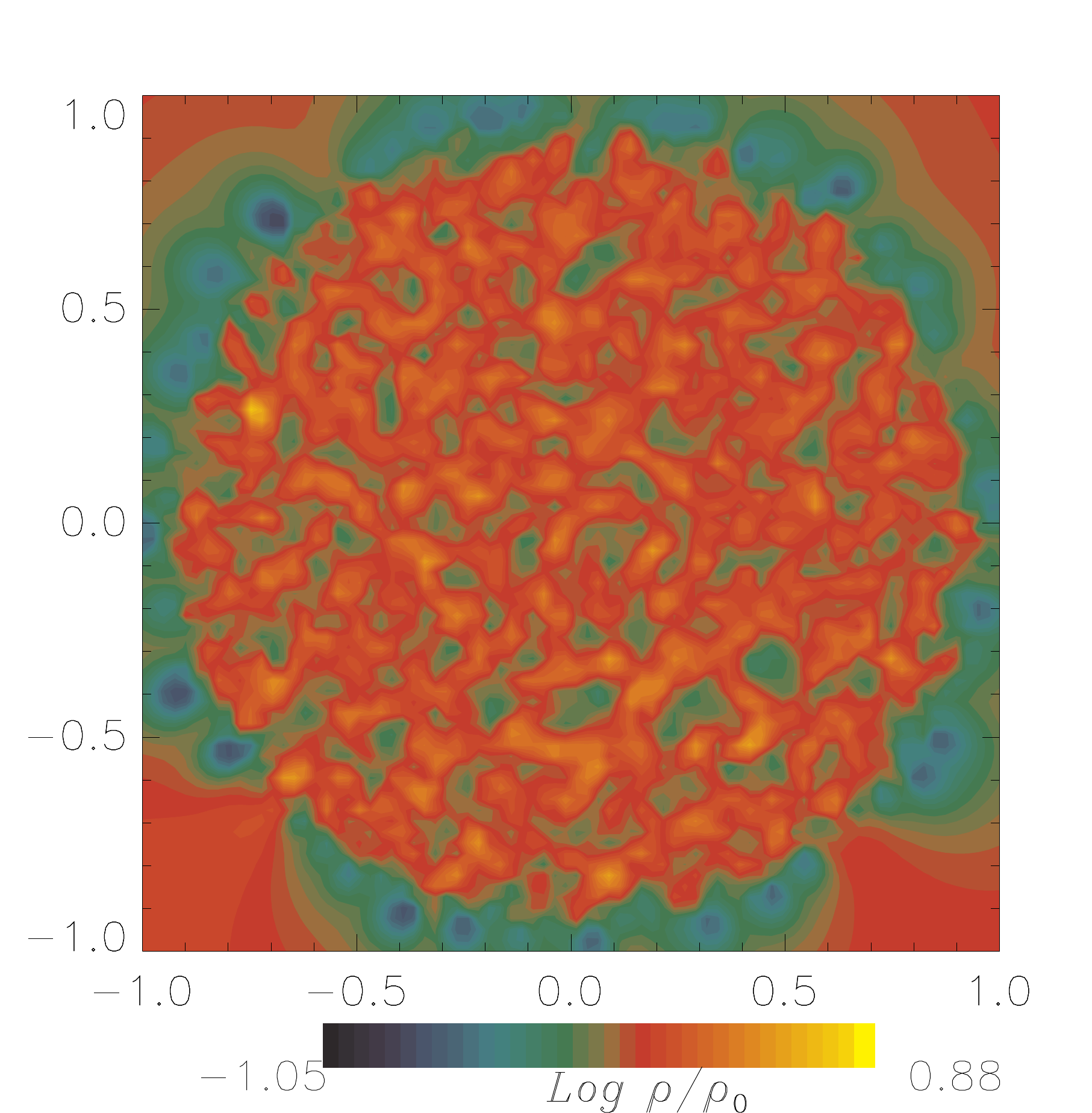} & \includegraphics[width=2 in]{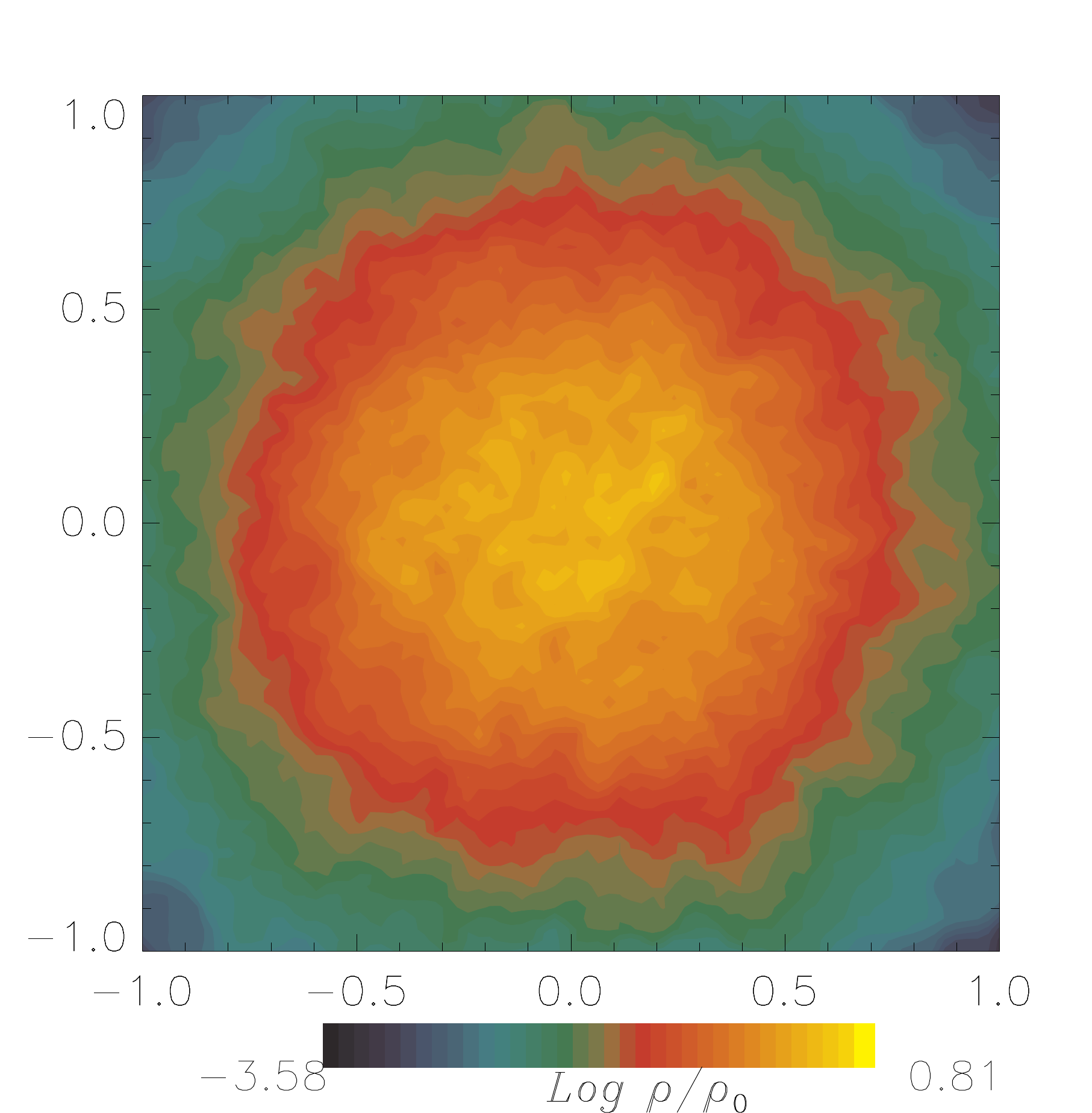} &
\includegraphics[width=2 in]{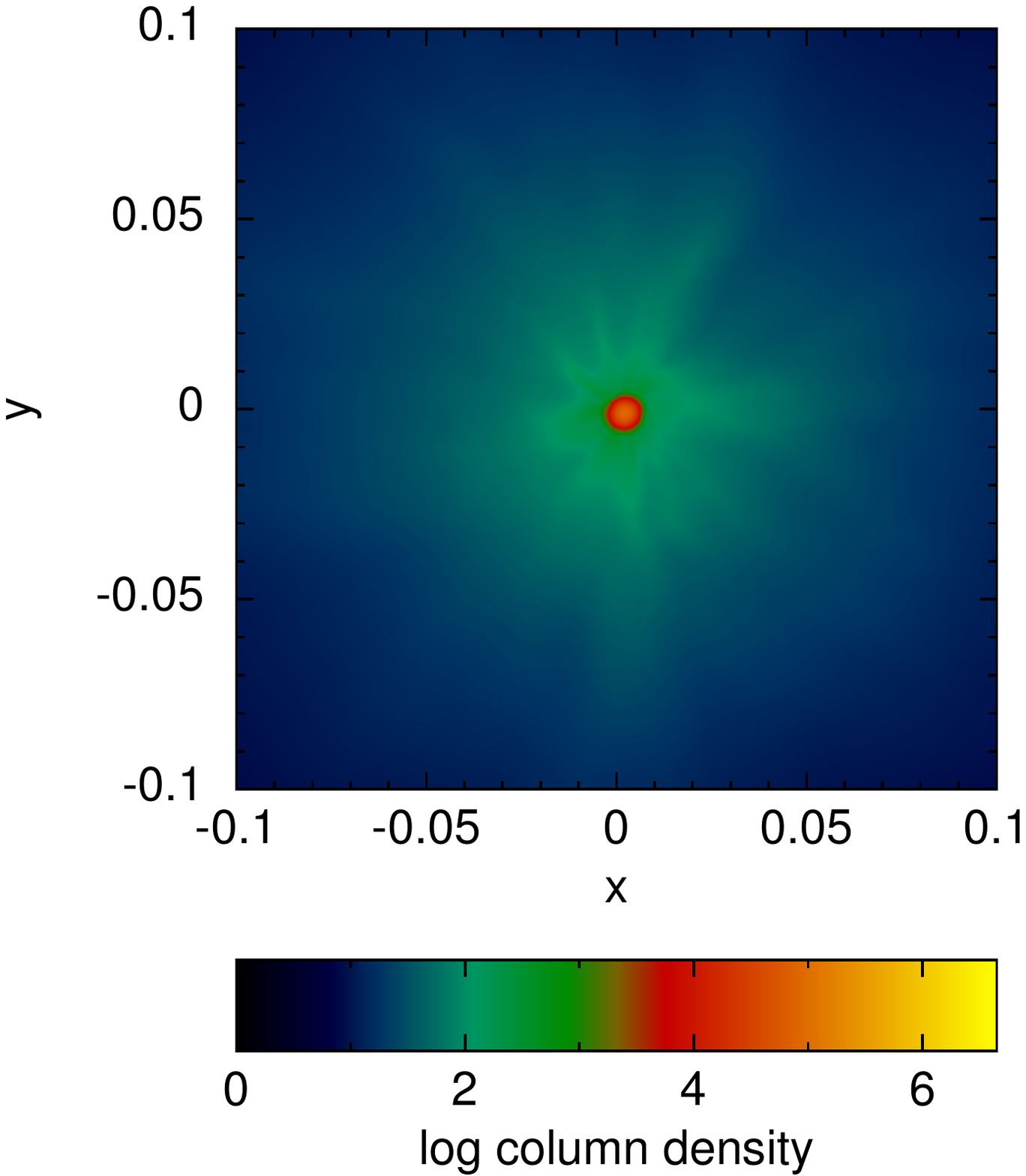}
\end{tabular}
\caption{\label{MosB2srmp}Iso-density  plots for
model 5.}
\end{figure}

\begin{figure}
\begin{tabular}{ccc}
\includegraphics[width=2 in]{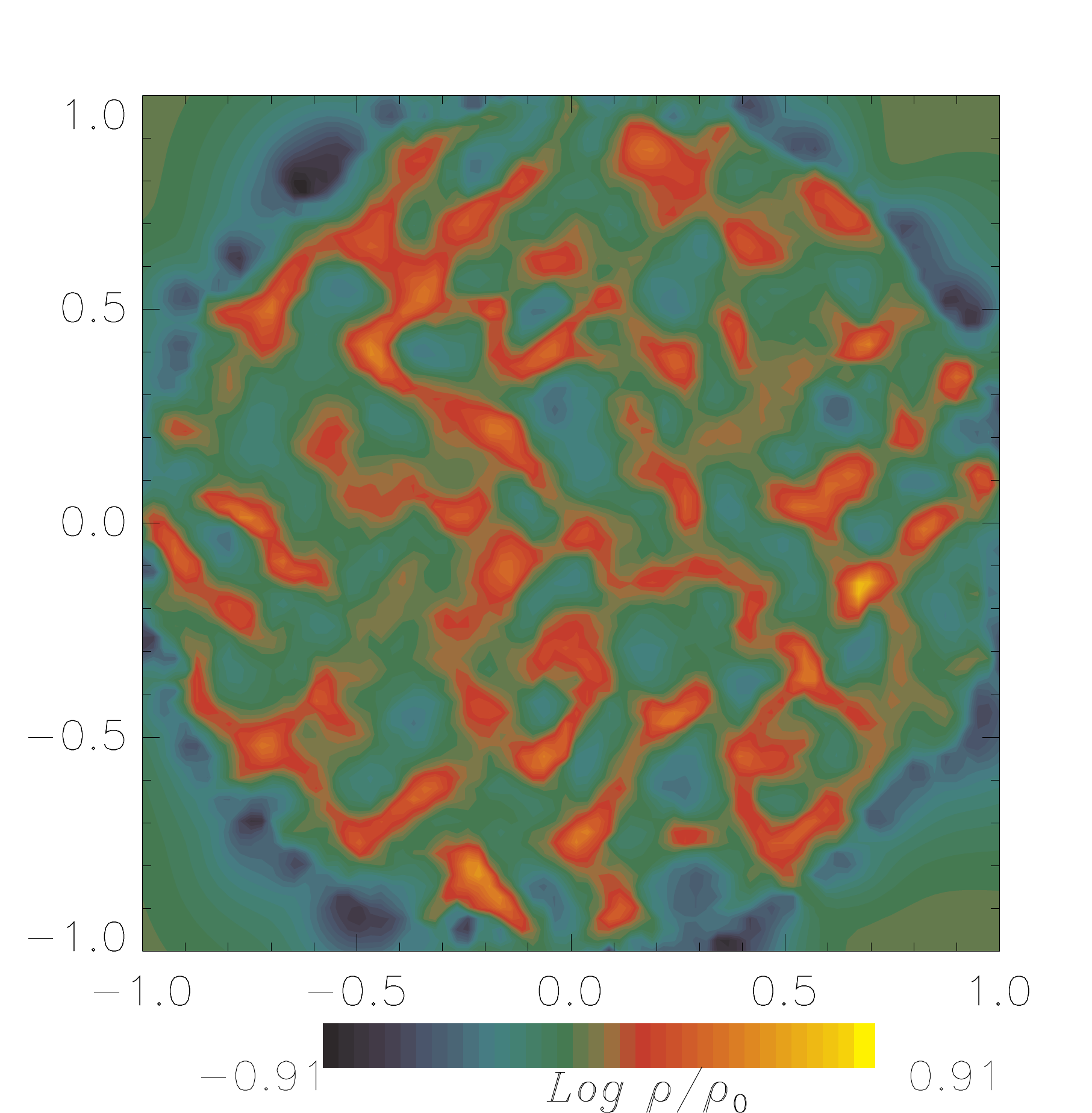} & \includegraphics[width=2 in]{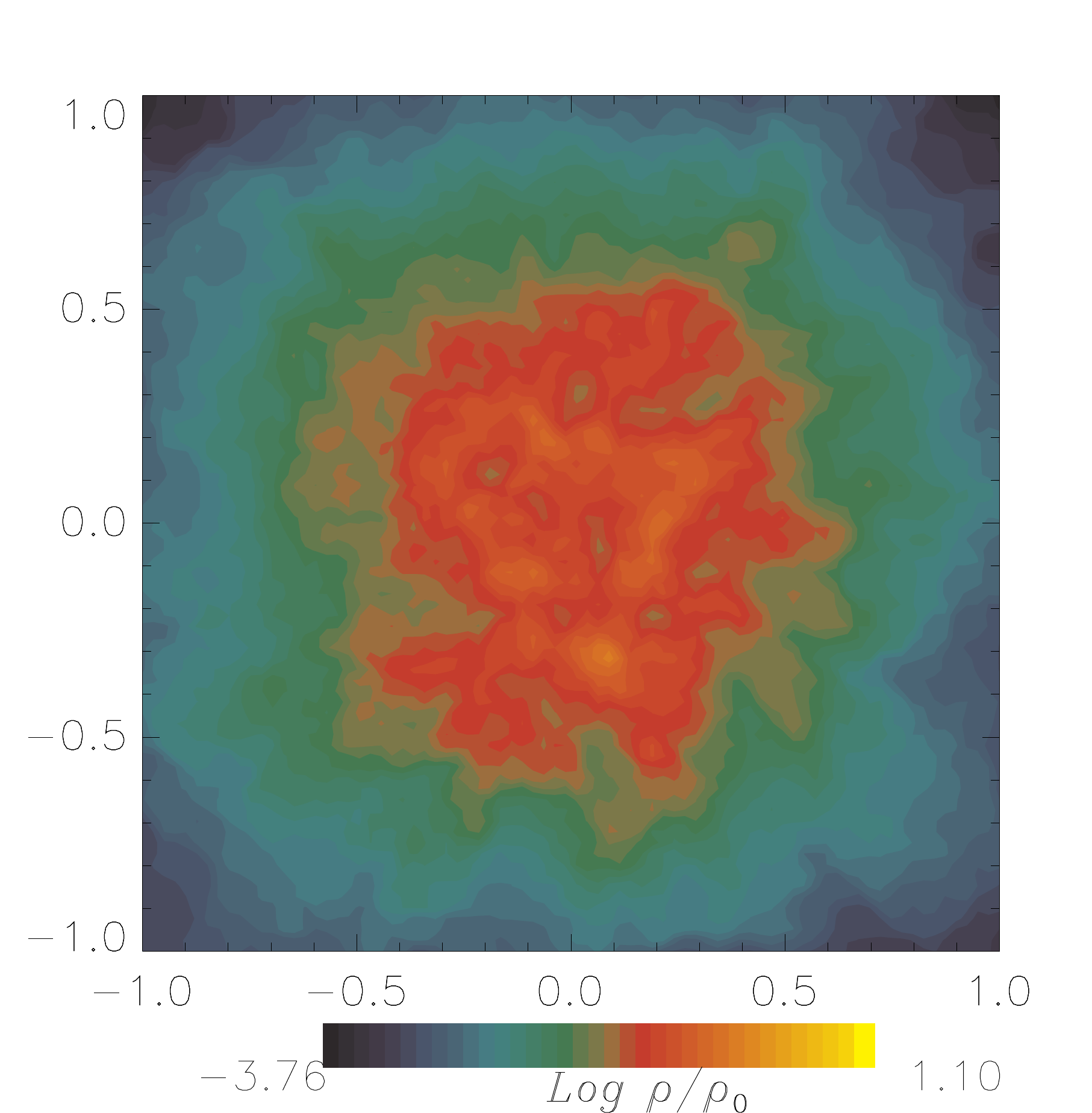} &
\includegraphics[width=2 in]{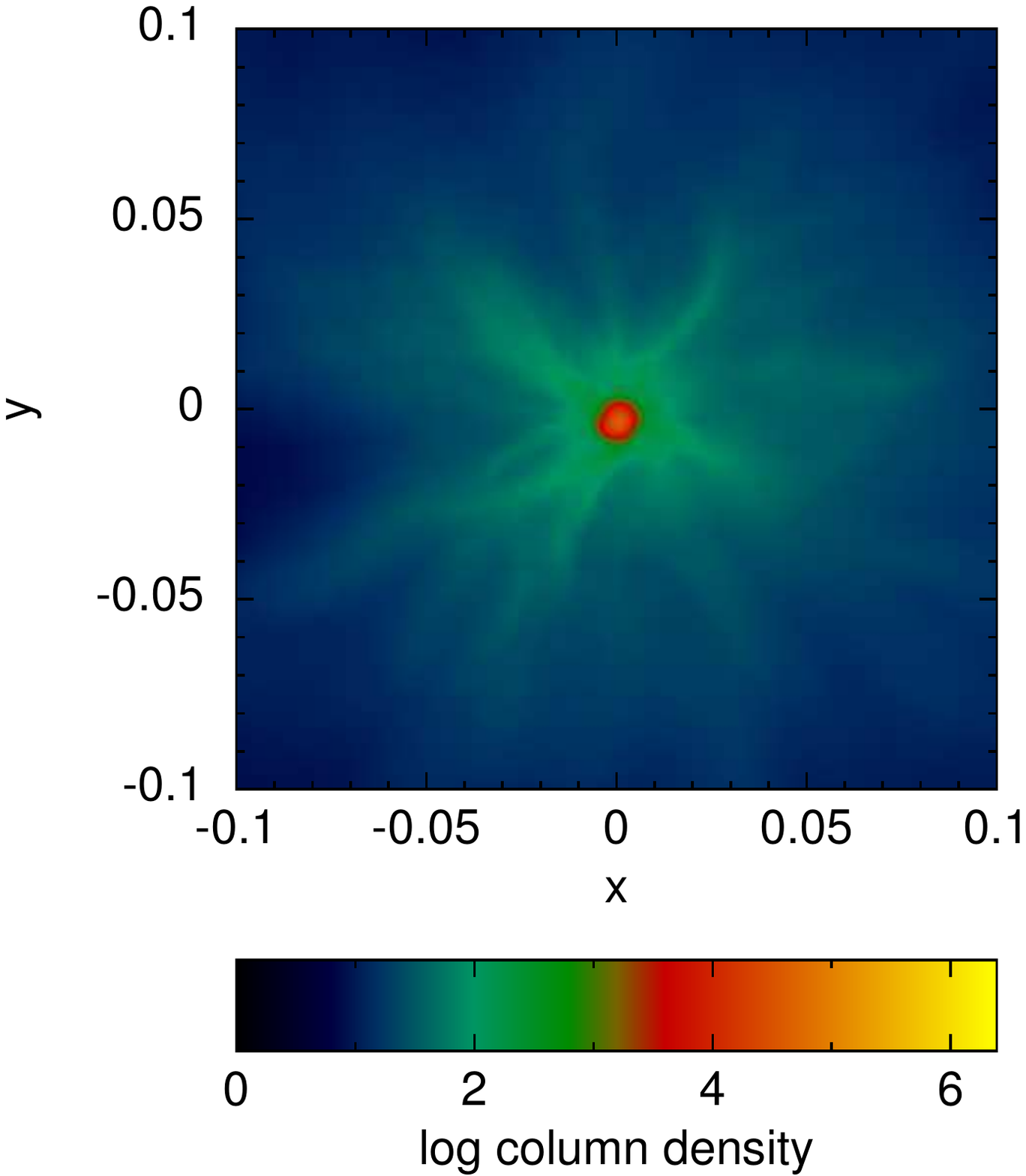}
\end{tabular}
\caption{\label{MosB2s2rmp} Iso-density plots for
model 6.}
\end{figure}

Here we re-calculate models 1--4, keeping all their parameters
unchanged except the number of grid elements of the Fourier mesh;
see column 3 of Table~\ref{tab:mod}. We first note that the number of
small over-density clumps formed at the initial snapshots increases,
but they are still homogeneously distributed across the core volume;
see the left panels of Figs.~\ref{MosB2srmp} and \ref{MosB2s2rmp} and
compare them with those of models 2--3.

\begin{figure}
\begin{tabular}{ccc}
\includegraphics[width=2 in]{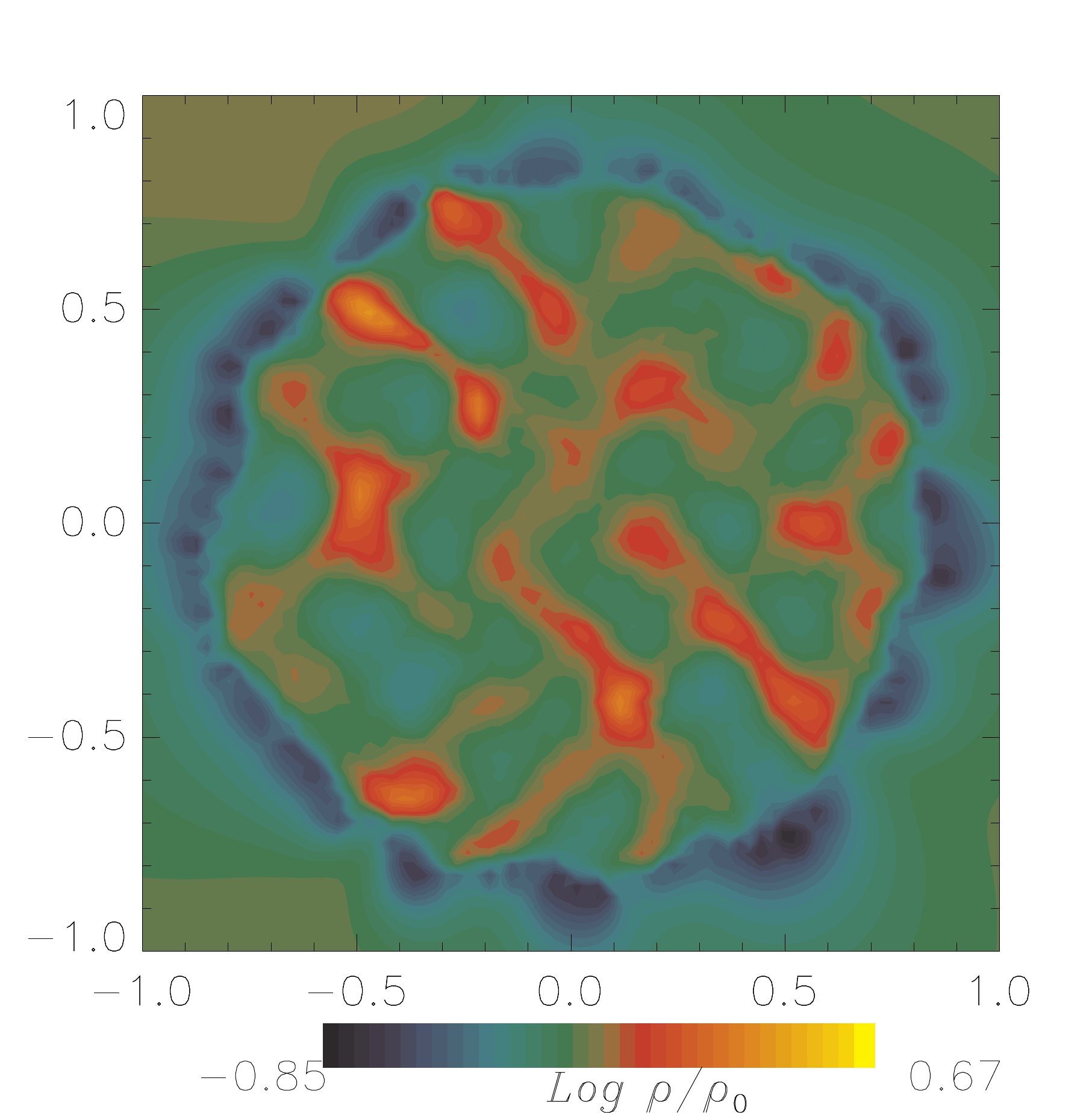} & \includegraphics[width=2 in]{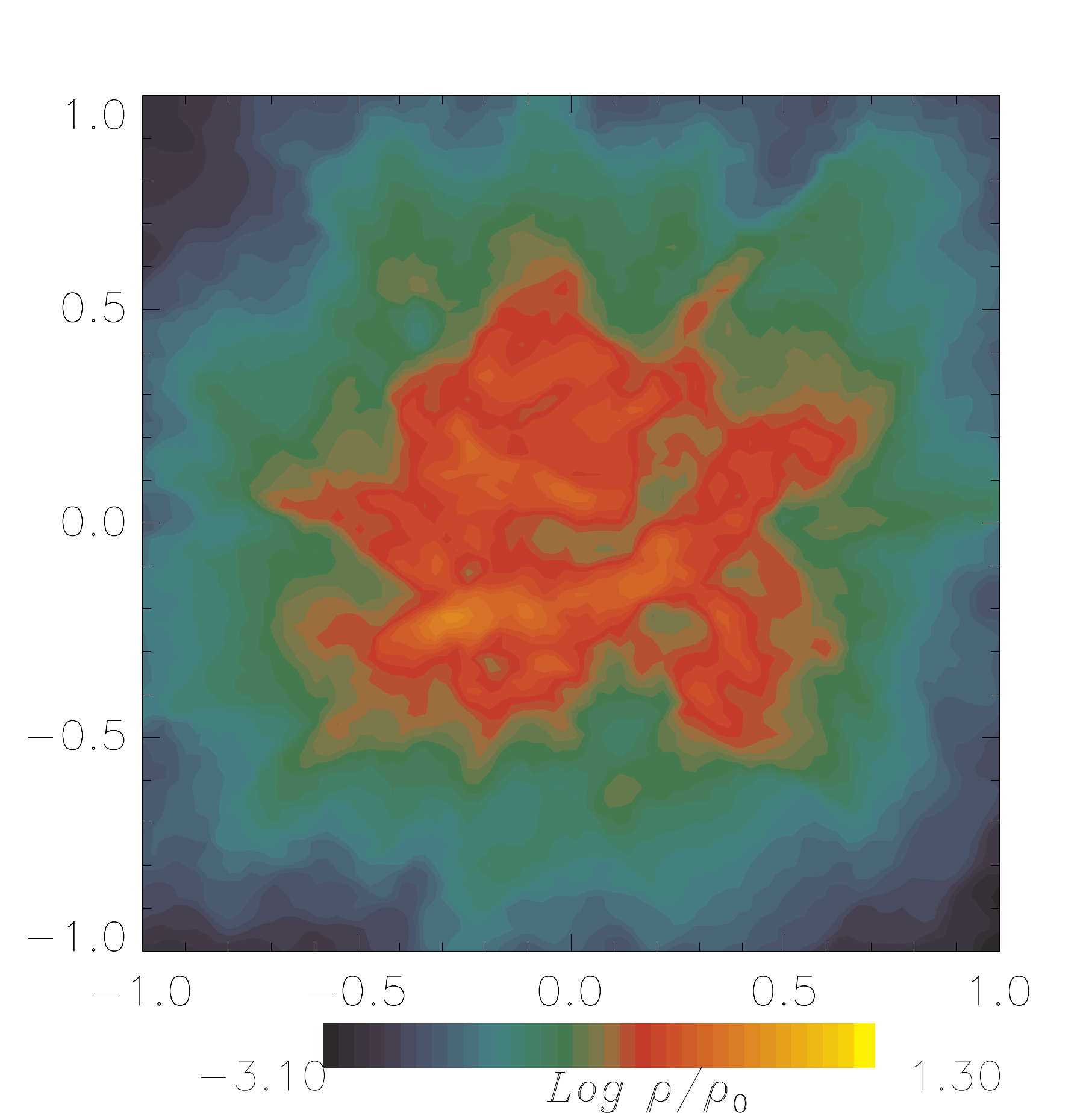} &
\includegraphics[width=2 in]{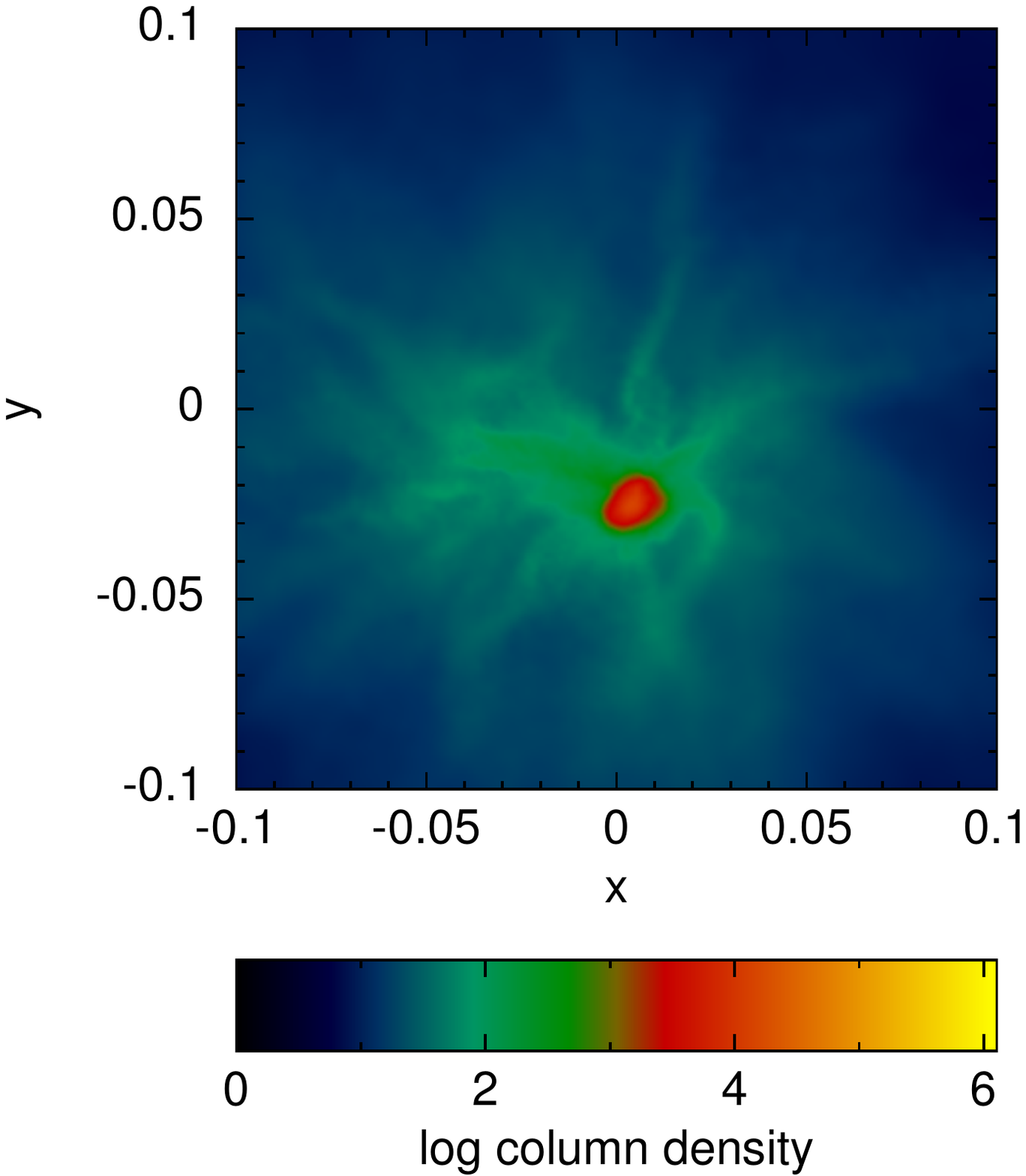}
\end{tabular}
\caption{\label{MosB2s3rmp} Iso-density plots for model 7.}
\end{figure}

The loss of homogeneity in the initial distribution of over-density clumps
is noticeable in model 7, as can be seen in Fig.~\ref{MosB2s3rmp}. A
significant reduction in the number of the initial over-density
clumps can also be seen in the initial snapshot of model 8; see the
left panel of Fig.~\ref{MosB2s3rmp}. However, by comparing this panel
with that of model 4, shown in Fig.~\ref{MosBs4rm}, we see that there
are a very few more over-density clumps than those formed in model
4.
\subsection{Models 9--11}
\label{subsec:results9-11}

\begin{figure}
\begin{tabular}{ccc}
\includegraphics[width=2 in]{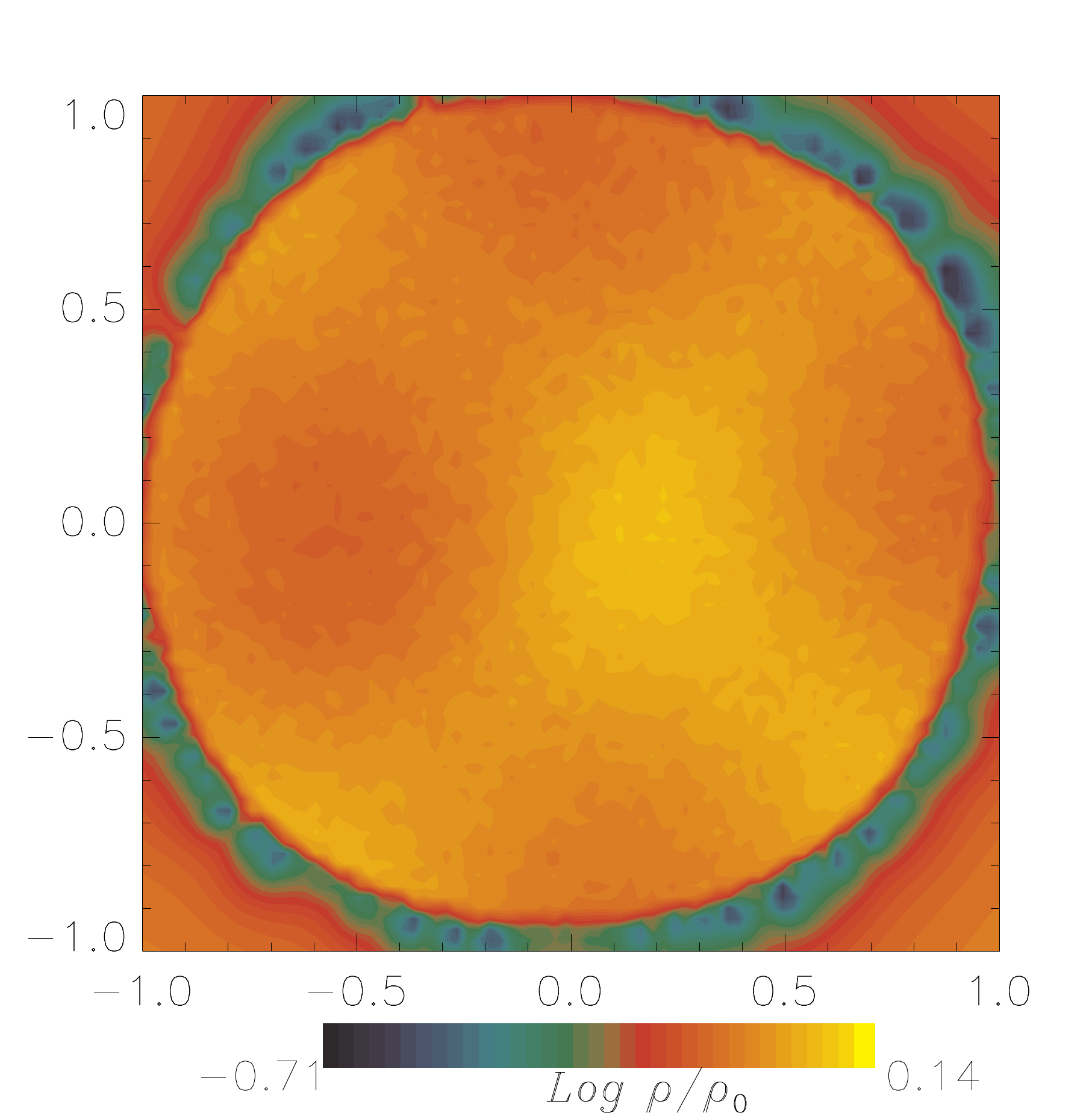} & \includegraphics[width=2 in]{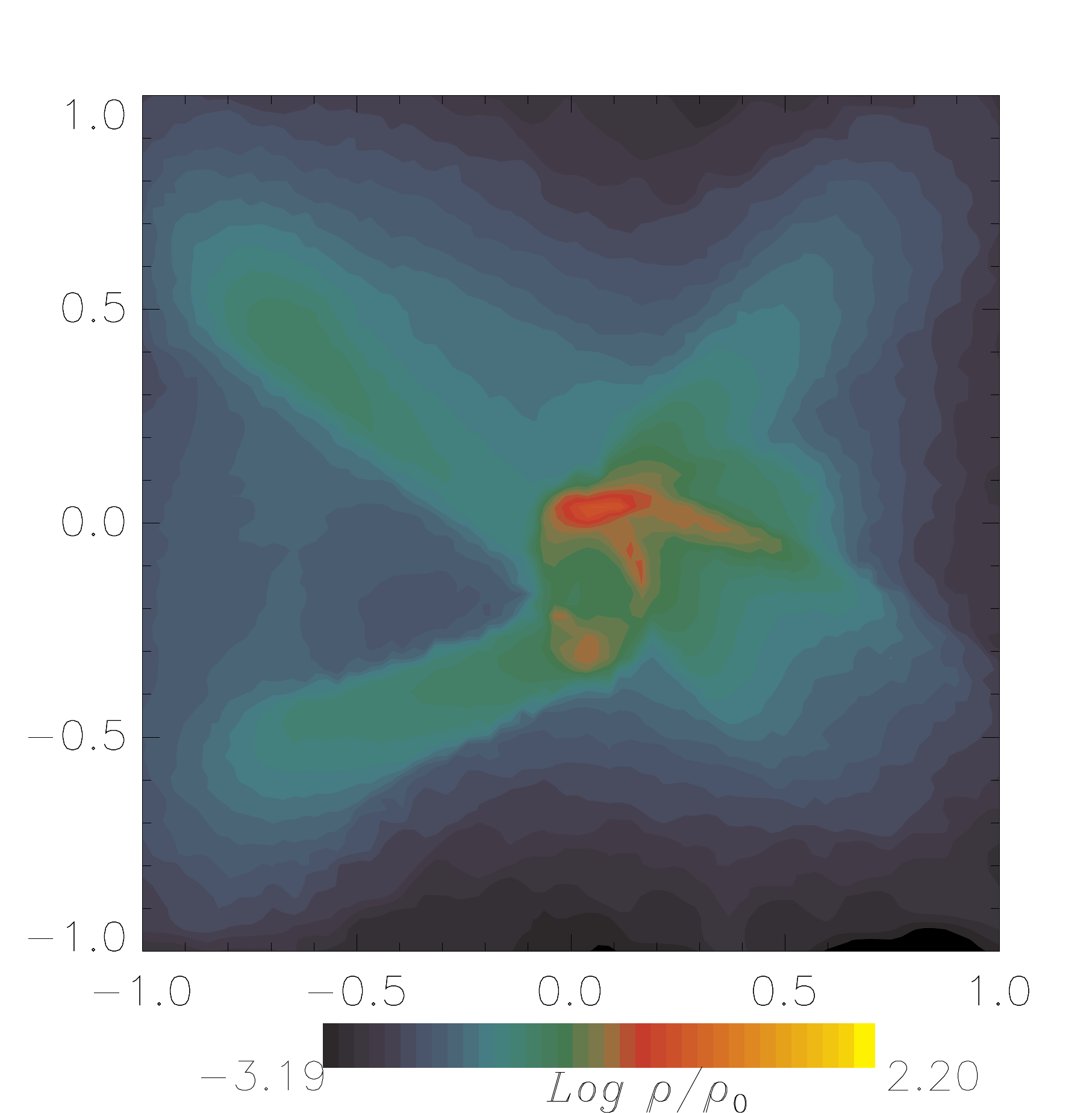} &
\includegraphics[width=2 in]{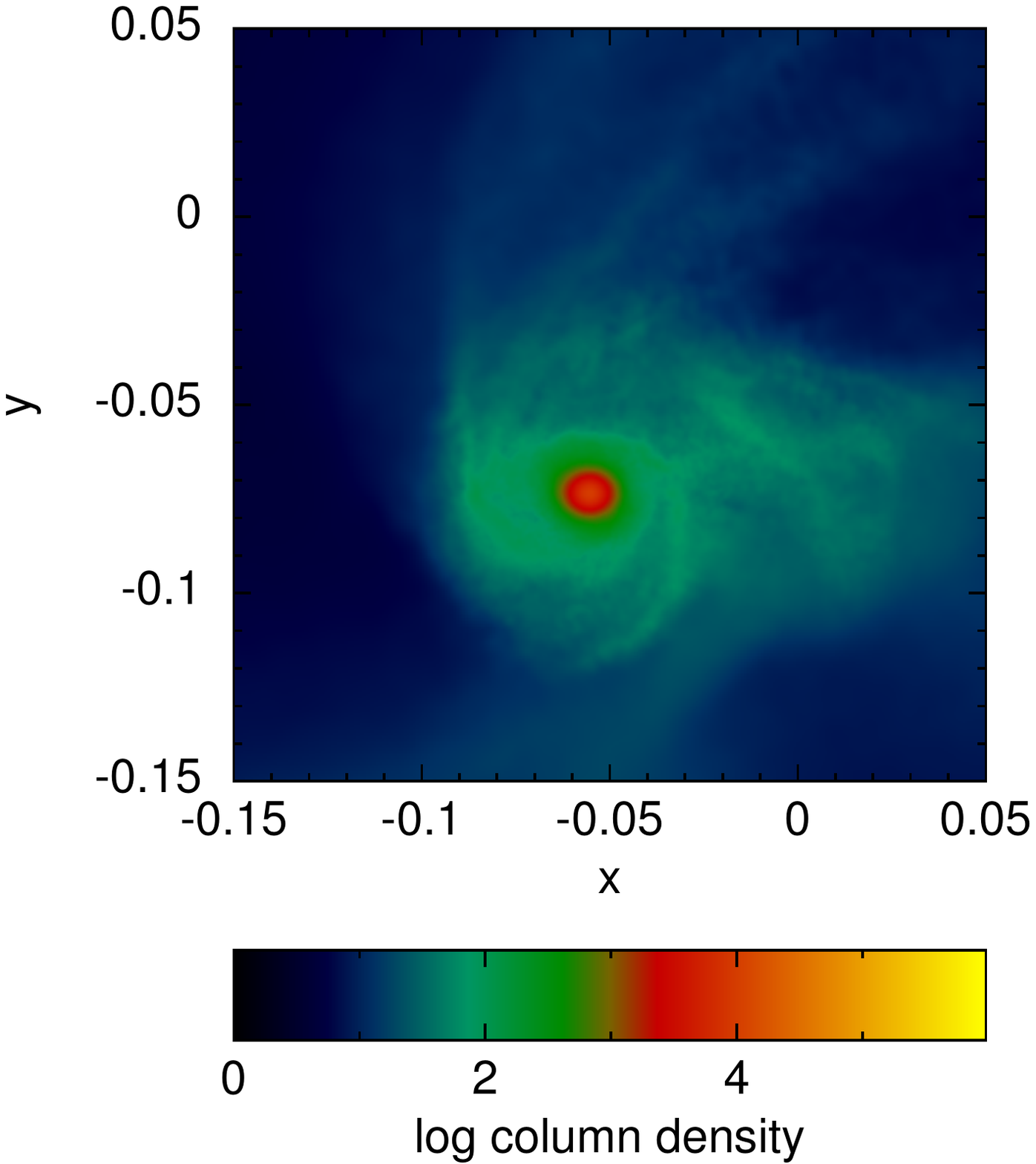}
\end{tabular}
\caption{\label{MosB-Osrm} Iso-density plots for
model 9.}
\end{figure}

\begin{figure}
\begin{tabular}{ccc}
\includegraphics[width=2 in]{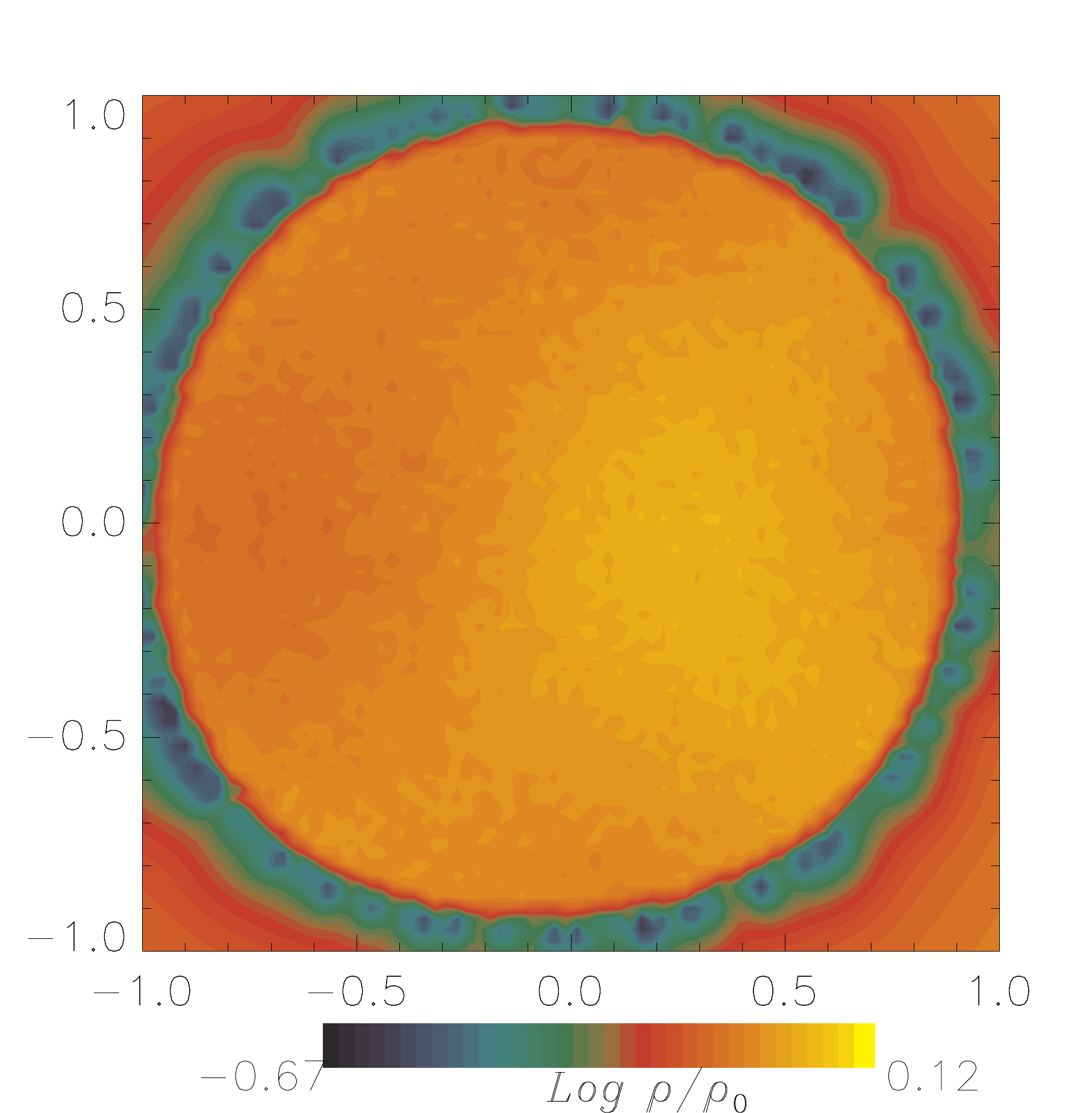} & \includegraphics[width=2 in]{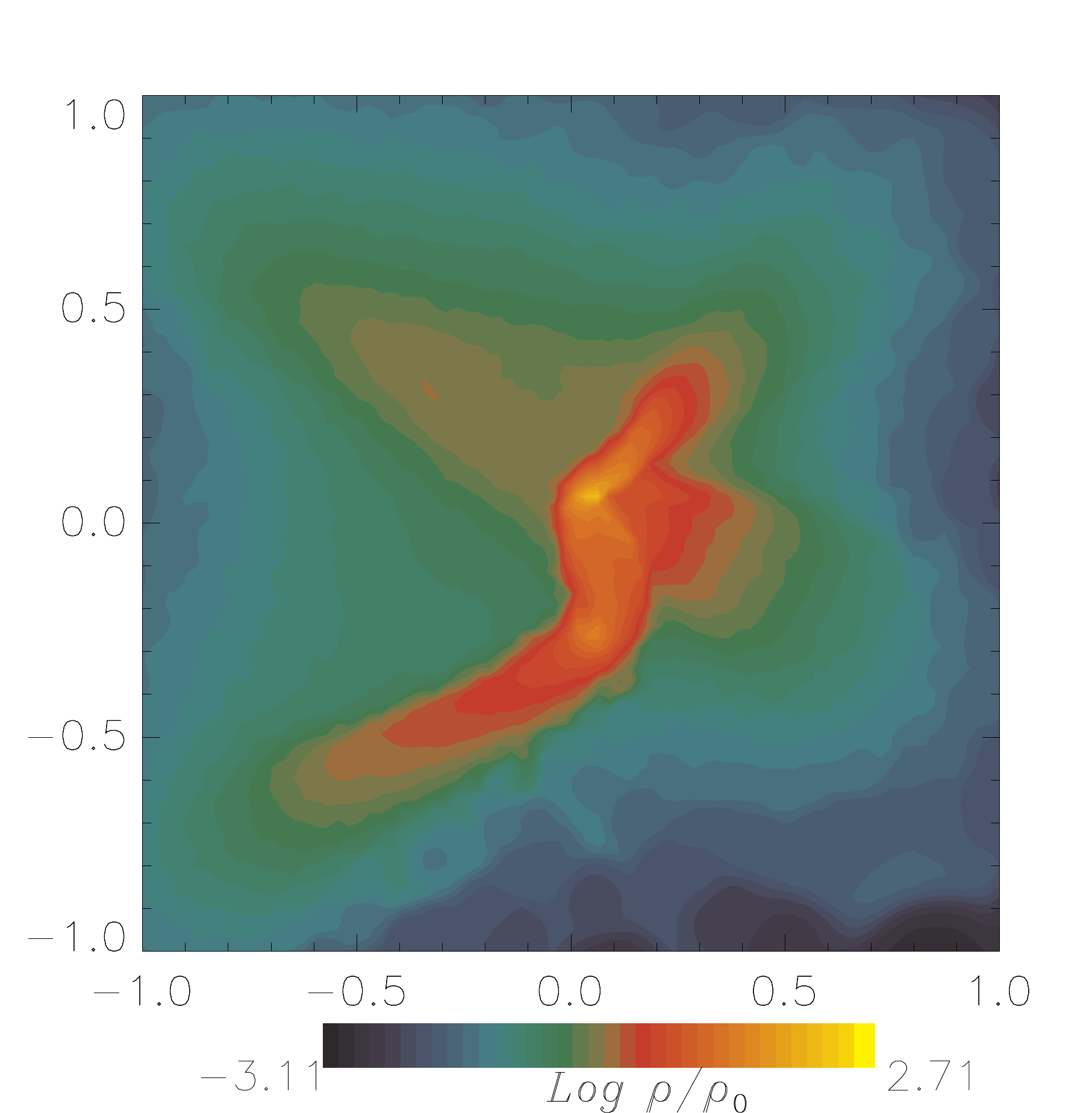} &
\includegraphics[width=2 in]{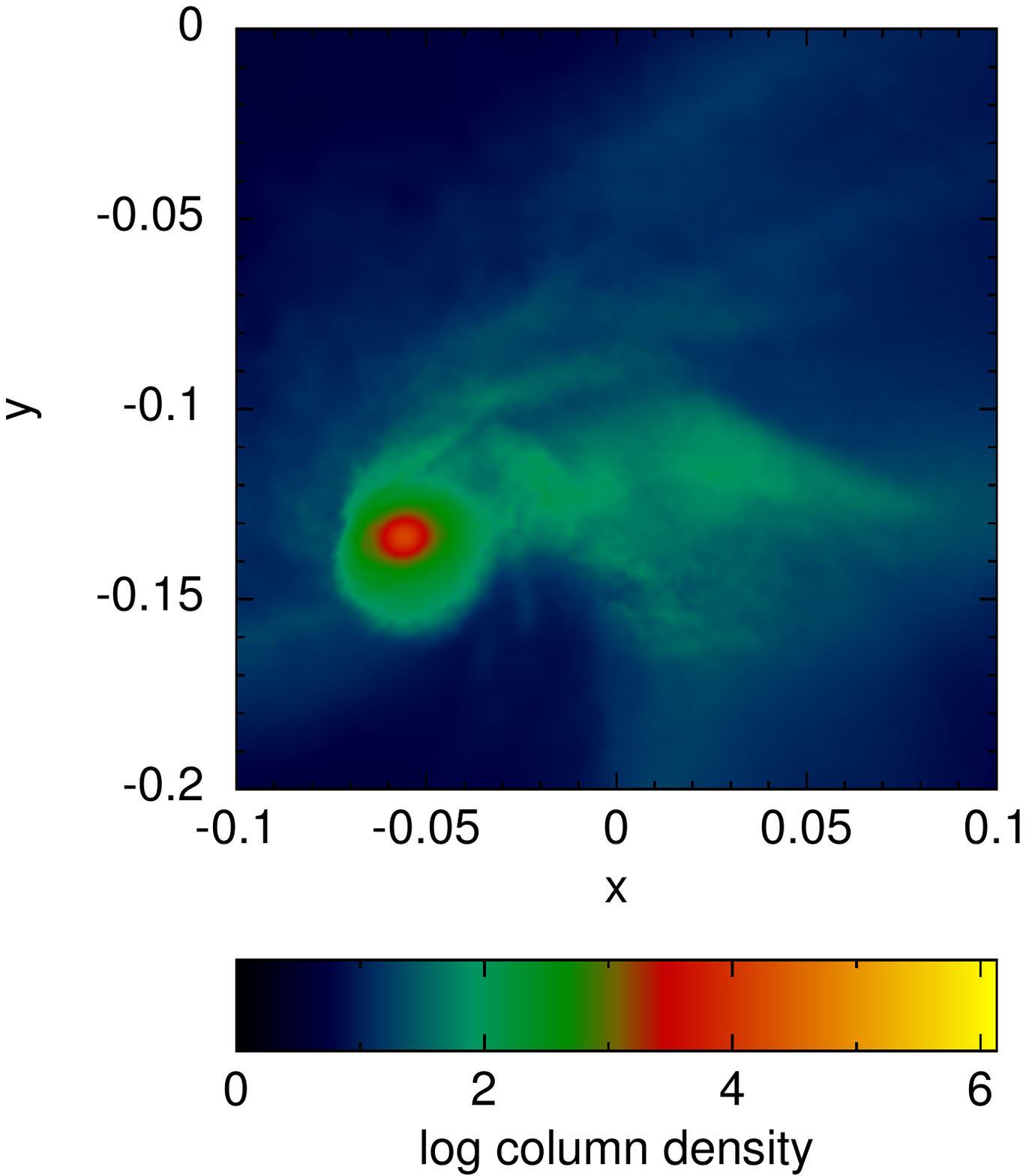}
\end{tabular}
\caption{\label{MosB-O2srm} Iso-density plots for
model 10.}
\end{figure}

\begin{figure}
\begin{tabular}{ccc}
\includegraphics[width=2 in]{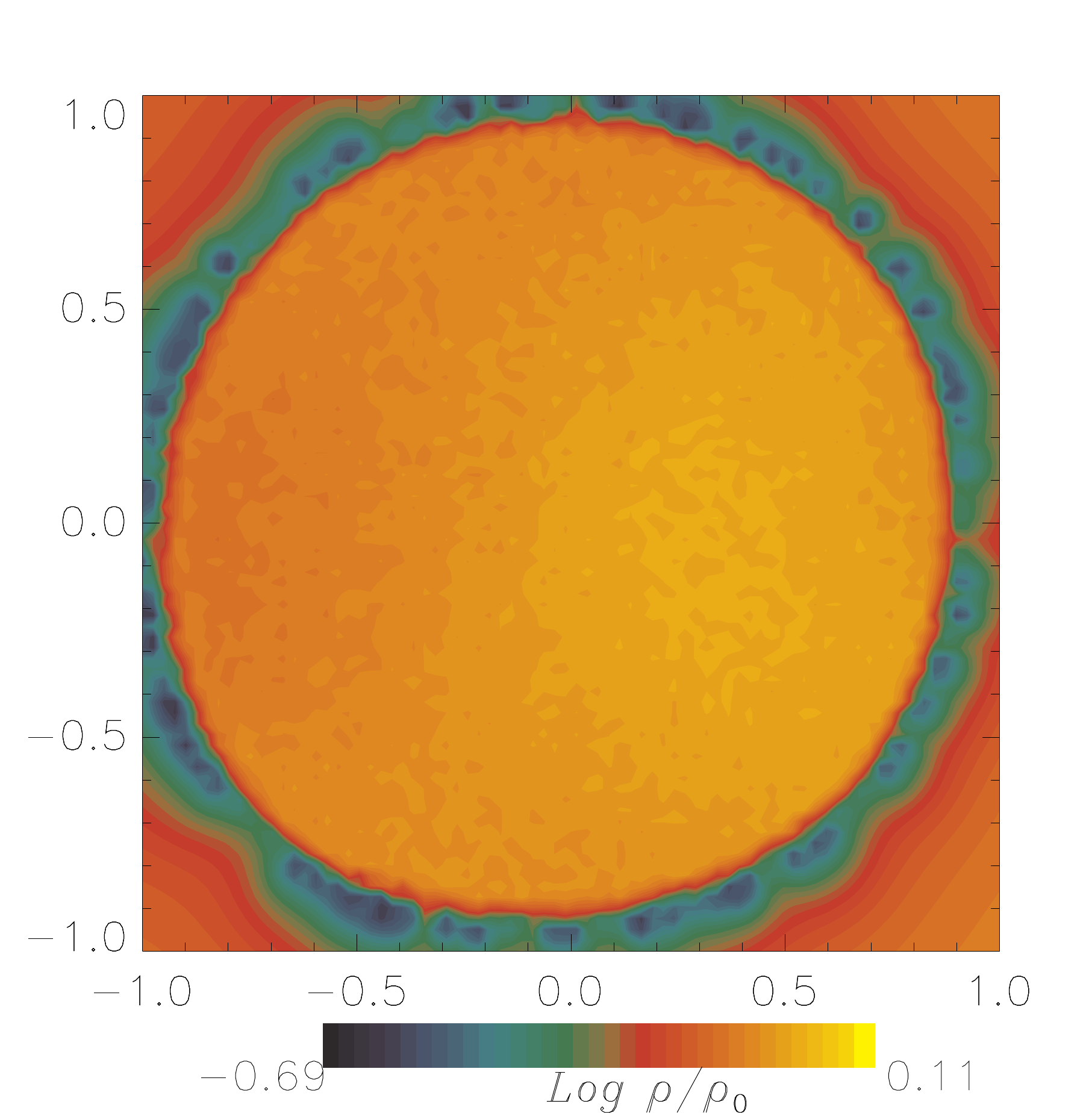} & \includegraphics[width=2 in]{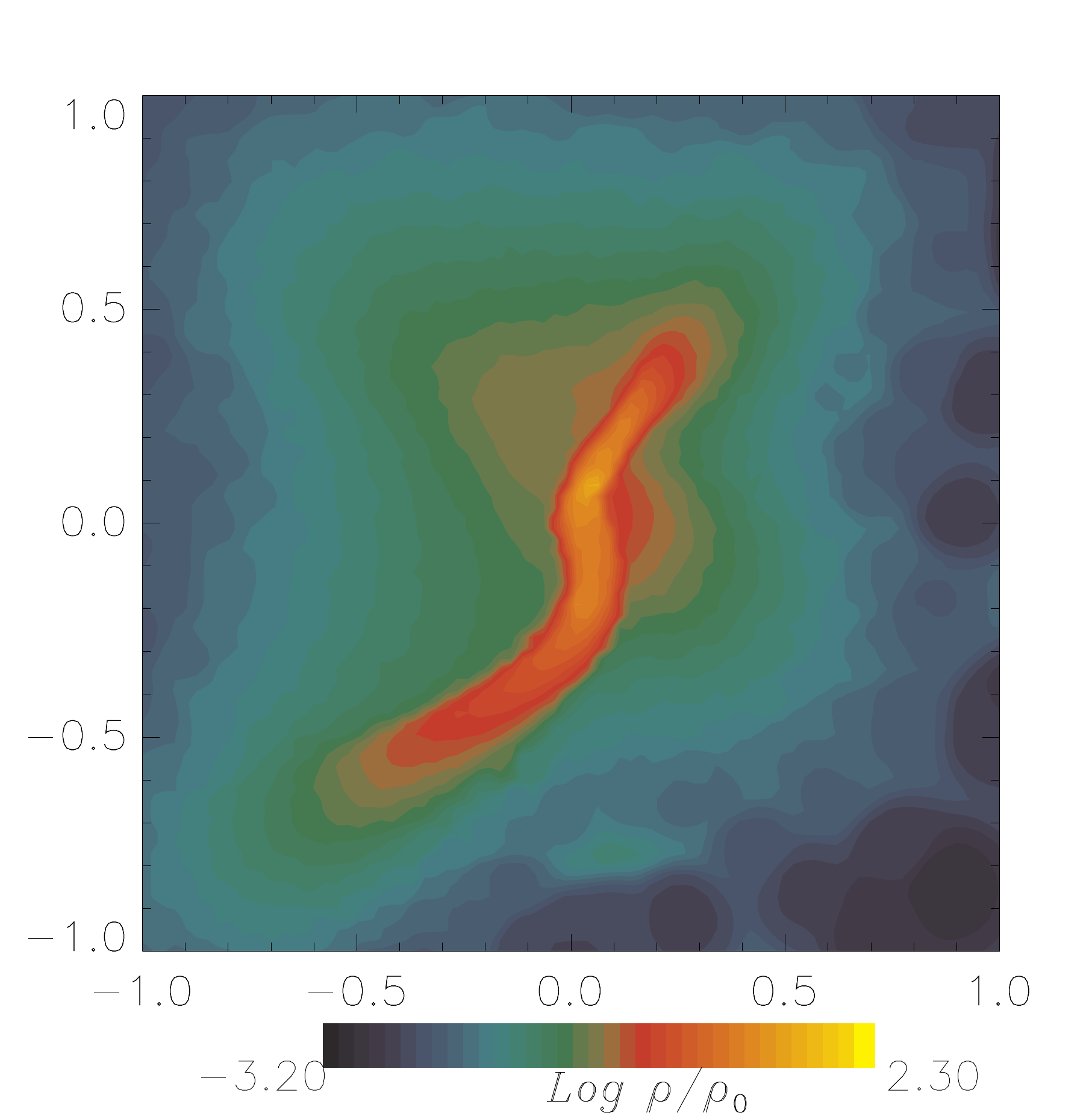} &
\includegraphics[width=2 in]{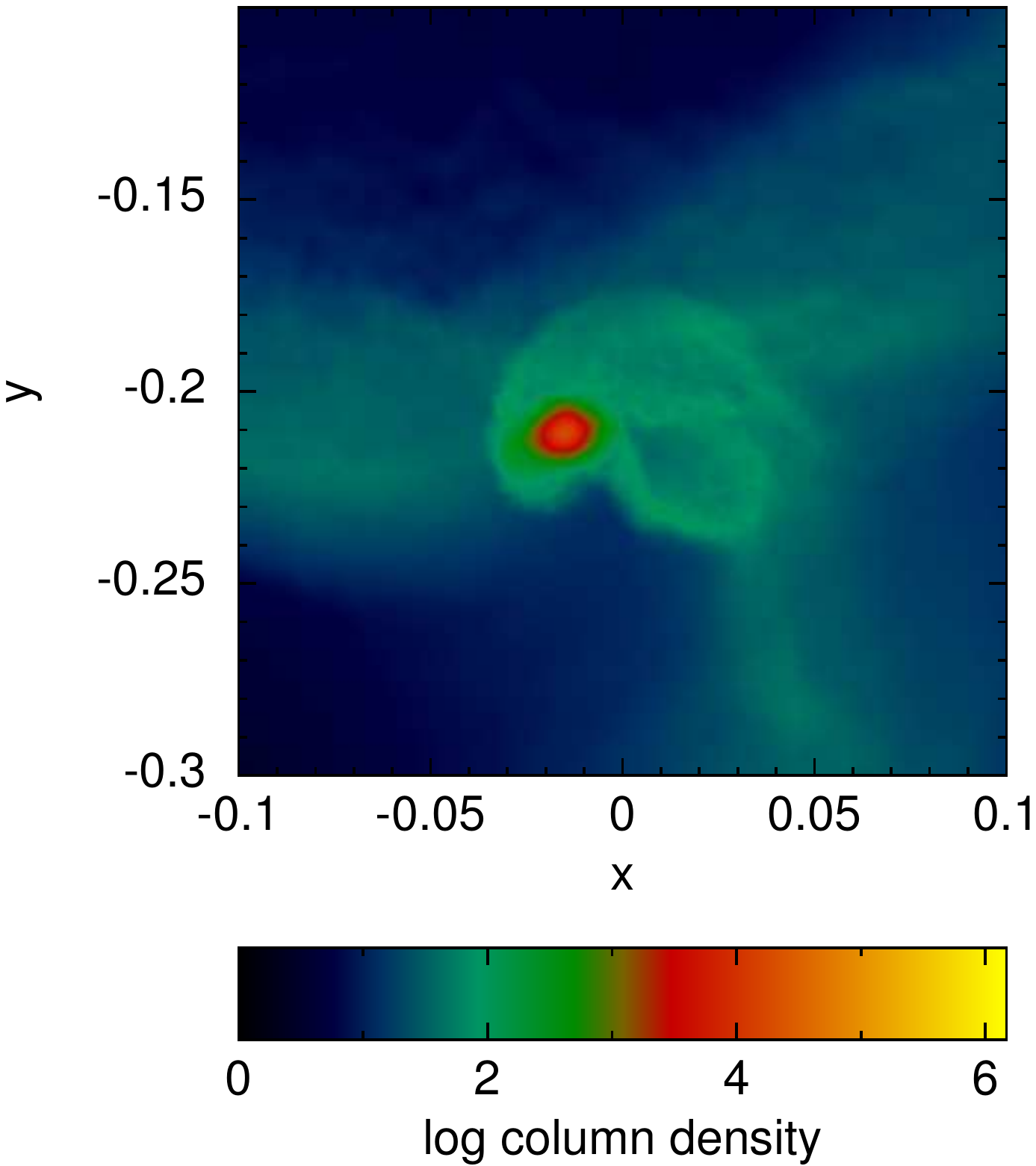}
\end{tabular}
\caption{\label{MosB-Os3rm} Iso-density plots for
model 11.}
\end{figure}

In these models, the wavelength of the initial perturbation far
exceeds the core radius; see column 2 of Table~\ref{tab:mod}. It can
be seen in the left panel of Fig.~\ref{MosB-Osrm} that only two
over-density clumps are clearly visible in the equatorial slice of
model 9. These perturbations are coupled to deform the central core
simultaneously in two orthogonal directions, see the panel in the middle
of Fig.~\ref{MosB-Osrm}. By looking at the right panel of
Fig.~\ref{MosB-Osrm}, we see the appearance of small spiral arms
around the protostar, indicating that the protostar has gained a net
angular momentum as a consequence of the small number of coupled
perturbations that acted upon it initially. For model 10, we see
that only one over-density is visible at the initial snapshot shown
in the left panel of  Fig.~\ref{MosB-O2srm}. For this reason, the
still forming protostar can be seen to be highly deformed, mainly along a
single direction; see the left and middle panels of Fig.~\ref{MosB-O2srm}.
Thus, the right panel of Fig.~\ref{MosB-O2srm} shows that the
still forming protostar drags a long tail of gas along this single
direction. The same behavior is mainly observed for model 11; see
Fig.~\ref{MosB-Os3rm}.
\subsection{Models 12--14}
\label{subsec:results12-14}

\begin{figure}
\begin{tabular}{ccc}
\includegraphics[width=2 in]{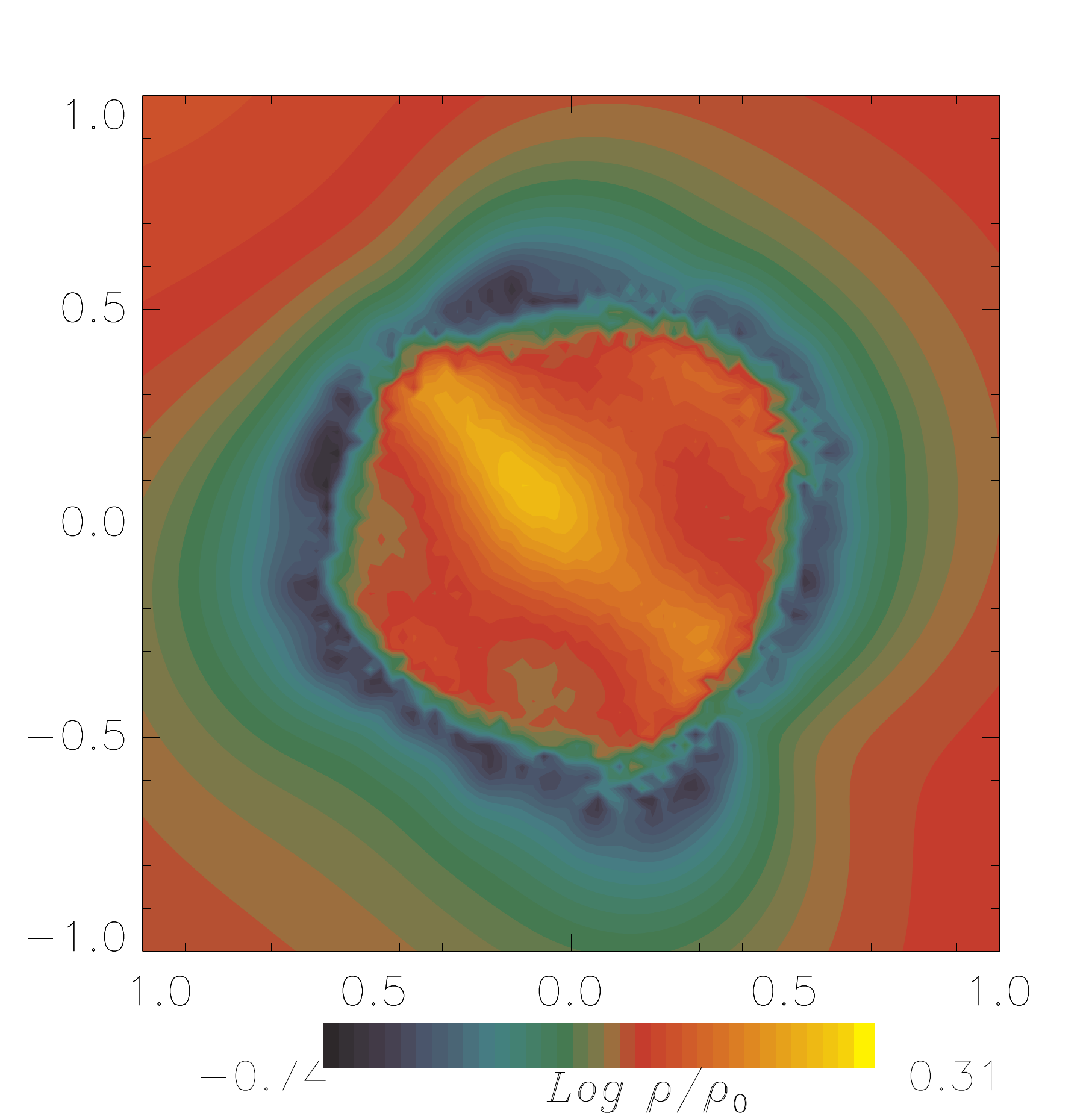} & \includegraphics[width=2 in]{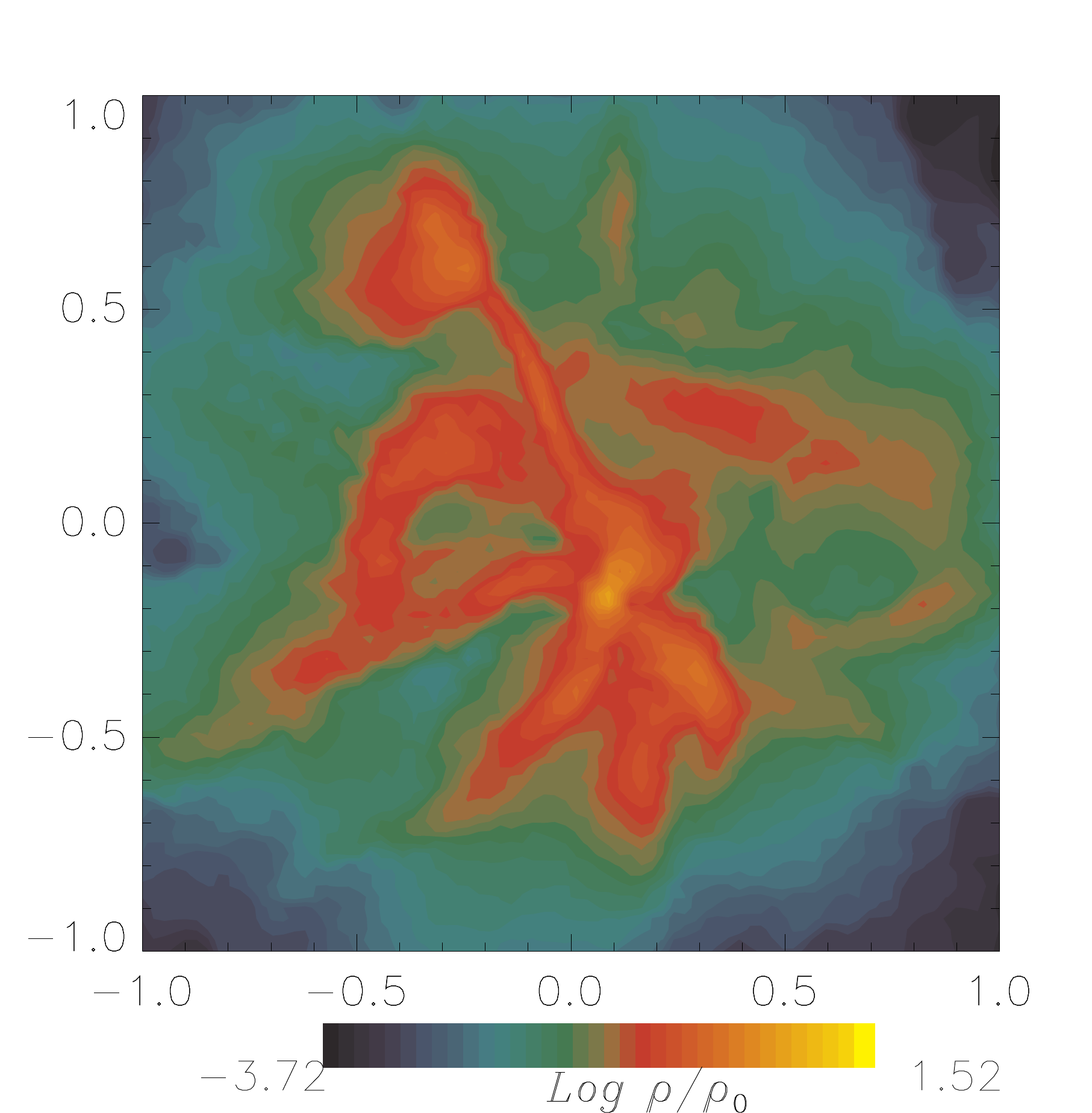} &
\includegraphics[width=2 in]{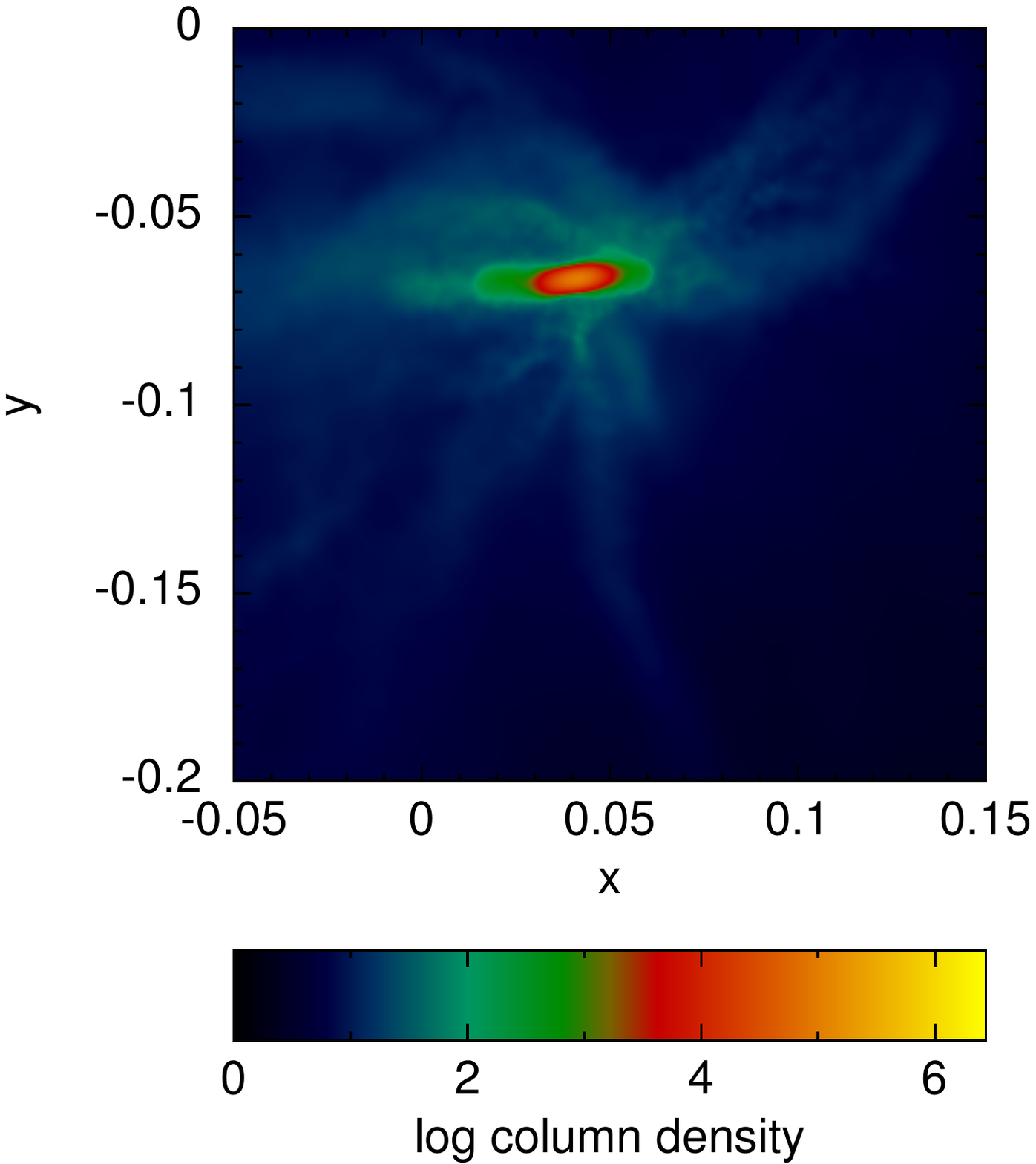}
\end{tabular}
\caption{\label{MosBOscTrp} Iso-density plots for
model 12.}
\end{figure}

For these models we use more Fourier modes to re-calculate the
models of Section~\ref{subsec:results9-11}. Let us examine model 12
shown in Fig.~\ref{MosBOscTrp} and compare it with model 9 shown in
Fig.~\ref{MosB-Osrm}. We first note that there is only one
perturbation mode visible in the left panel of Fig.~\ref{MosBOscTrp}; despite this, there
are a few dominant directions
along which the initial over-density grows; see the middle panel of
Fig.~\ref{MosBOscTrp}. The velocity of the collapsing peak is significantly
greater in model 12 than that observed in model 9; see column 5 of
Table~\ref{tab:mod}. Because of this, it is very likely that the
protostar formed in model 12 has gained a net angular momentum
higher than that of the protostar of model 9, as is suggested by
looking at the right panel of Fig.~\ref{MosBOscTrp}.

\begin{figure}
\begin{tabular}{ccc}
\includegraphics[width=2 in]{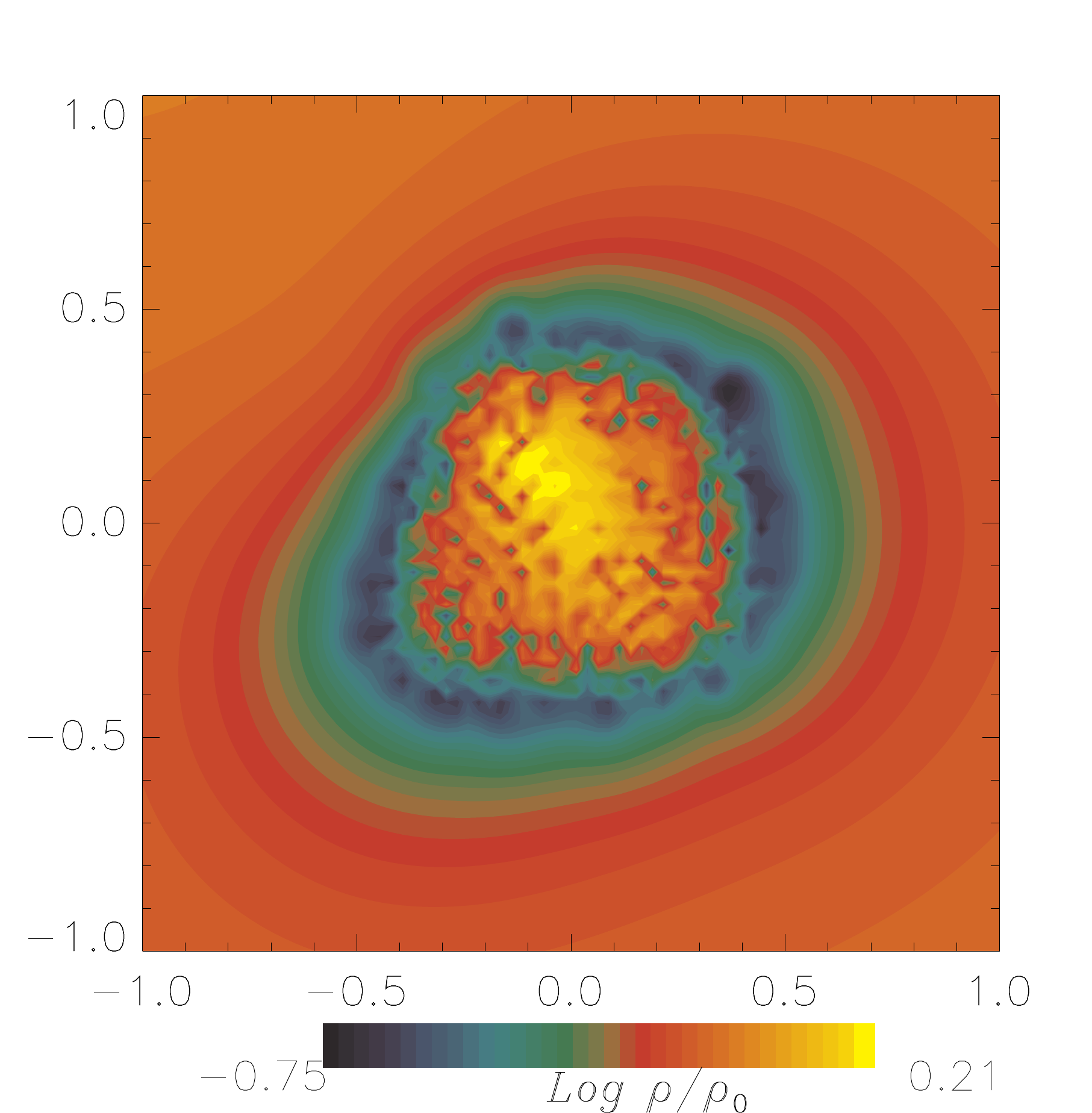} & \includegraphics[width=2 in]{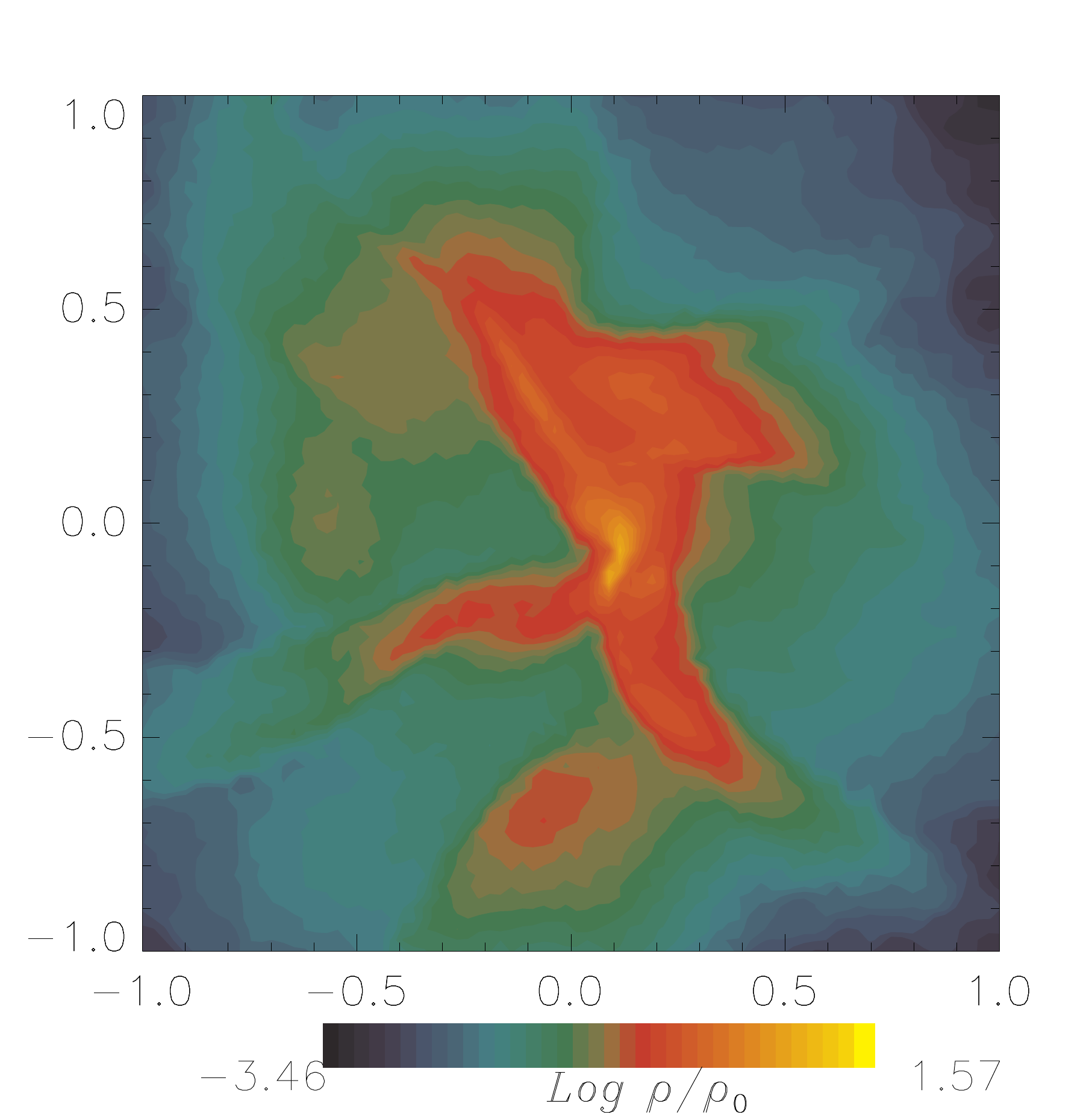} &
\includegraphics[width=2 in]{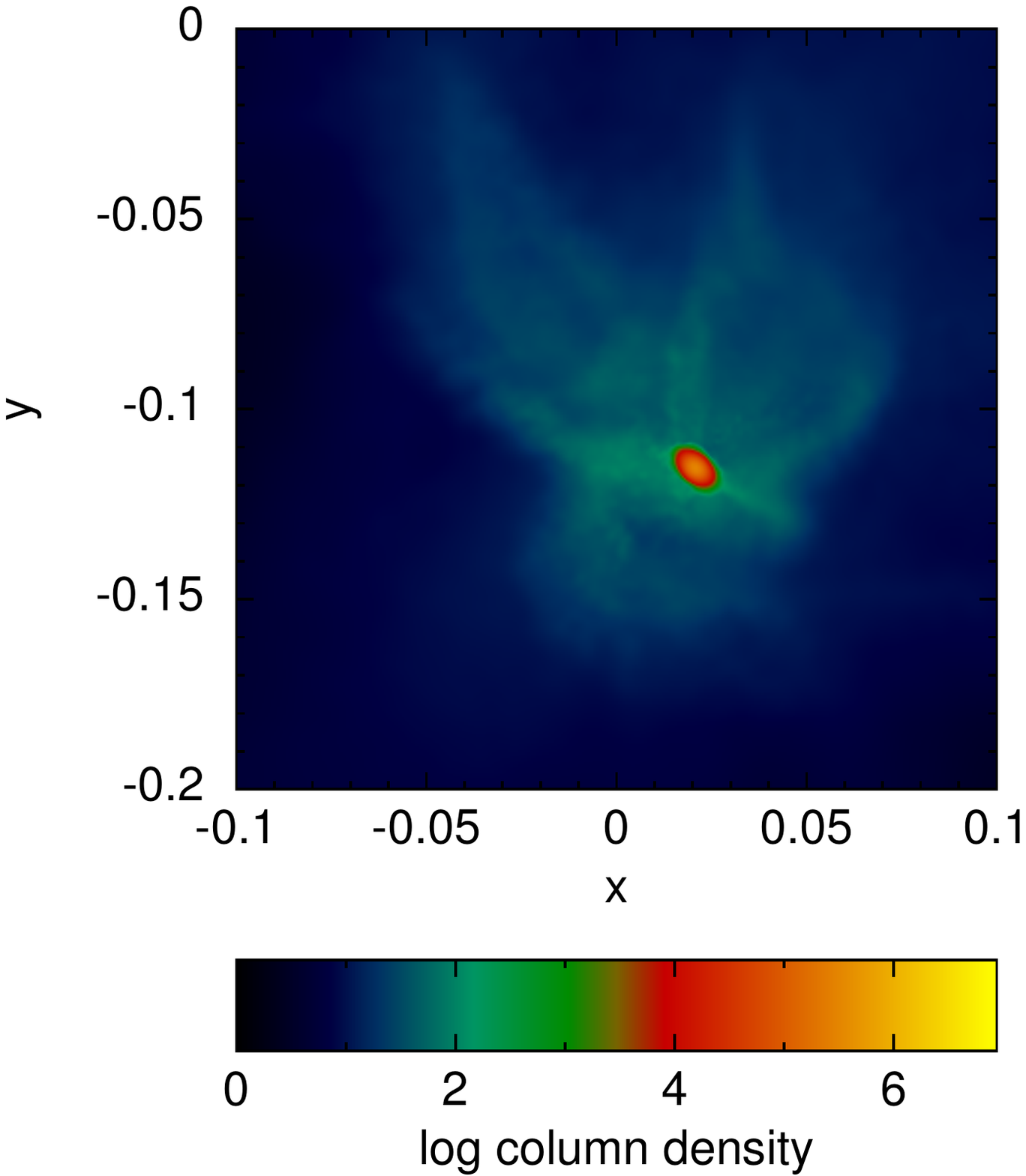}
\end{tabular}
\caption{\label{MosBOscT2rp} Iso-density  plots for
model 13.}
\end{figure}

The size of the initial over-density clump induced by the
perturbation mode of model 13 looks smaller than that induced in
model 10; compare Fig.~\ref{MosBOscT2rp} with Fig.~\ref{MosB-O2srm}.

In model 14 we observe the occurrence of
fragmentation of the central core region, so that a binary
protostar system is formed.

\subsection{Models 15--18}
\label{subsec:results15-18}

\begin{figure}
\begin{tabular}{ccc}
\includegraphics[width=2 in]{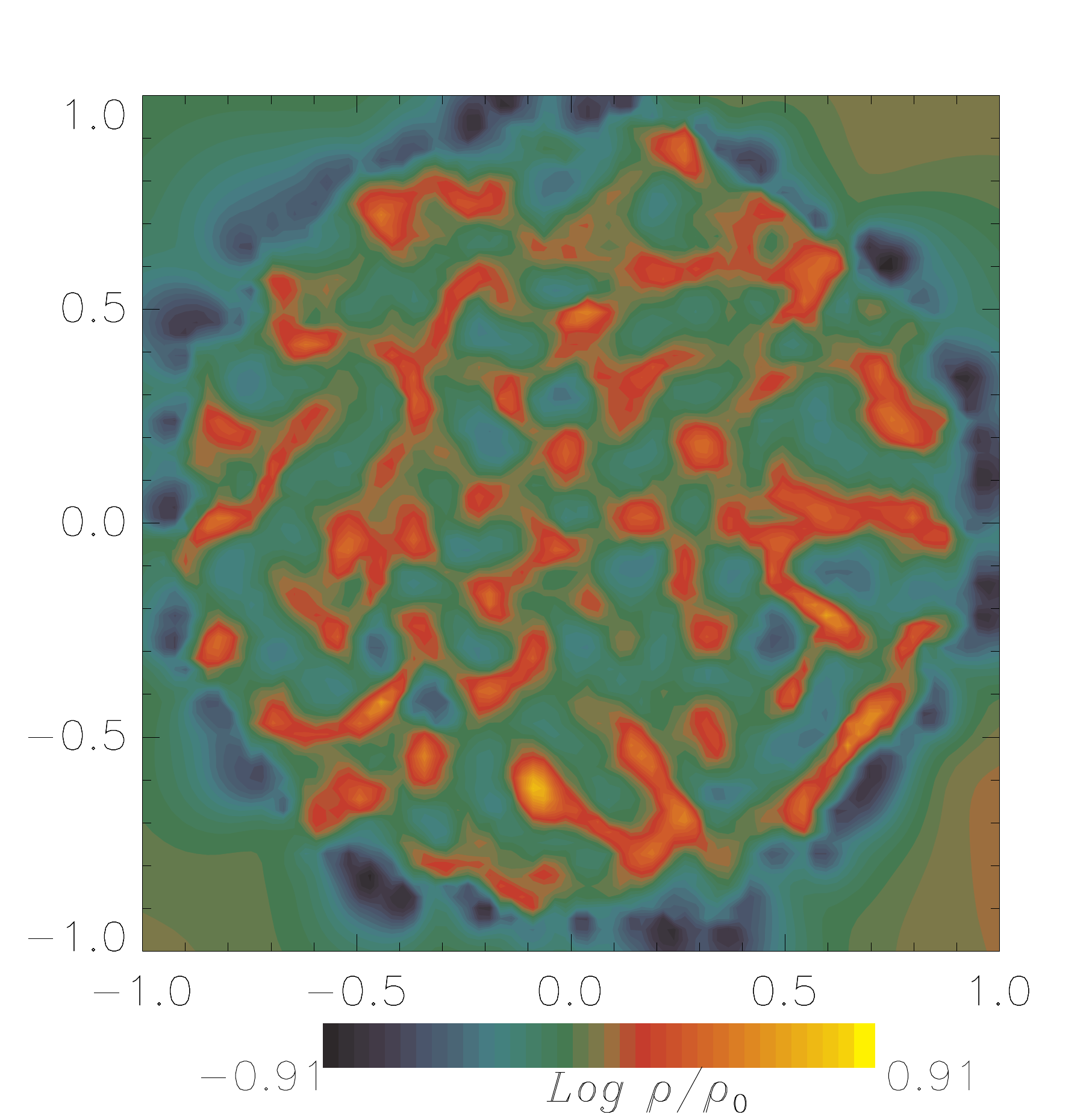} & \includegraphics[width=2 in]{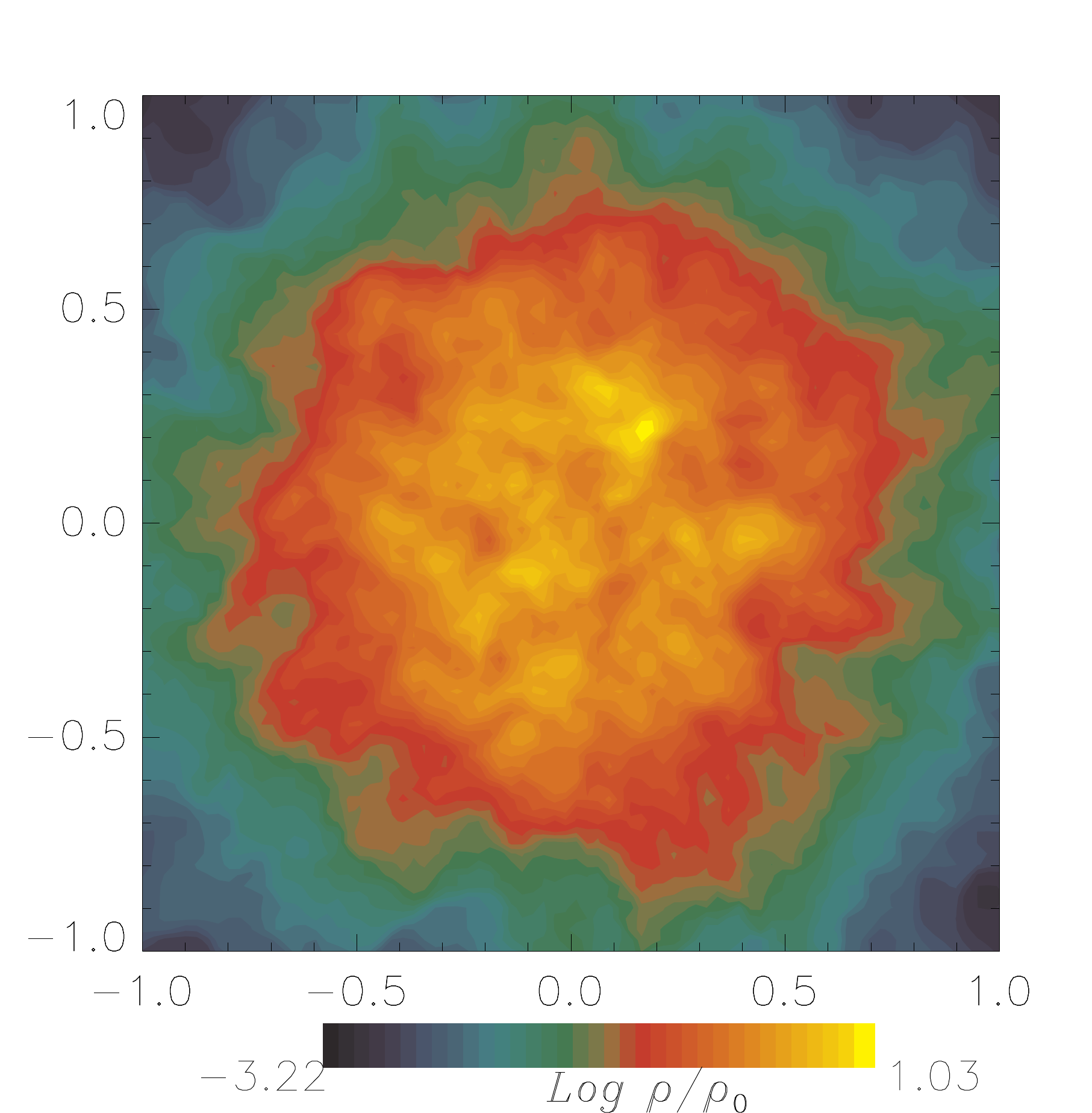} &
\includegraphics[width=2 in]{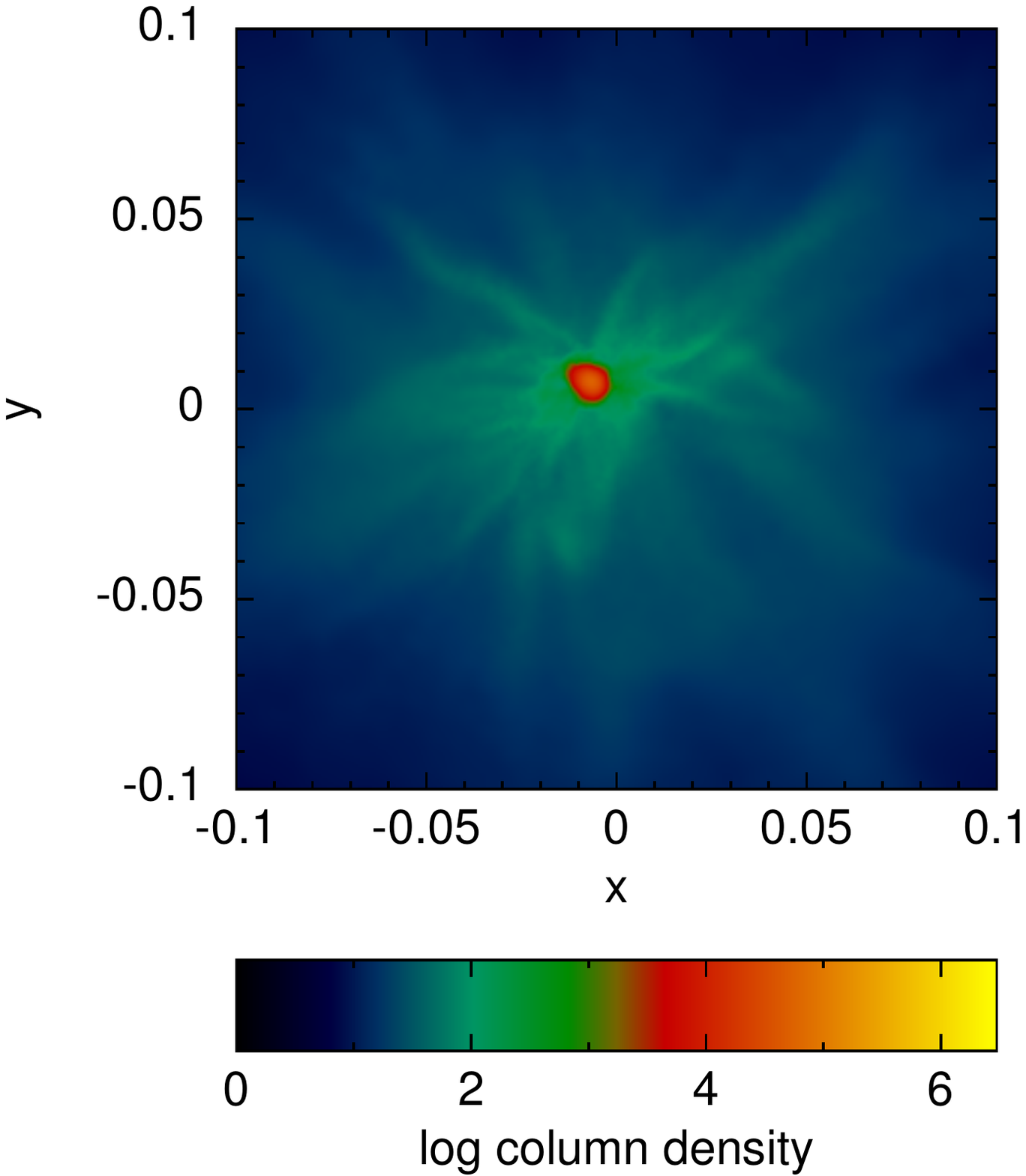}
\end{tabular}
\caption{\label{MosBsrm}Iso-density  plots for model
1.}
\end{figure}

\begin{figure}
\begin{tabular}{ccc}
\includegraphics[width=2 in]{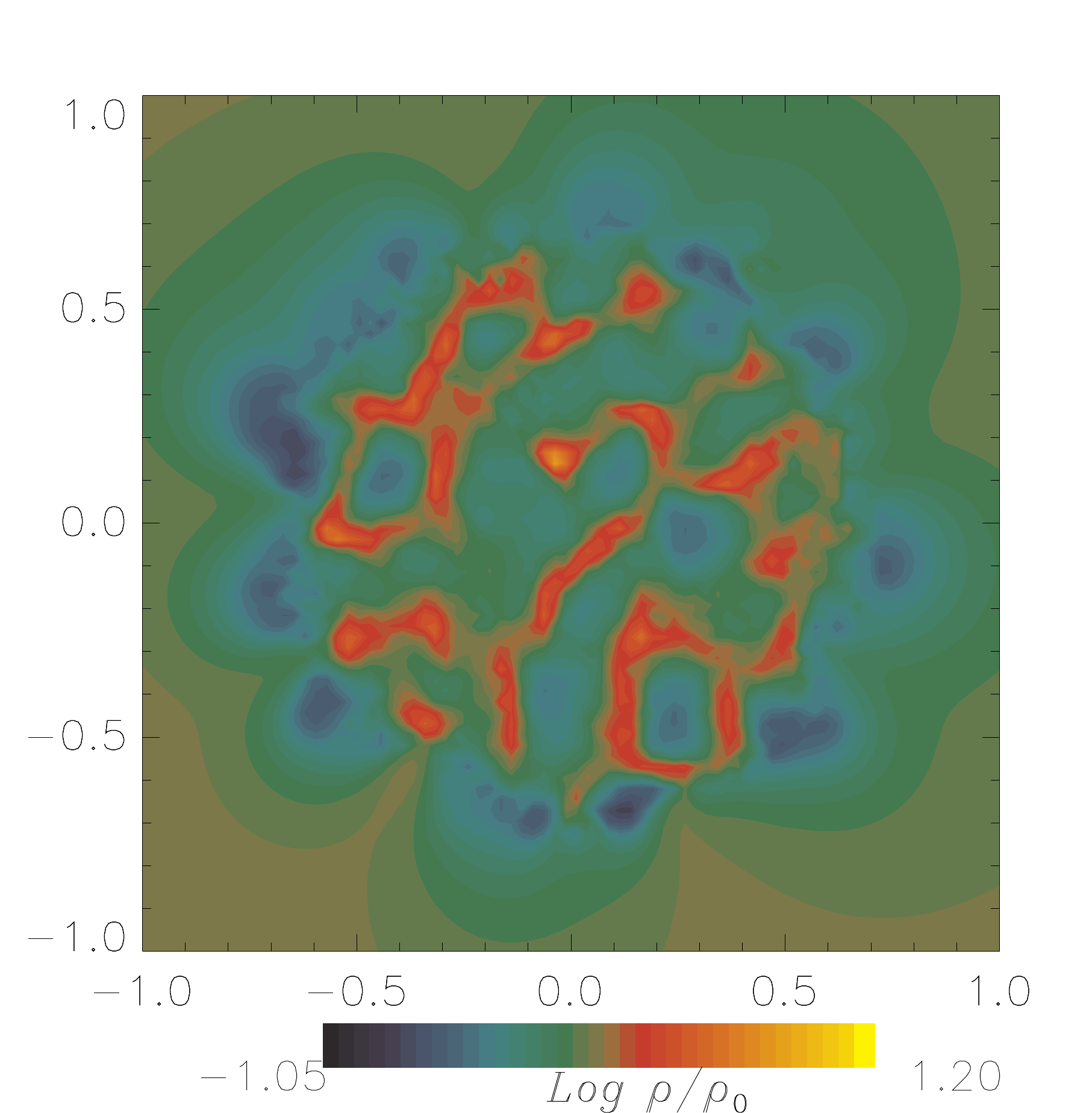} & \includegraphics[width=2 in]{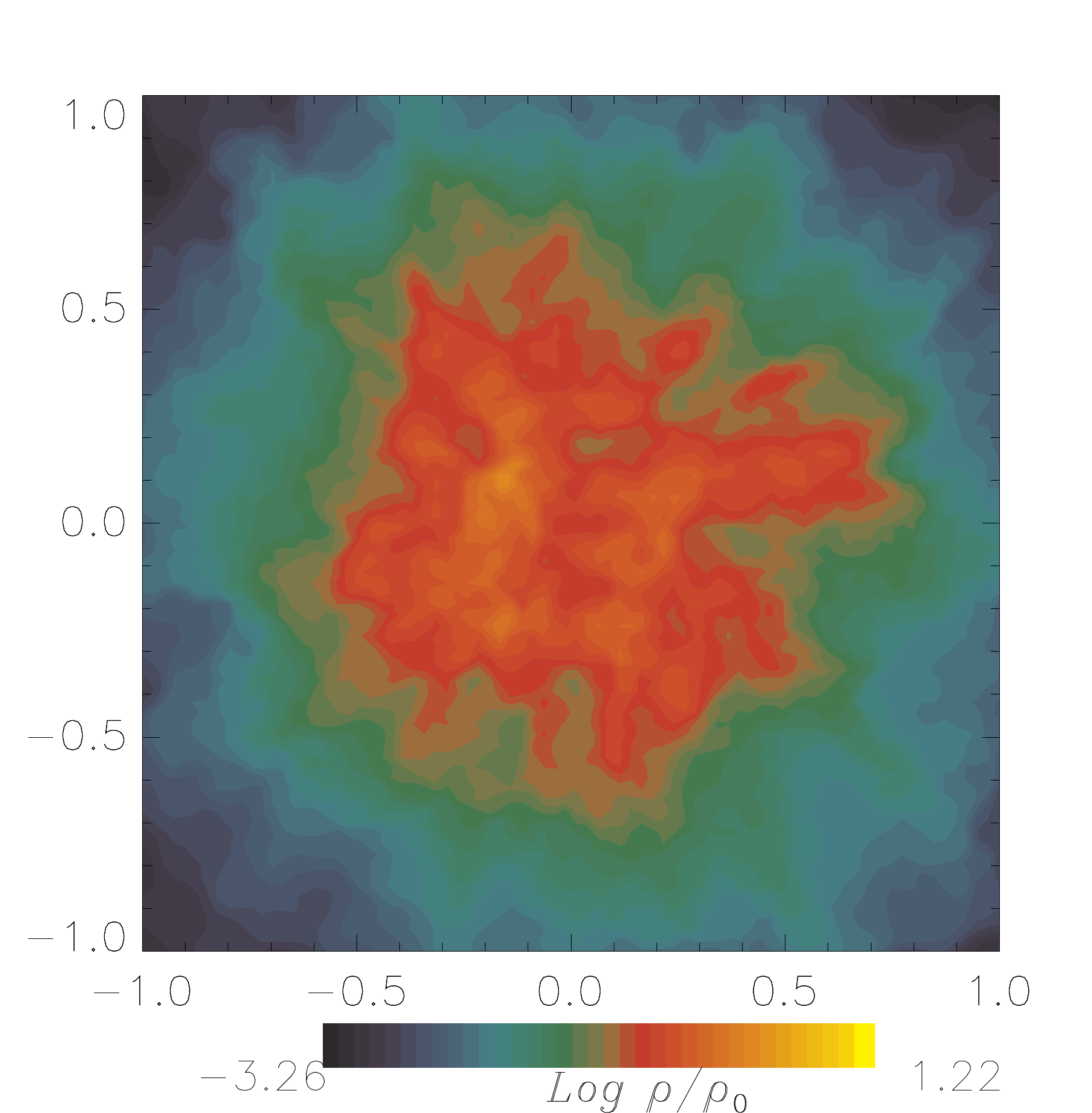} &
\includegraphics[width=2 in]{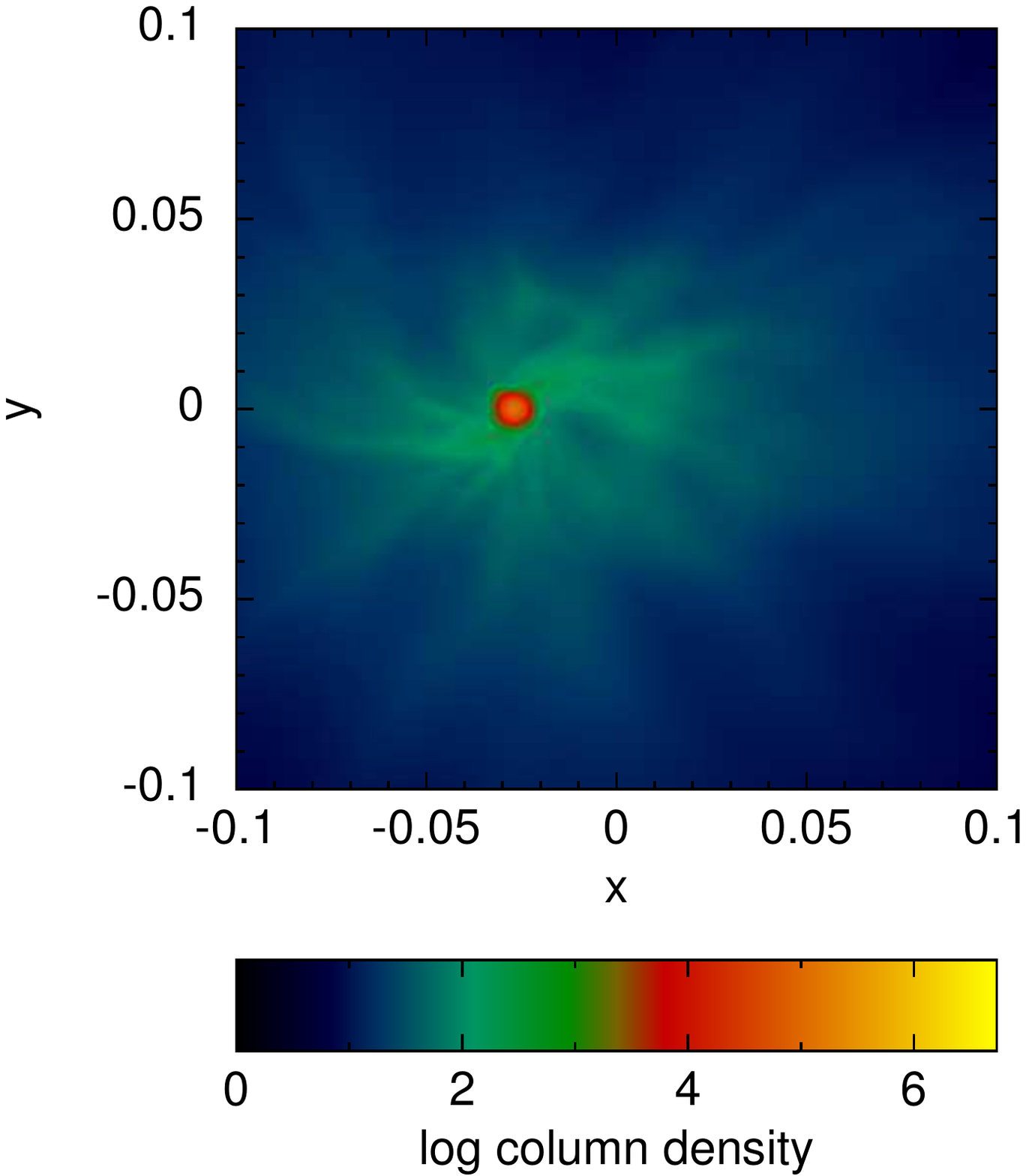}
\end{tabular}
\caption{\label{Seg1} Iso-density  plots for model
15.}
\end{figure}

We consider now the set of models generated with the curl-free
turbulence spectrum. In the left panel of Fig.~\ref{Seg1}, we
see that the number of over-density clumps formed initially in model
15 is smaller than the number observed in model 1, shown in the left
panel of Fig.~\ref{MosBsrm}. The clumps seen in Fig.~\ref{Seg1}
are also in fact larger that those seen in Fig.~\ref{MosBsrm}.
Despite this, we still see that the curl-free turbulence
spectrum creates a homogeneous distribution of over-density clumps.
At the time 0.67 t$_{ff}$, when the strong collapse is yet to begin,
the spherical symmetry of the collapsing core is already visible in
the middle panel of Fig.~\ref{Seg1}. For this reason, we see that
the resulting protostar accretes gas from the remaining core almost
isotropically; see the right panel of Fig.~\ref{Seg1}.

\begin{figure}
\begin{tabular}{ccc}
\includegraphics[width=2 in]{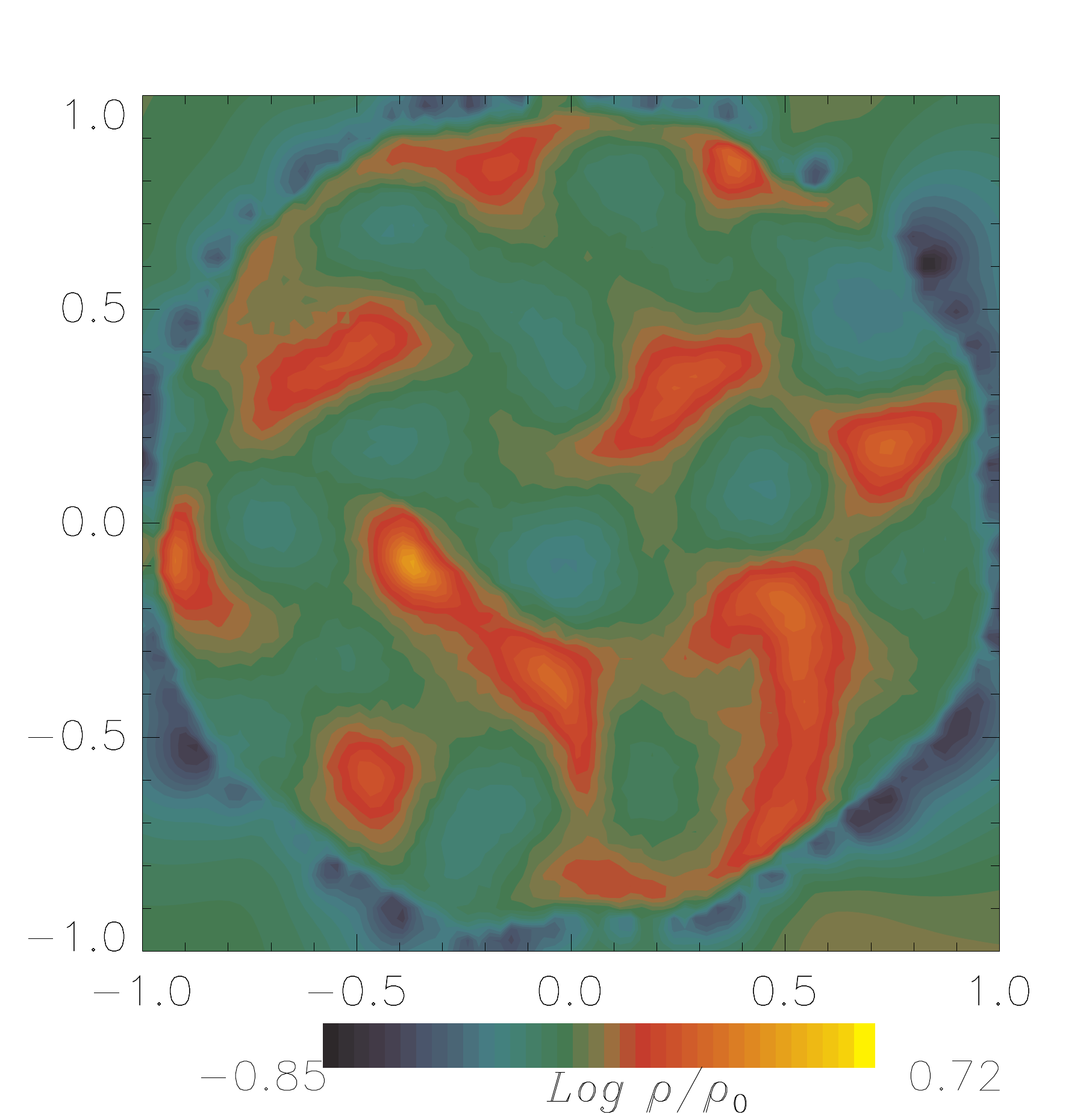} & \includegraphics[width=2 in]{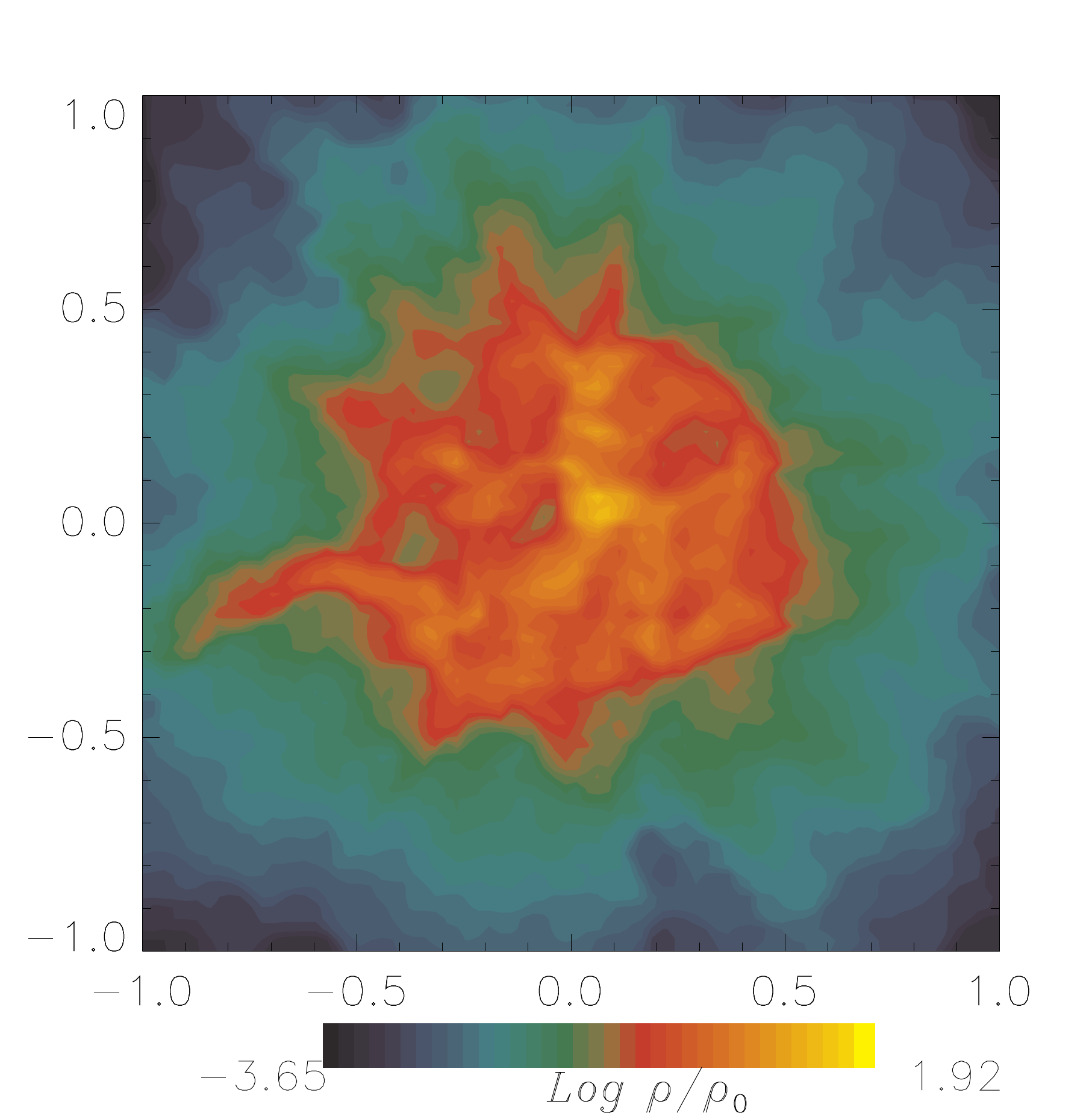} &
\includegraphics[width=2 in]{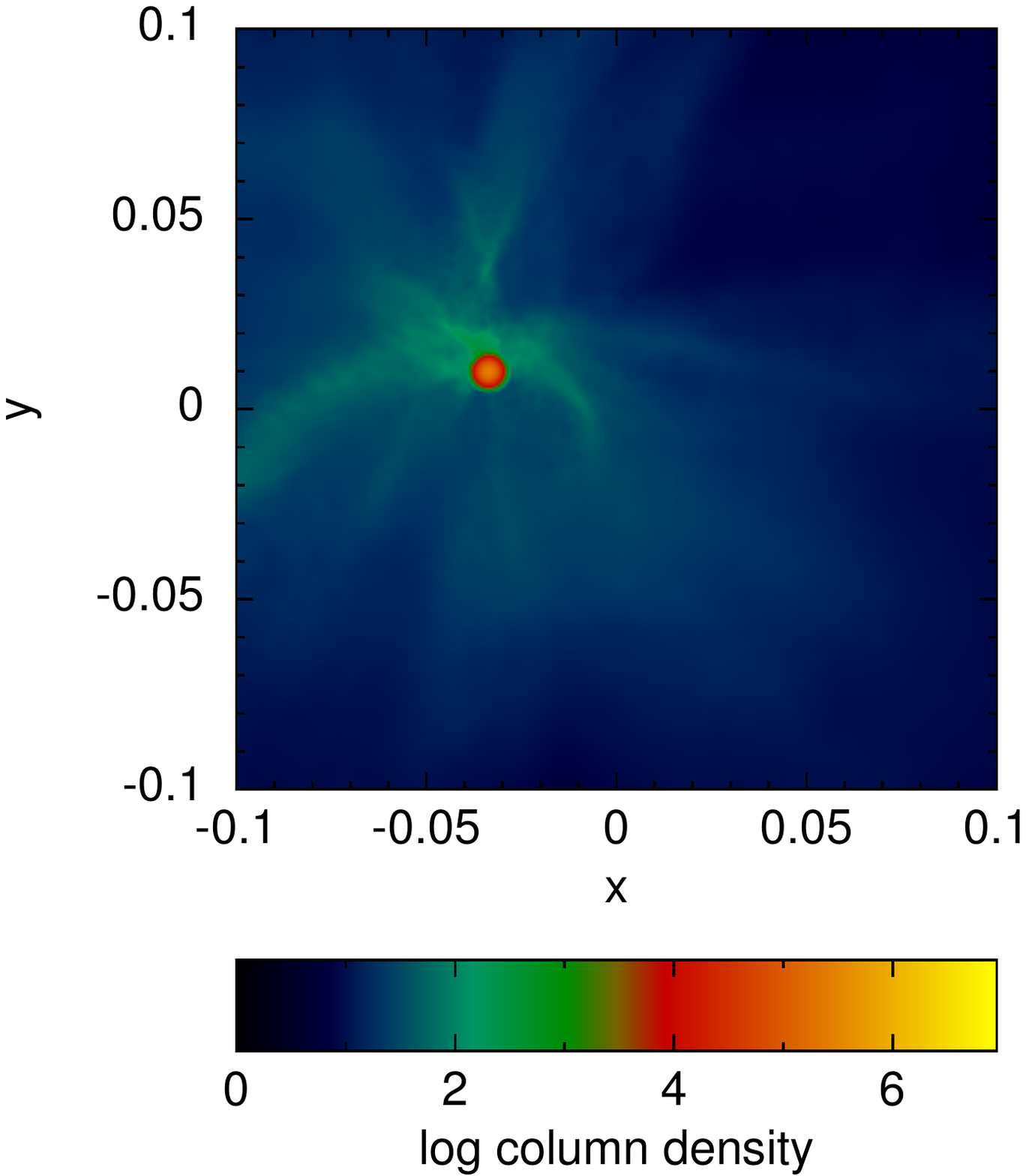}
\end{tabular}
\caption{\label{Seg2} Iso-density  plots for model
16.}
\end{figure}

\begin{figure}
\begin{tabular}{ccc}
\includegraphics[width=2 in]{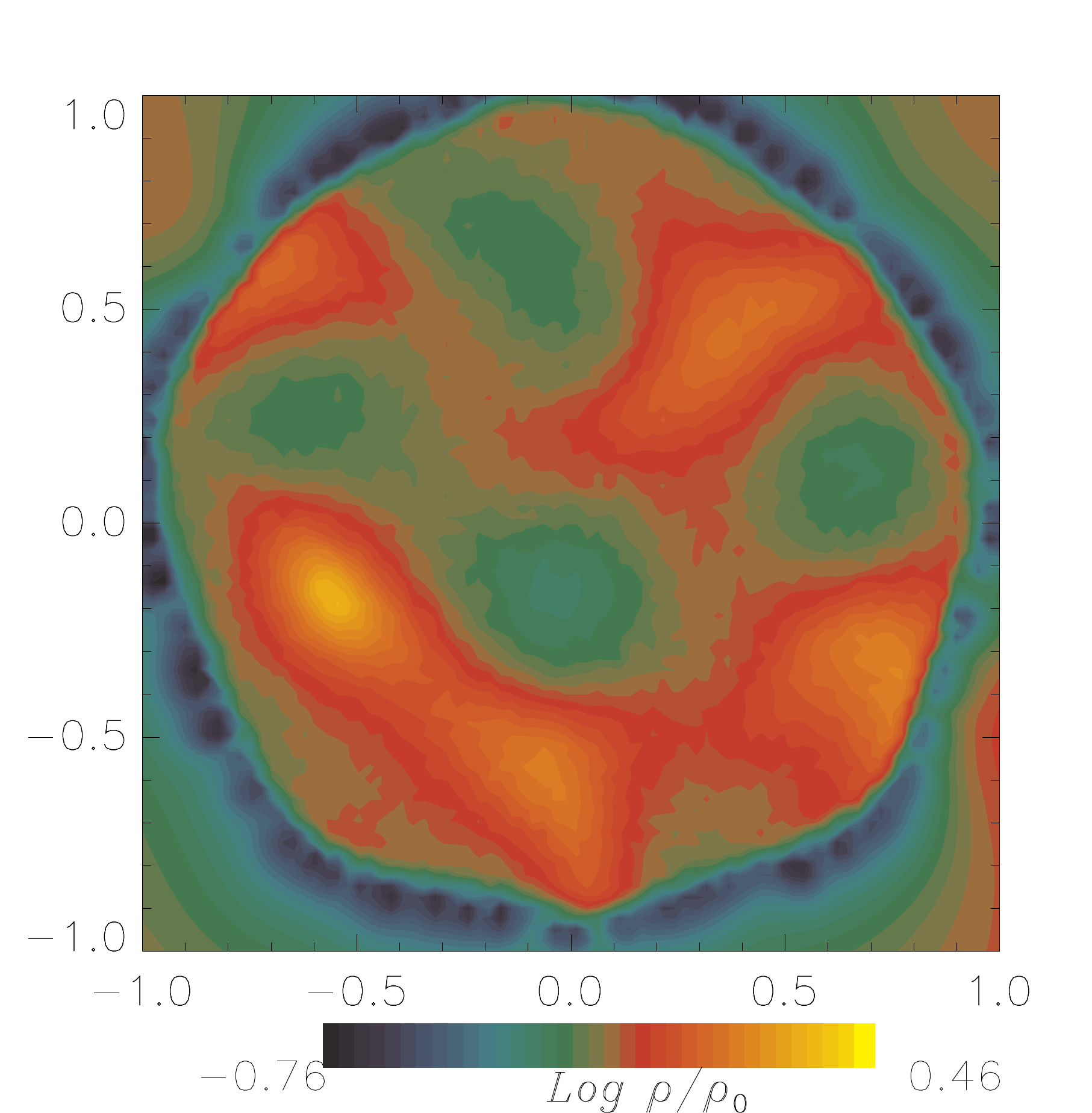} & \includegraphics[width=2 in]{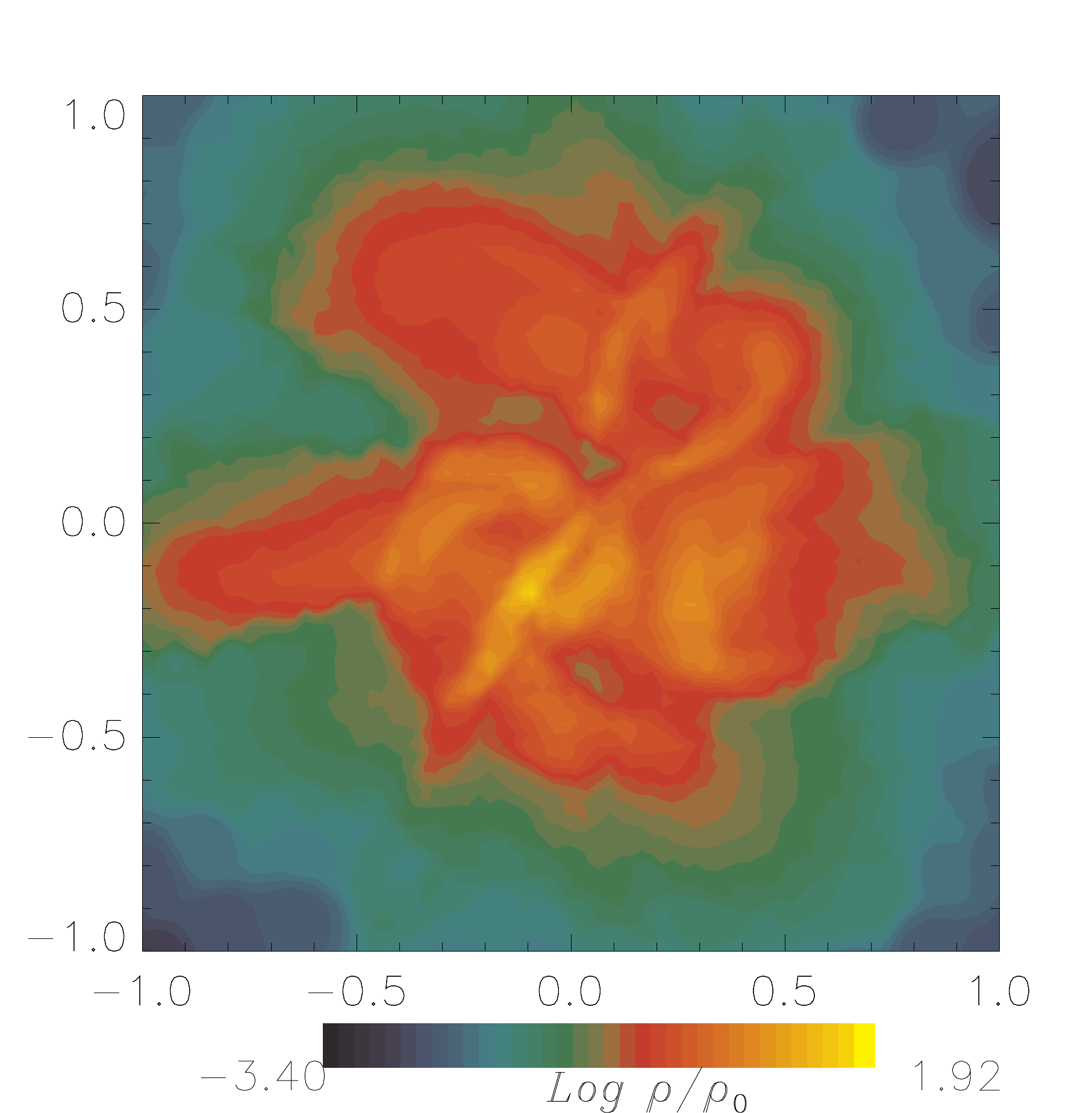} &
\includegraphics[width=2 in]{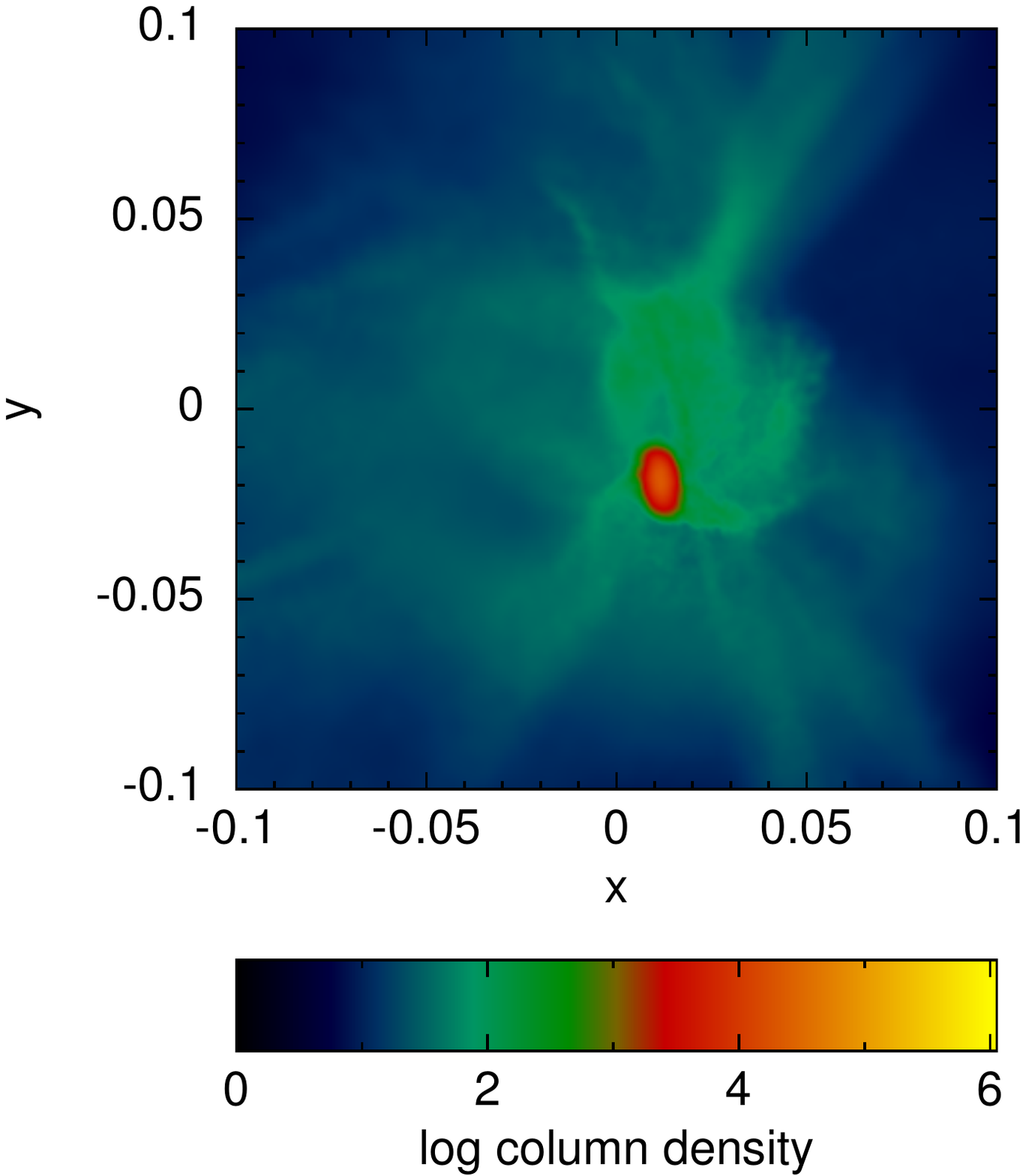}
\end{tabular}
\caption{\label{Seg3} Iso-density  plots for model
17.}
\end{figure}

\begin{figure}
\begin{tabular}{ccc}
\includegraphics[width=2 in]{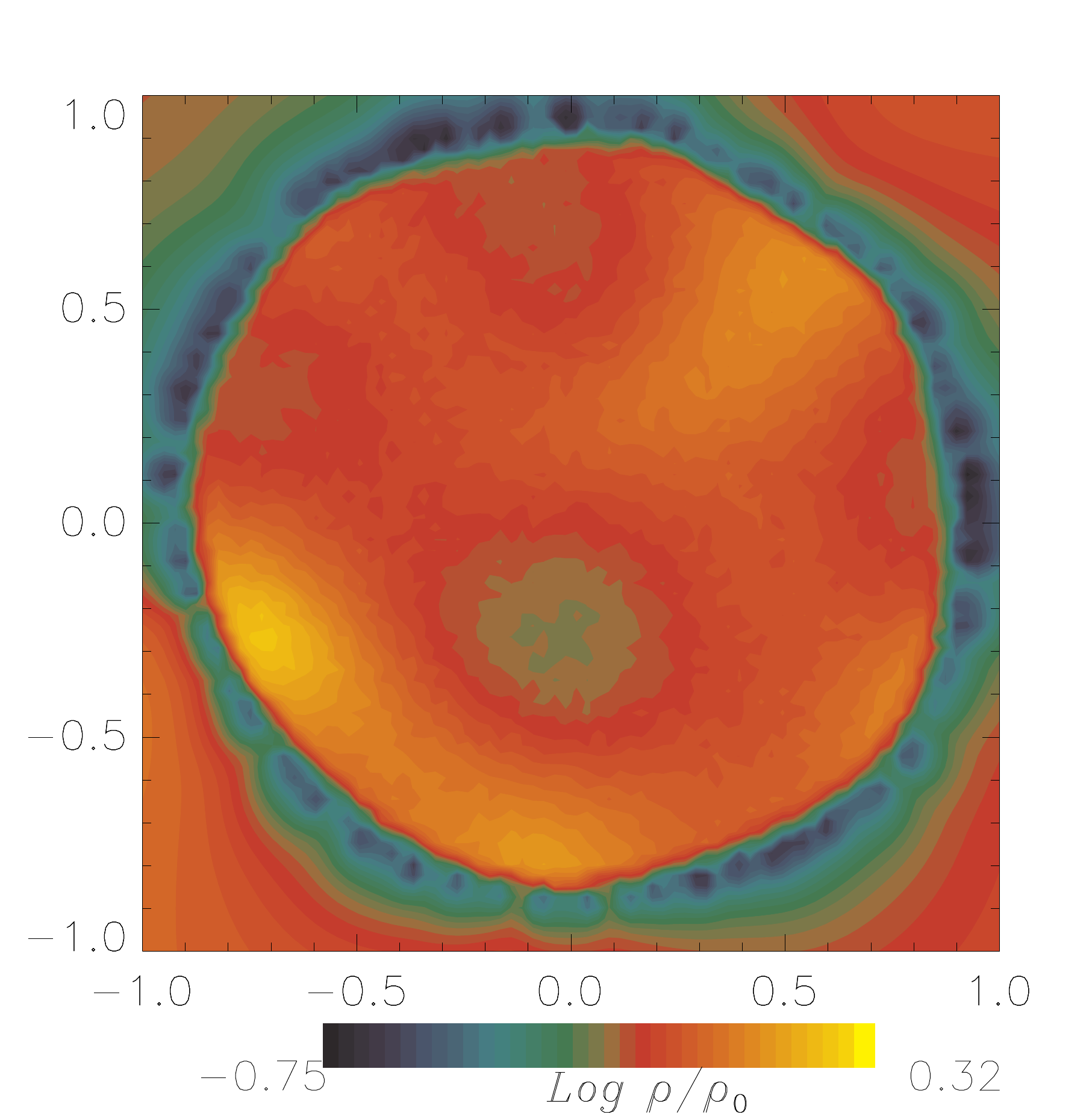} & \includegraphics[width=2 in]{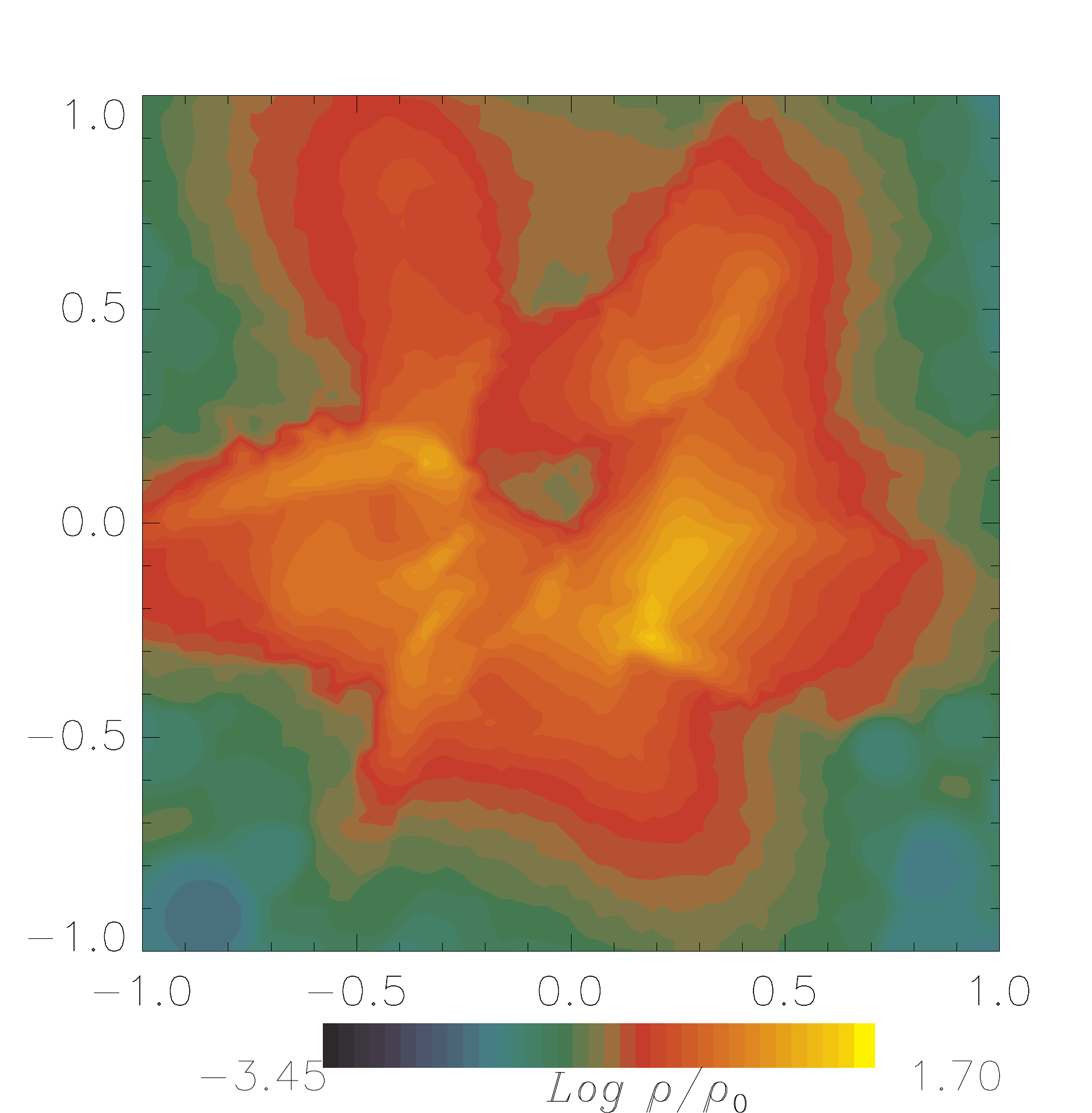} &
\includegraphics[width=2 in]{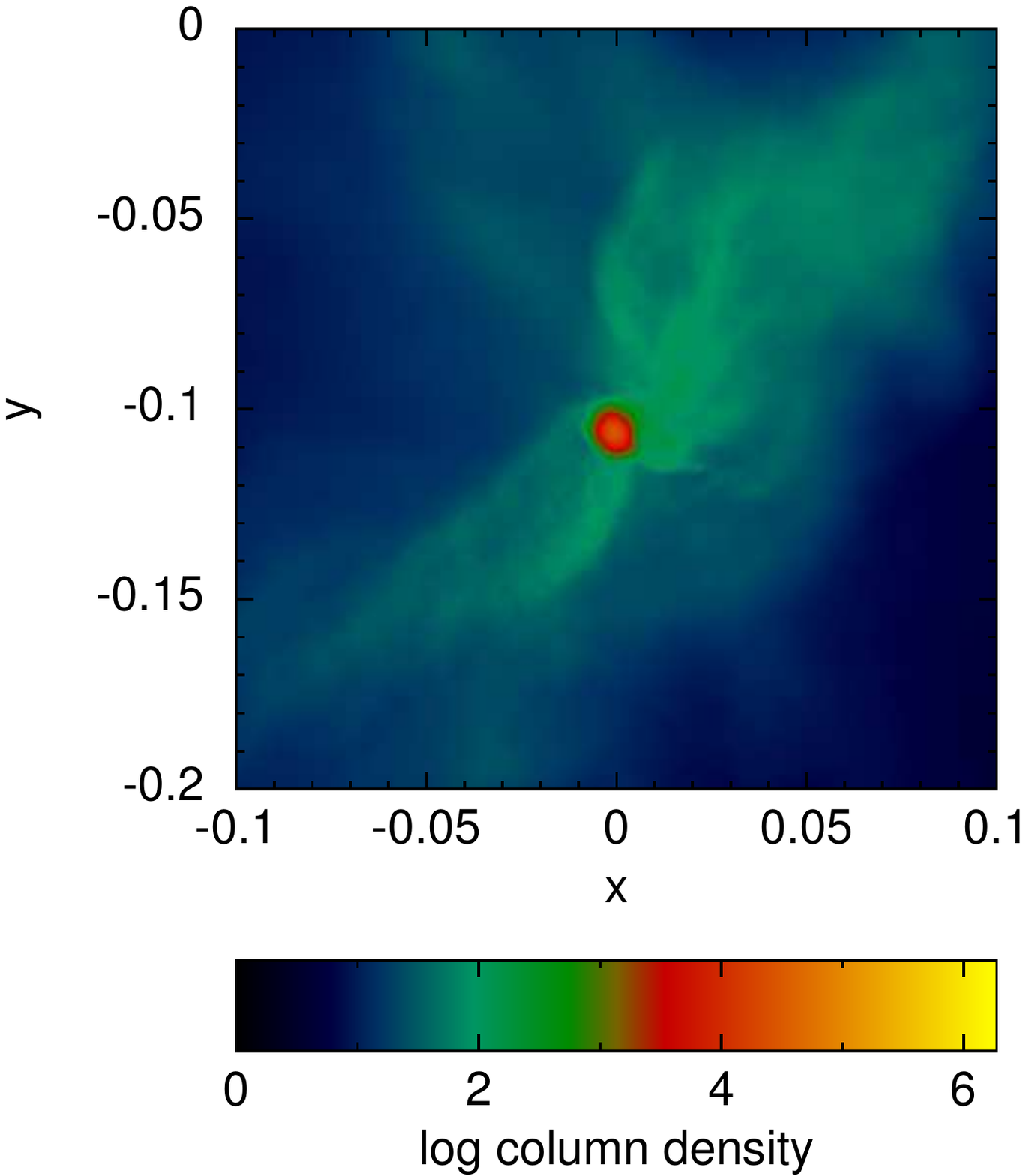}
\end{tabular}
\caption{\label{Seg4} Iso-density  plots for model
18.}
\end{figure}

When the size of the Fourier mesh begins to increase in terms of the
core radius, which is shown in models 16--18, then the initial
over-density clumps are formed thicker and their number is smaller
than those seen in model 15; see the left panels of
Figs.~\ref{Seg2}, \ref{Seg3} and \ref{Seg4}. As a consequence,
the homogeneity and isotropy of the over-density clump
distribution is slightly lessened, and then some clumps grow forming
large filaments, as can be observed in the right panels of
Figs.~\ref{Seg3} and \ref{Seg4}. All these models finish with a
single protostar formed in the collapsed central region.

By comparing the left panels of Figs.~\ref{Seg2}, \ref{Seg3} and
\ref{Seg4}, corresponding to models 16--18, with those of models 2--4,
shown in Fig.~\ref{MosBs2rm}, \ref{MosBs3rm} and \ref{MosBs4rm}, we
conclude that when the Fourier mesh begins to increase, the models
generated with a divergence-free turbulence lose the
homogeneity and isotropy of the initial over-density clump
distribution earlier than those models generated with a curl-free
turbulence.
\subsection{Models 19--22}
\label{subsec:results19-22}

\begin{figure}
\begin{tabular}{ccc}
\includegraphics[width=2 in]{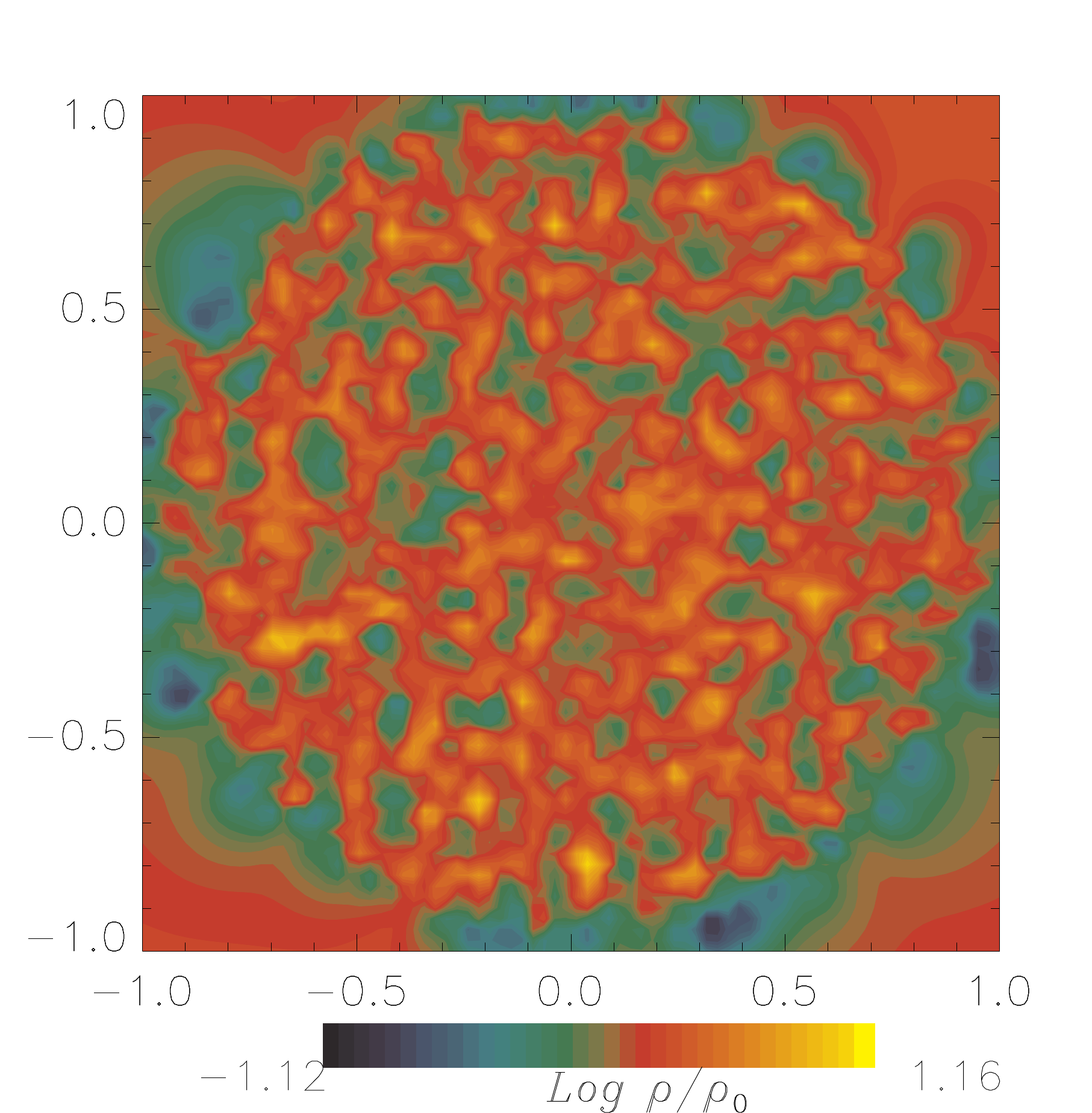} & \includegraphics[width=2 in]{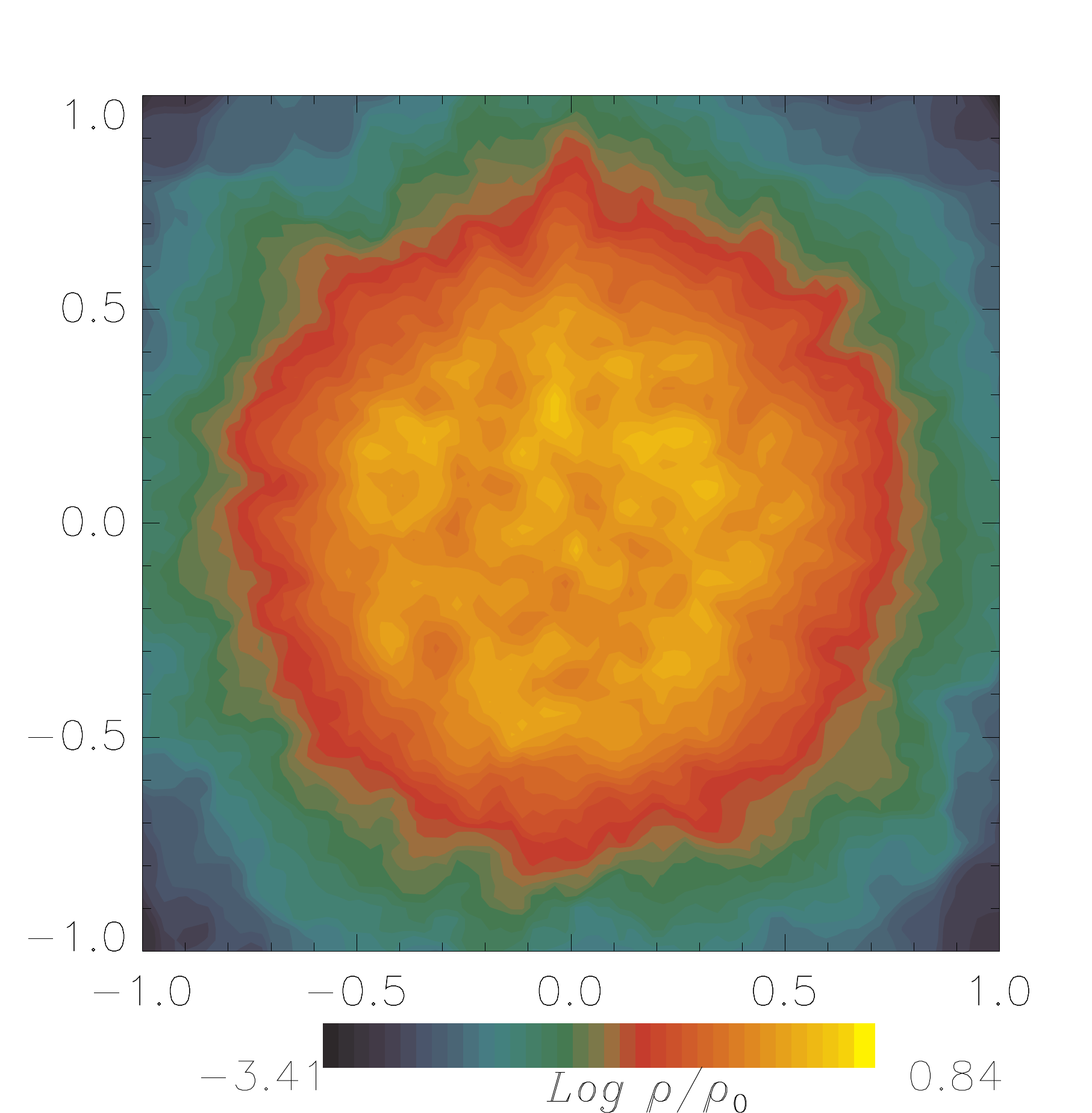} &
\includegraphics[width=2 in]{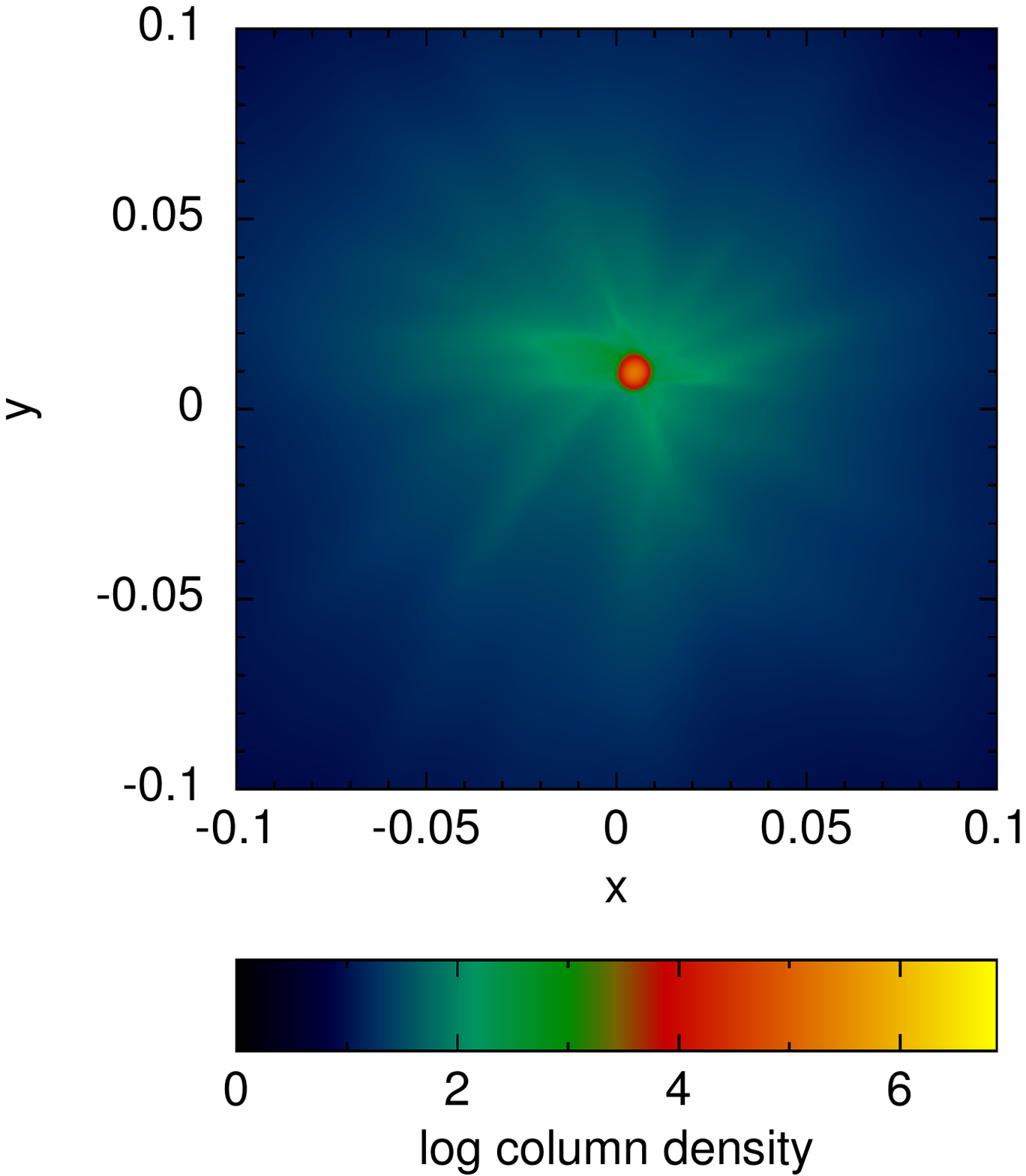}
\end{tabular}
\caption{\label{Seg5} Iso-density  plots for model
19.}
\end{figure}

\begin{figure}
\begin{tabular}{ccc}
\includegraphics[width=2 in]{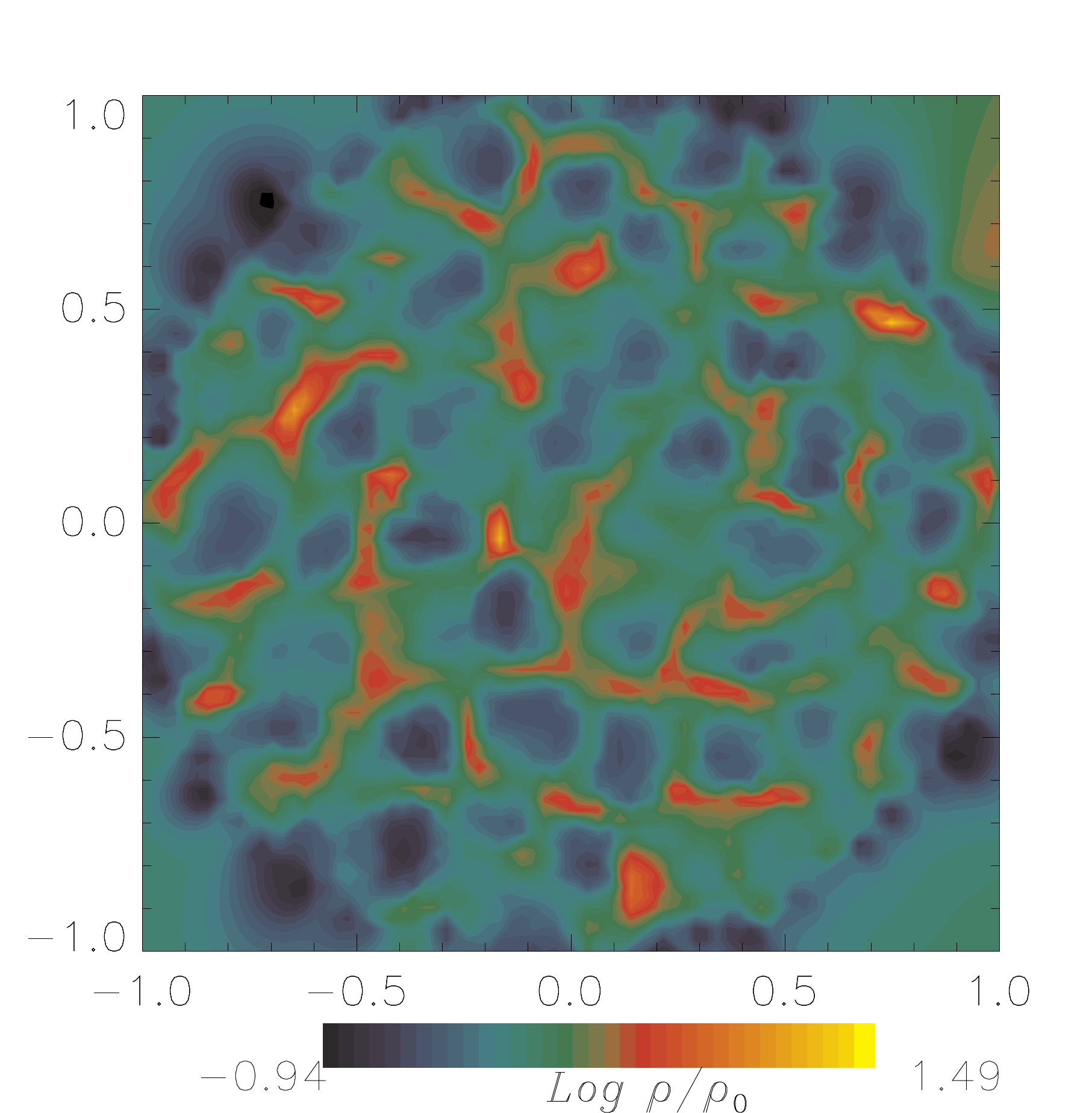} & \includegraphics[width=2 in]{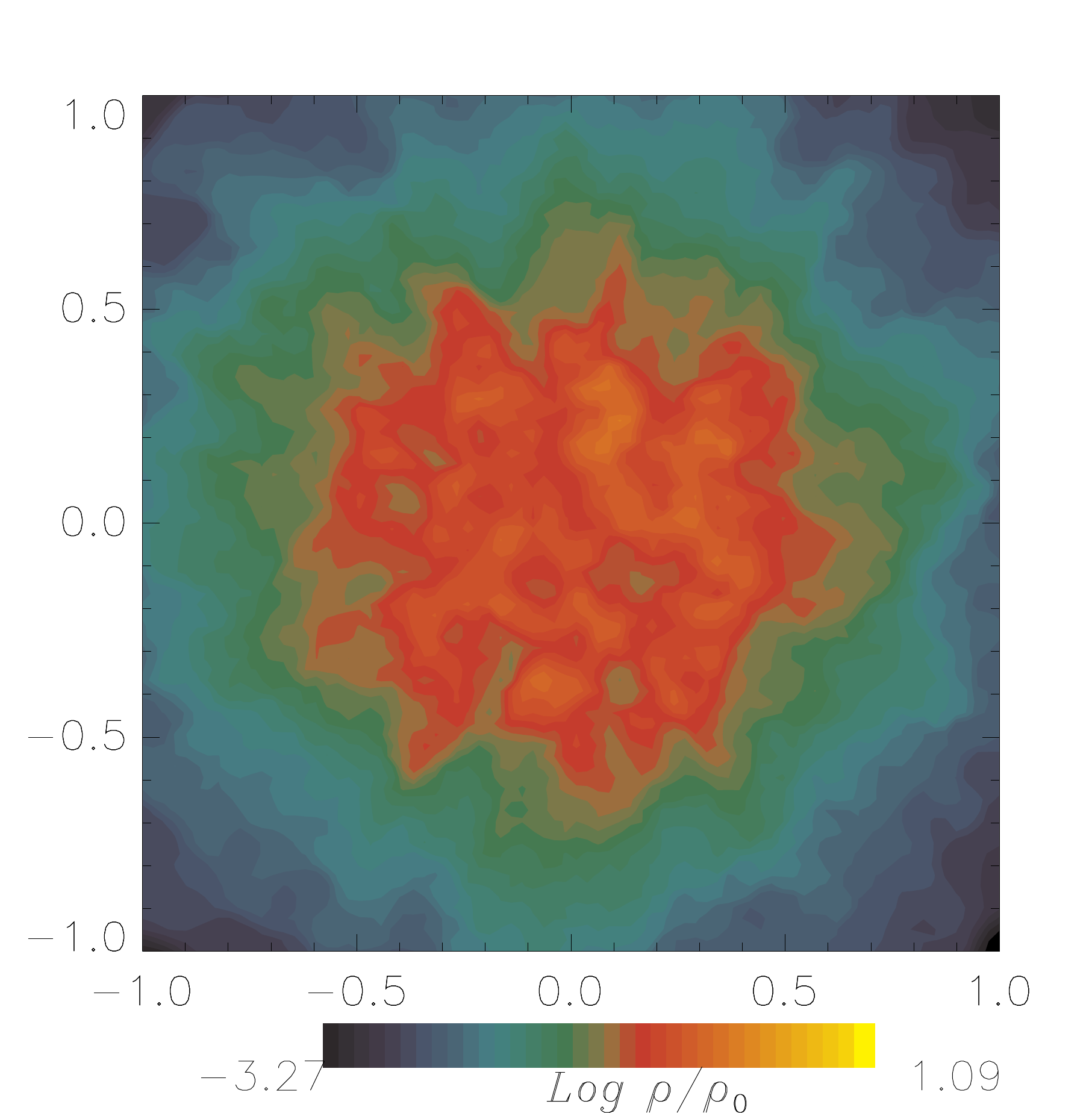} &
\includegraphics[width=2 in]{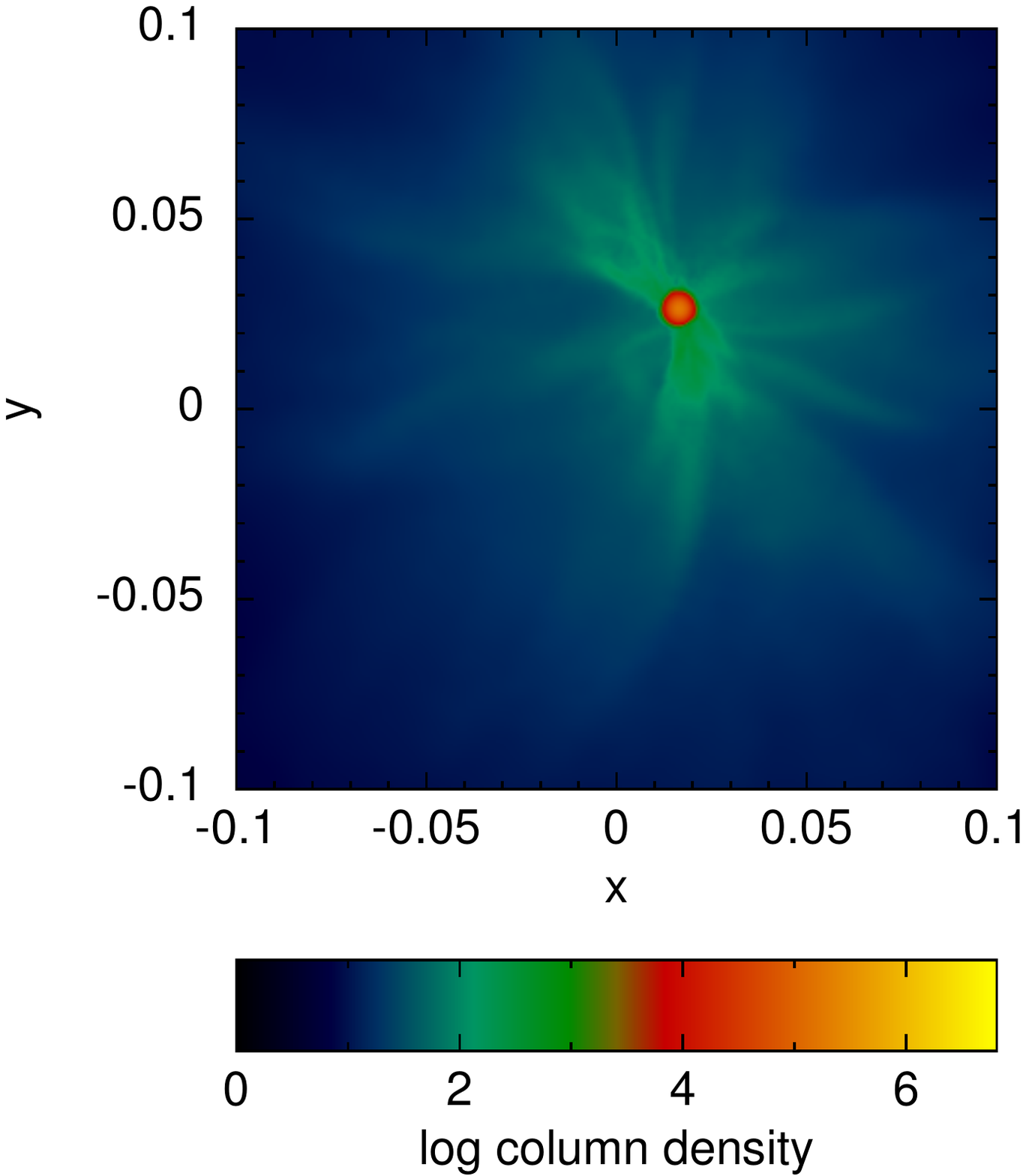}
\end{tabular}
\caption{\label{Seg6} Iso-density  plots for model
20.}
\end{figure}

\begin{figure}
\begin{tabular}{ccc}
\includegraphics[width=2 in]{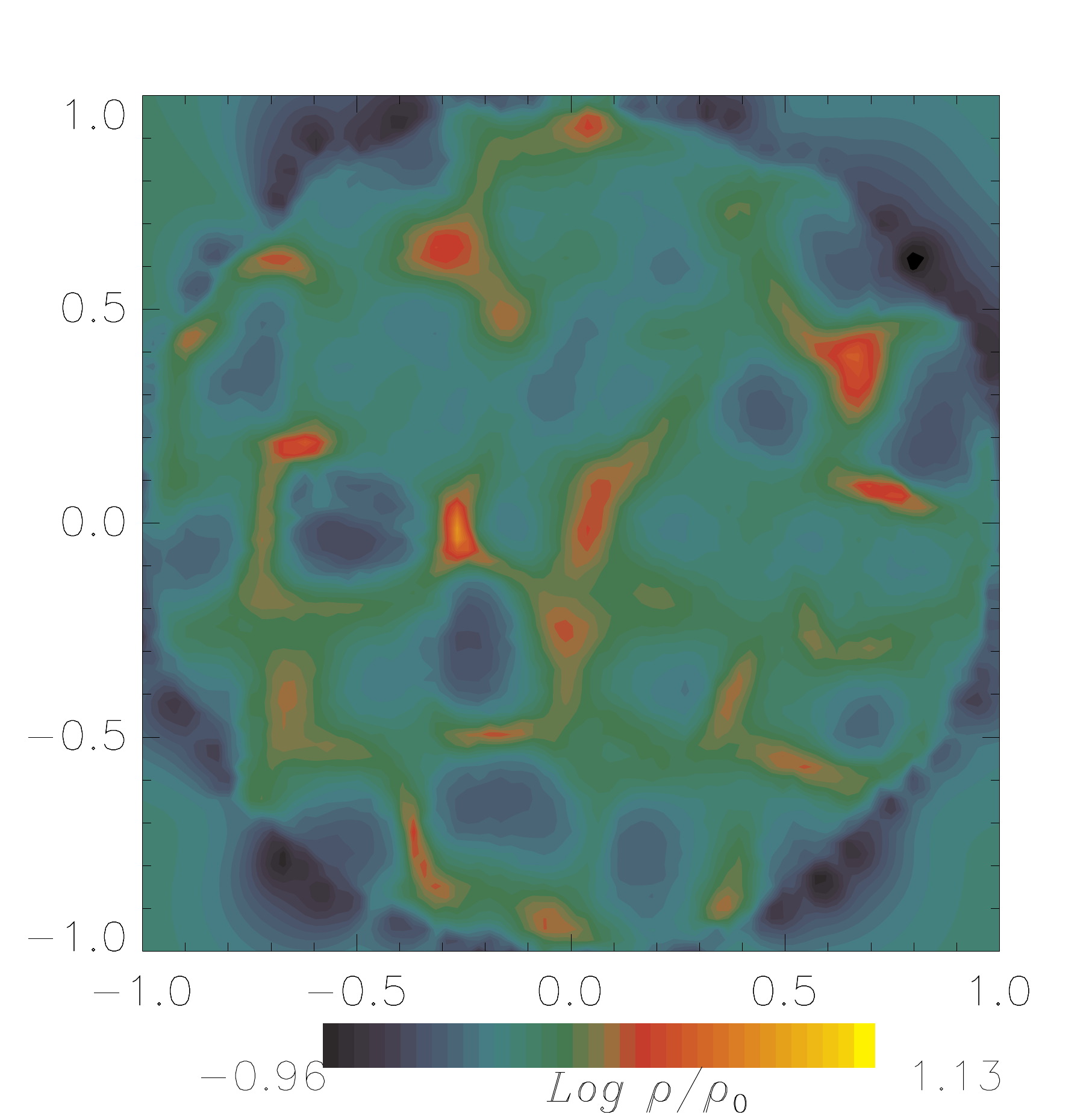} & \includegraphics[width=2 in]{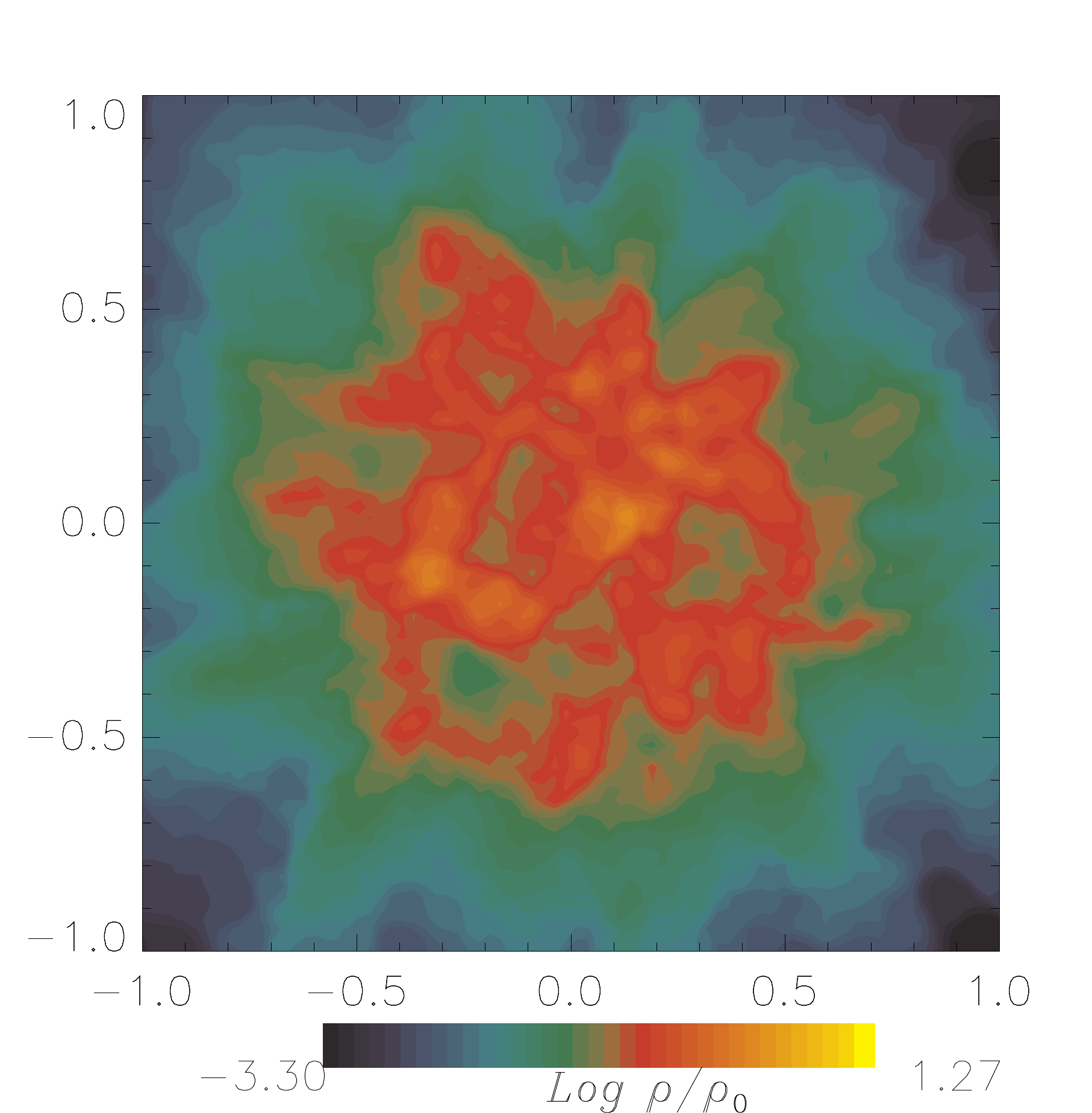} &
\includegraphics[width=2 in]{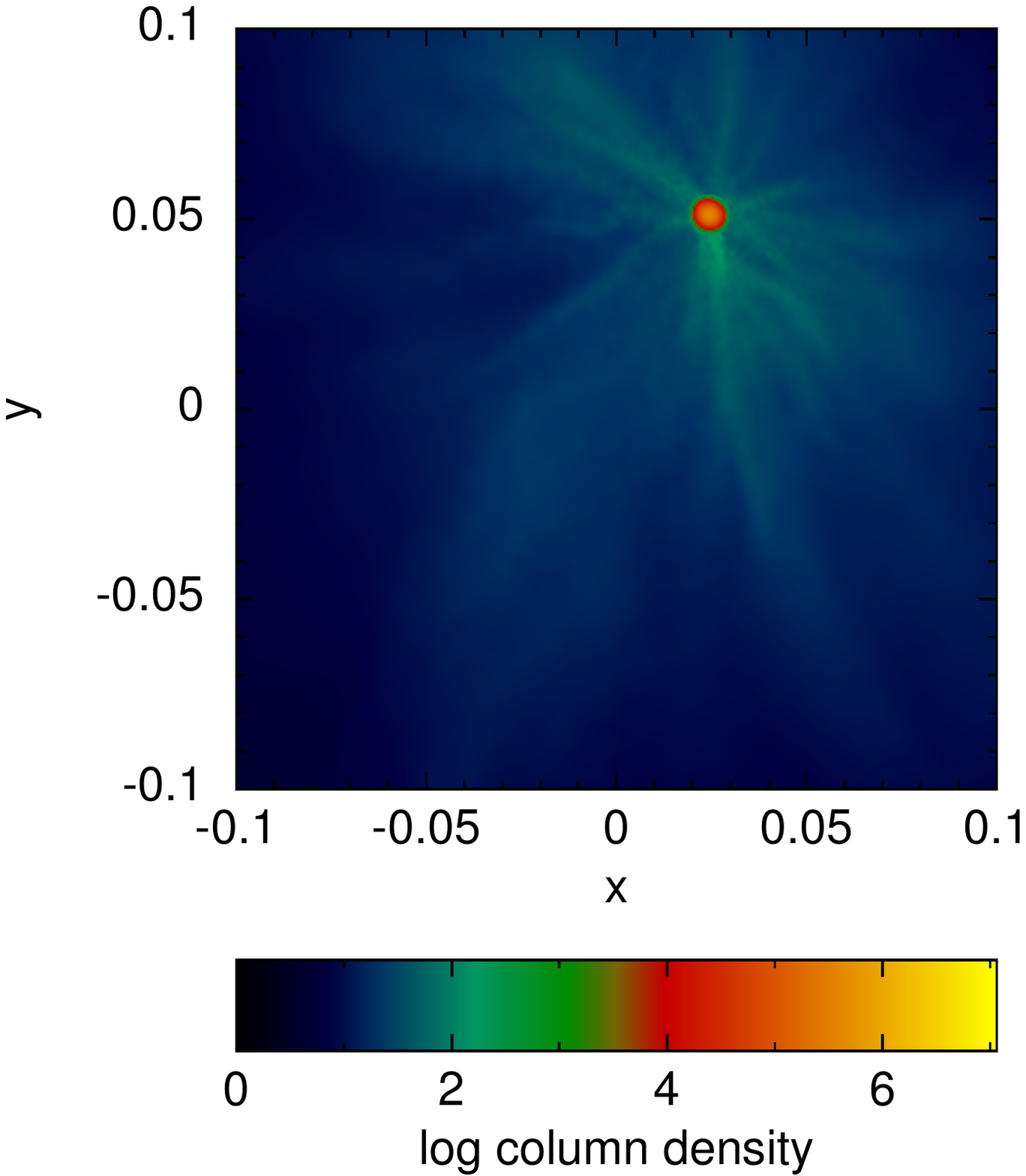}
\end{tabular}
\caption{\label{Seg7} Iso-density  plots for model
21.}
\end{figure}

\begin{figure}
\begin{tabular}{ccc}
\includegraphics[width=2 in]{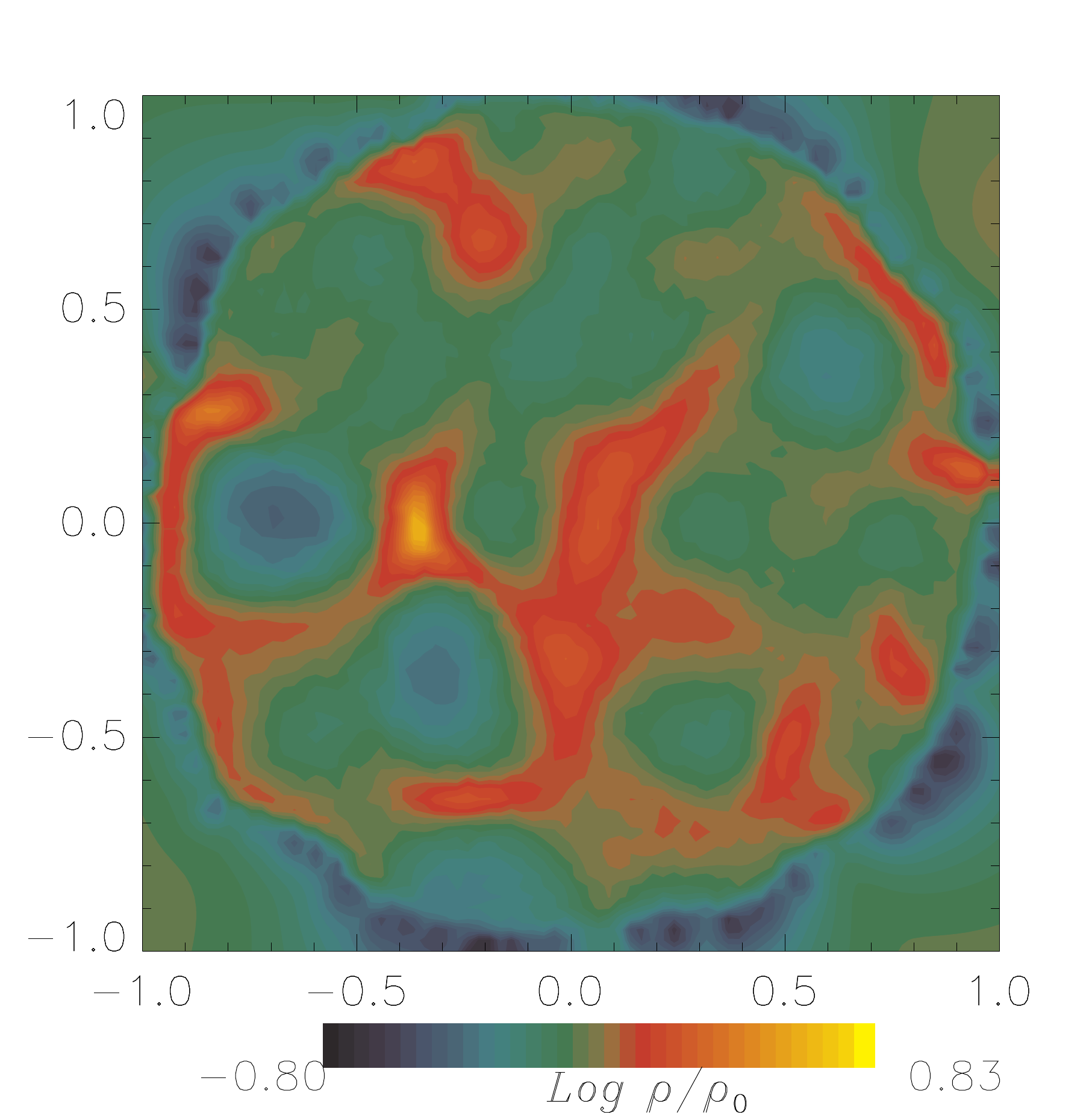} & \includegraphics[width=2 in]{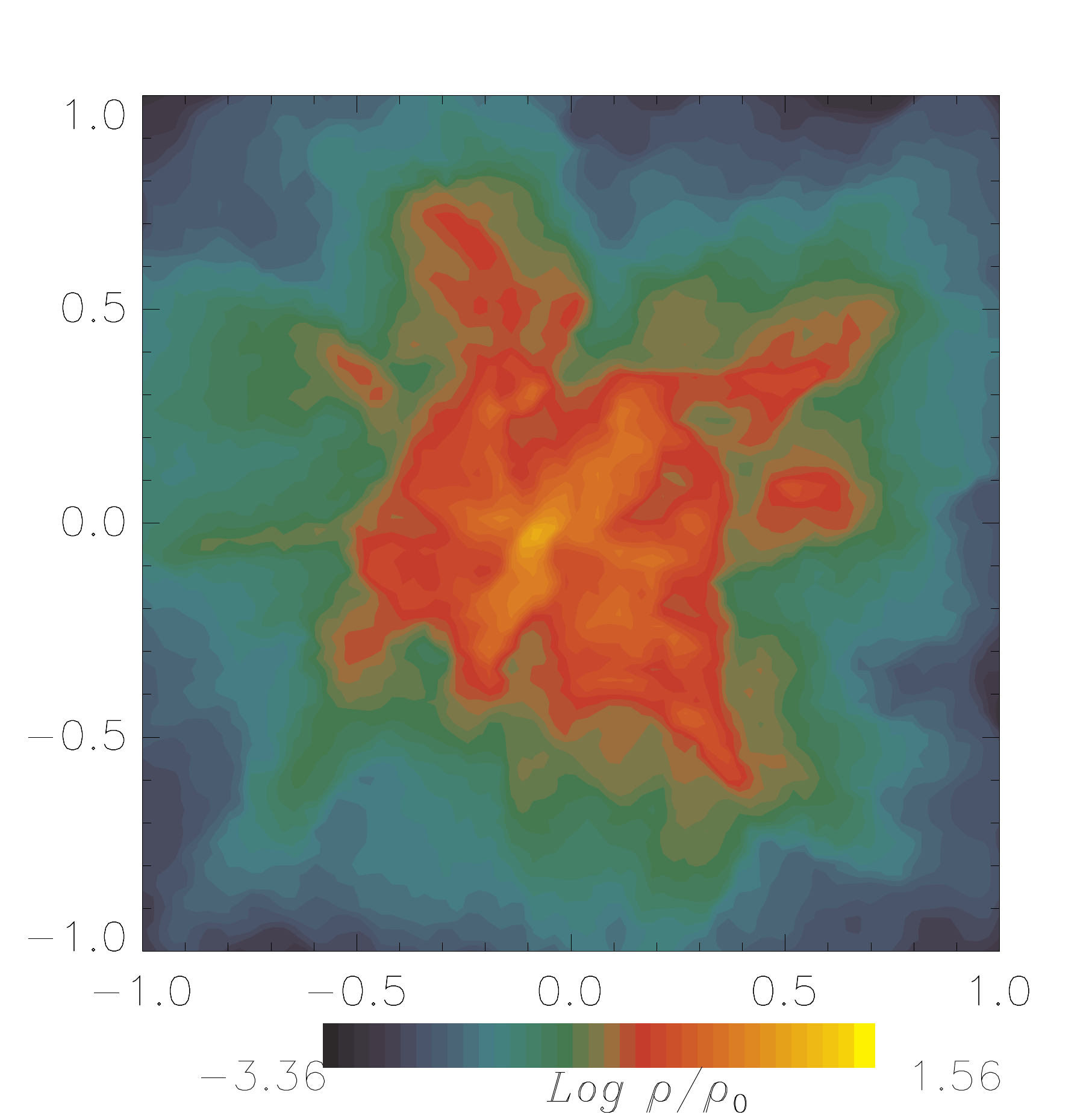} &
\includegraphics[width=2 in]{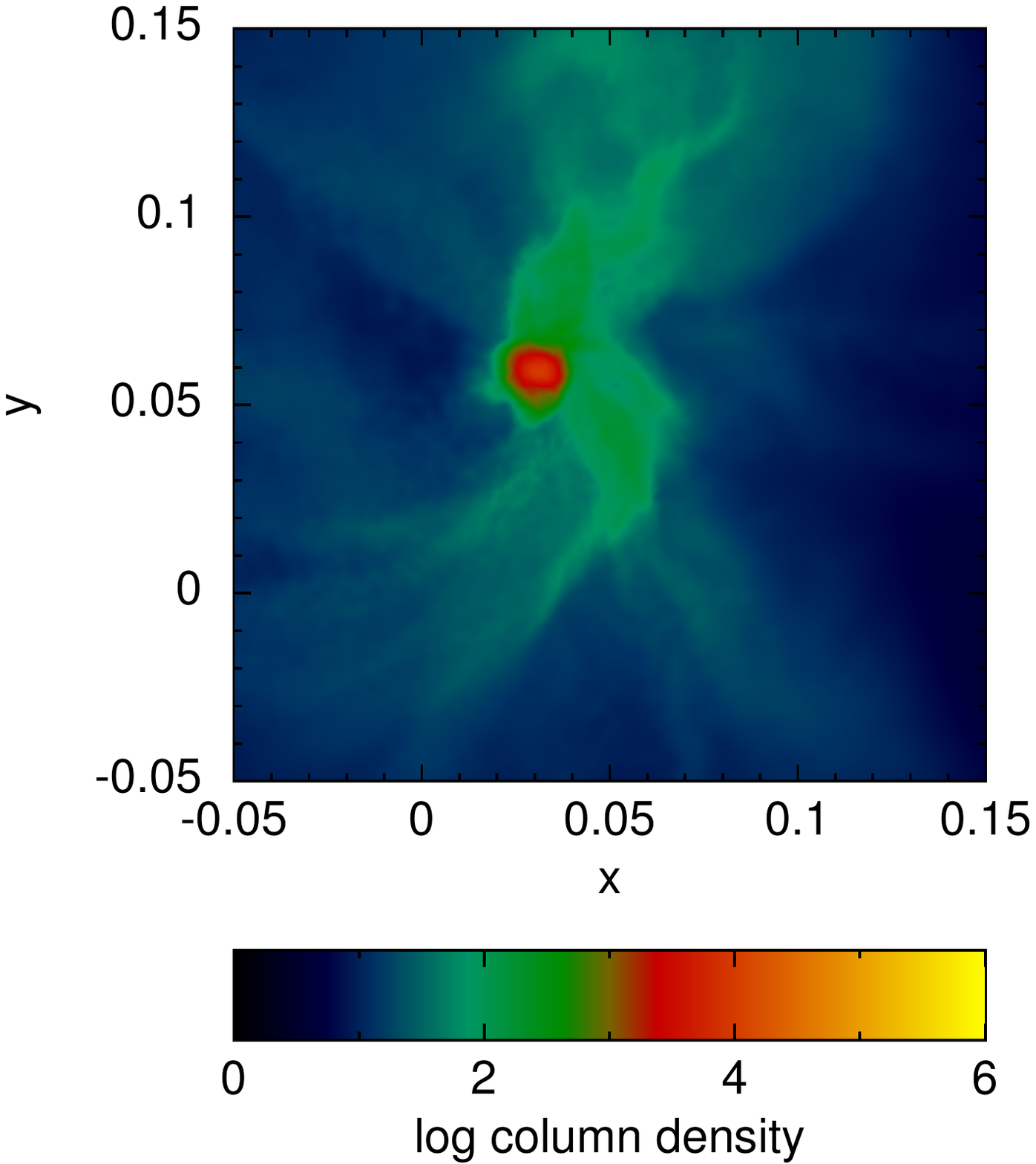}
\end{tabular}
\caption{ \label{Seg8} Iso-density  plots for model
22.}
\end{figure}

When the number of Fourier modes increases, then the number of
over-density clumps induced initially also increases, so that an
homogeneous distribution of small length over-density clumps is
formed; see the left panel of Fig.~\ref{Seg5} and compare it with
the  left panel of Fig.~\ref{Seg1}. When the size of the
Fourier mesh increases, which is the case of models 20--22, then
we observe the formation of a smaller number of over-density clumps; see
the left panels Figs.~\ref{Seg6} and \ref{Seg7} and compare
them with those of Figs.~\ref{Seg2} and \ref{Seg3}. However,
for model 22, the length of the over-density clumps begins also to
increase; see the left panel of Fig.~\ref{Seg8} and compare it
with that of Fig.~\ref{Seg4} of model 18.

We thus conclude that with a larger number of modes, the loss of the
homogeneity and isotropy of the initial over-density clump distribution is more delayed;
see the left panels of Figs.~\ref{Seg6}, \ref{Seg7}, and \ref{Seg8}.

\begin{figure}
\begin{tabular}{ccc}
\includegraphics[width=2 in]{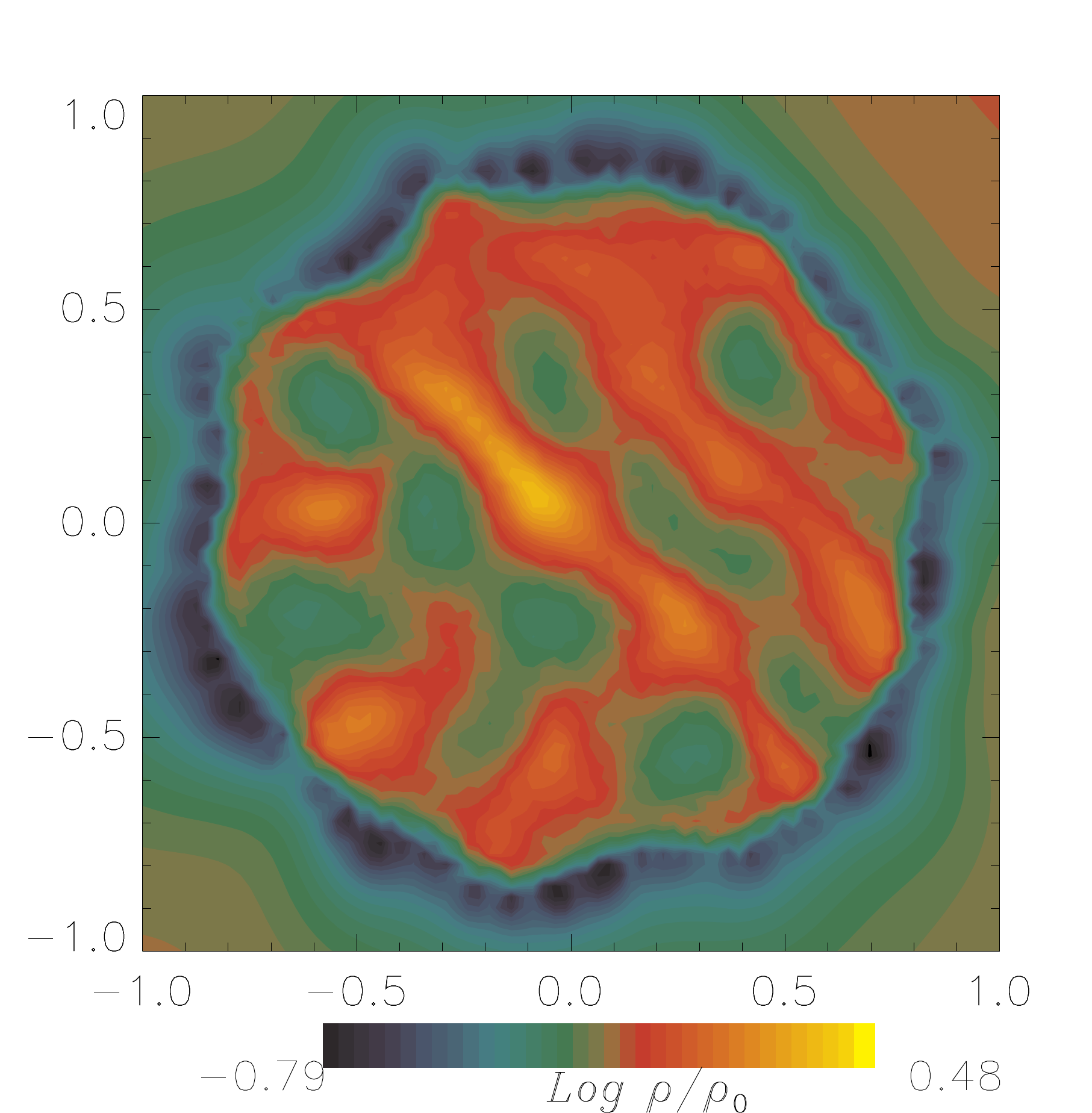} & \includegraphics[width=2 in]{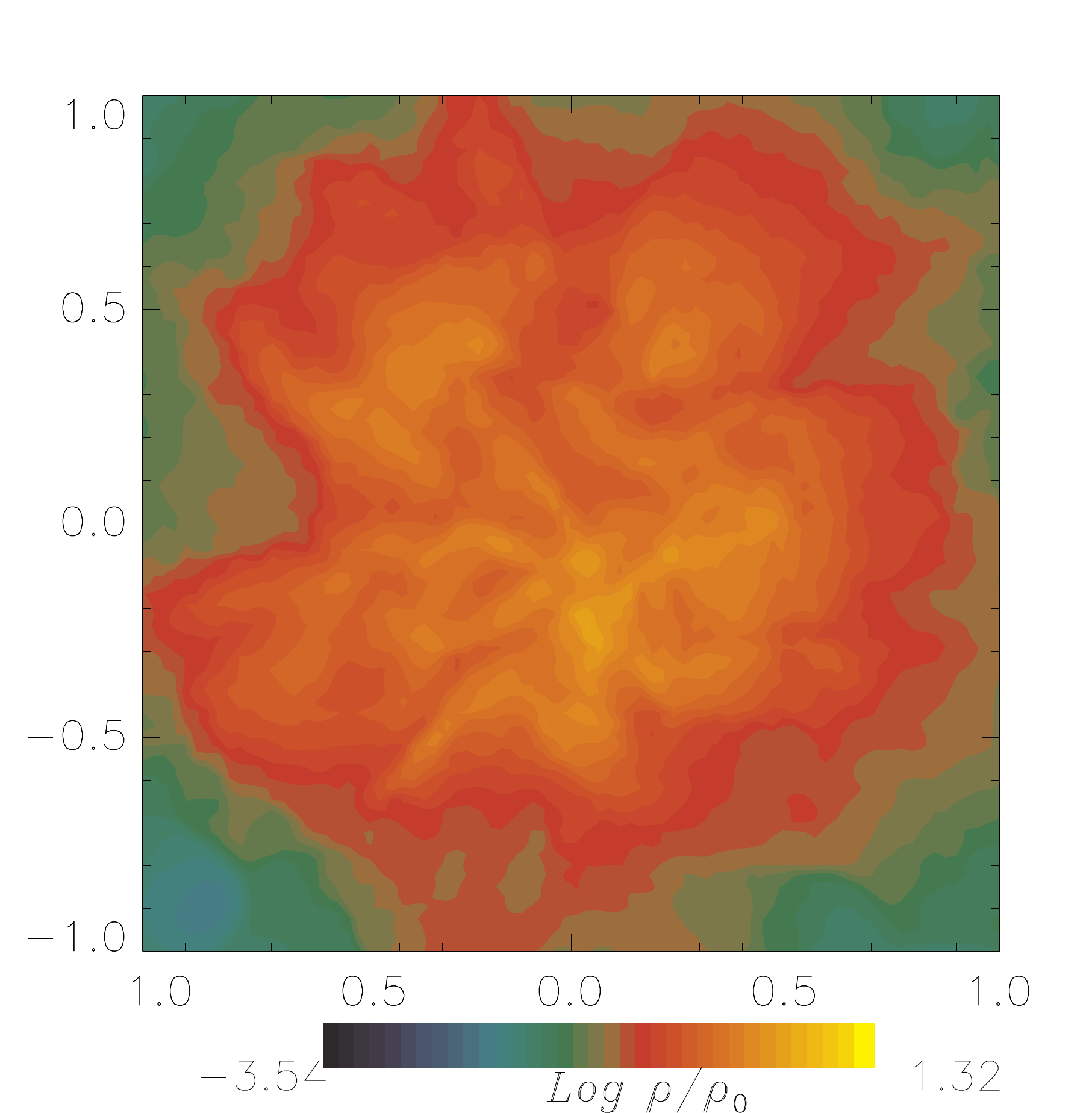} &
\includegraphics[width=2 in]{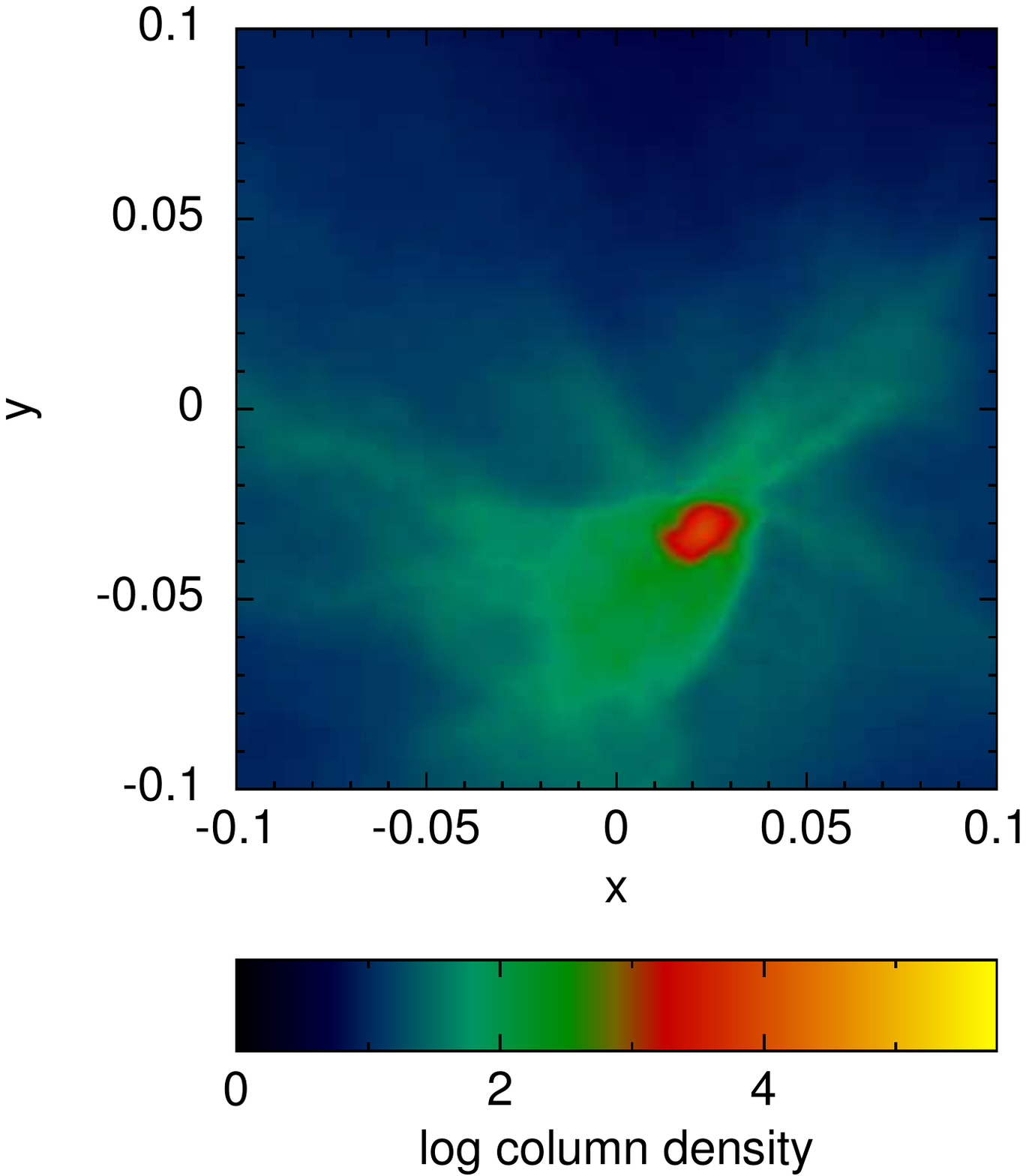}
\end{tabular}
\caption{\label{MosB2s4rmp} Iso-density plots for
model 8.}
\end{figure}

We do not see any significant difference between the initial panels of
models 19--20 shown in Figs.~\ref{Seg5}--\ref{Seg6} and those of models 5--6
shown in Figs.~\ref{MosB2srmp}--\ref{MosB2s2rmp}, so that the two types of
turbulence spectra are almost indistinguishable. However, by comparing the initial
over-density distribution of clumps for models 21 and 22, shown in Figs.~\ref{Seg7} and
\ref{Seg8}, with those of models 7 and 8, shown in Figs.~\ref{MosB2s3rmp} and \ref{MosB2s4rmp},
we see that the over-density clumps are in general larger for the divergence-free turbulence type
than those generated with the curl-free turbulence type.
\subsection{Models 23--25}
\label{subsec:results23-25}

\begin{figure}
\begin{tabular}{ccc}
\includegraphics[width=2 in]{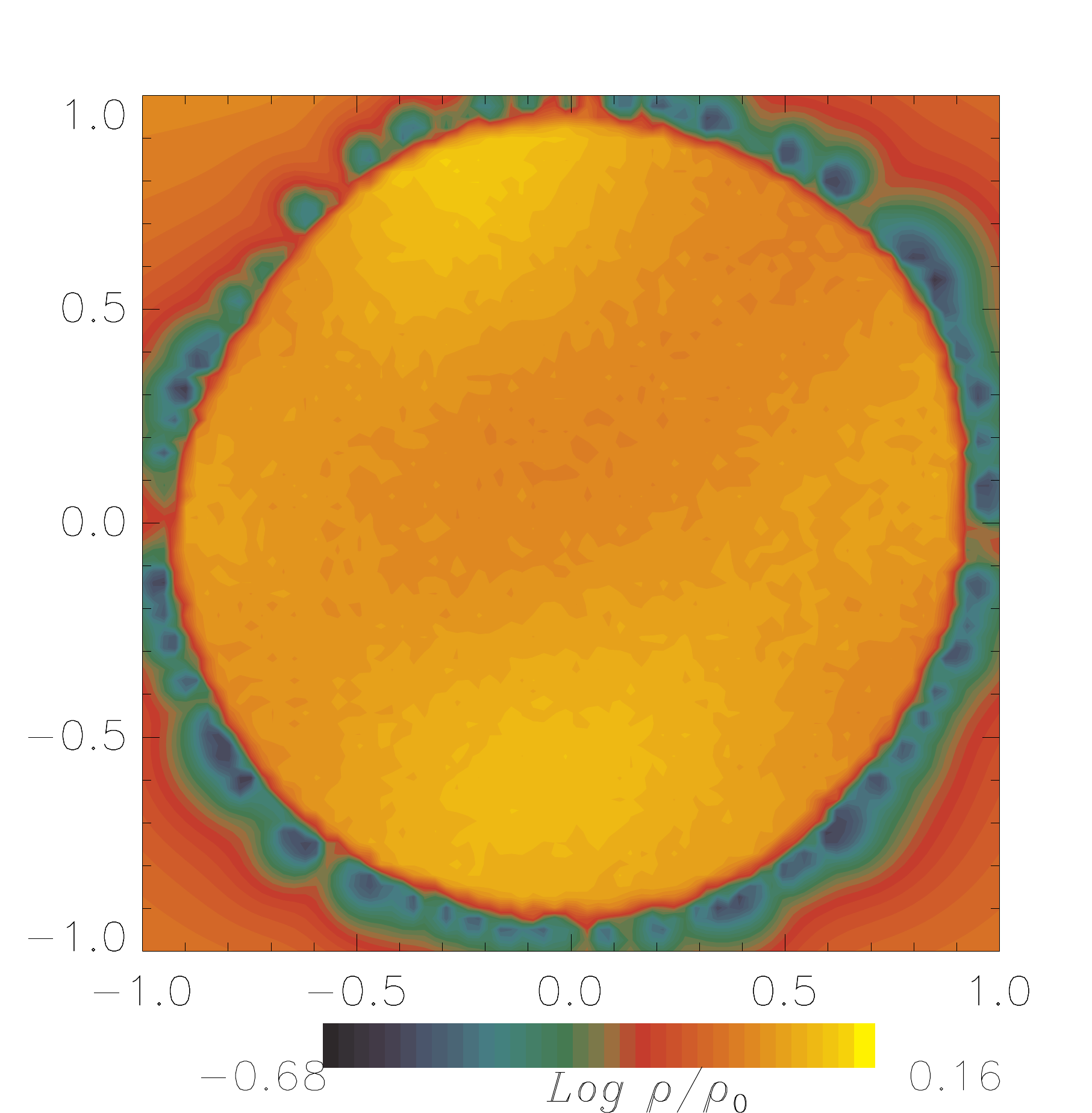} & \includegraphics[width=2 in]{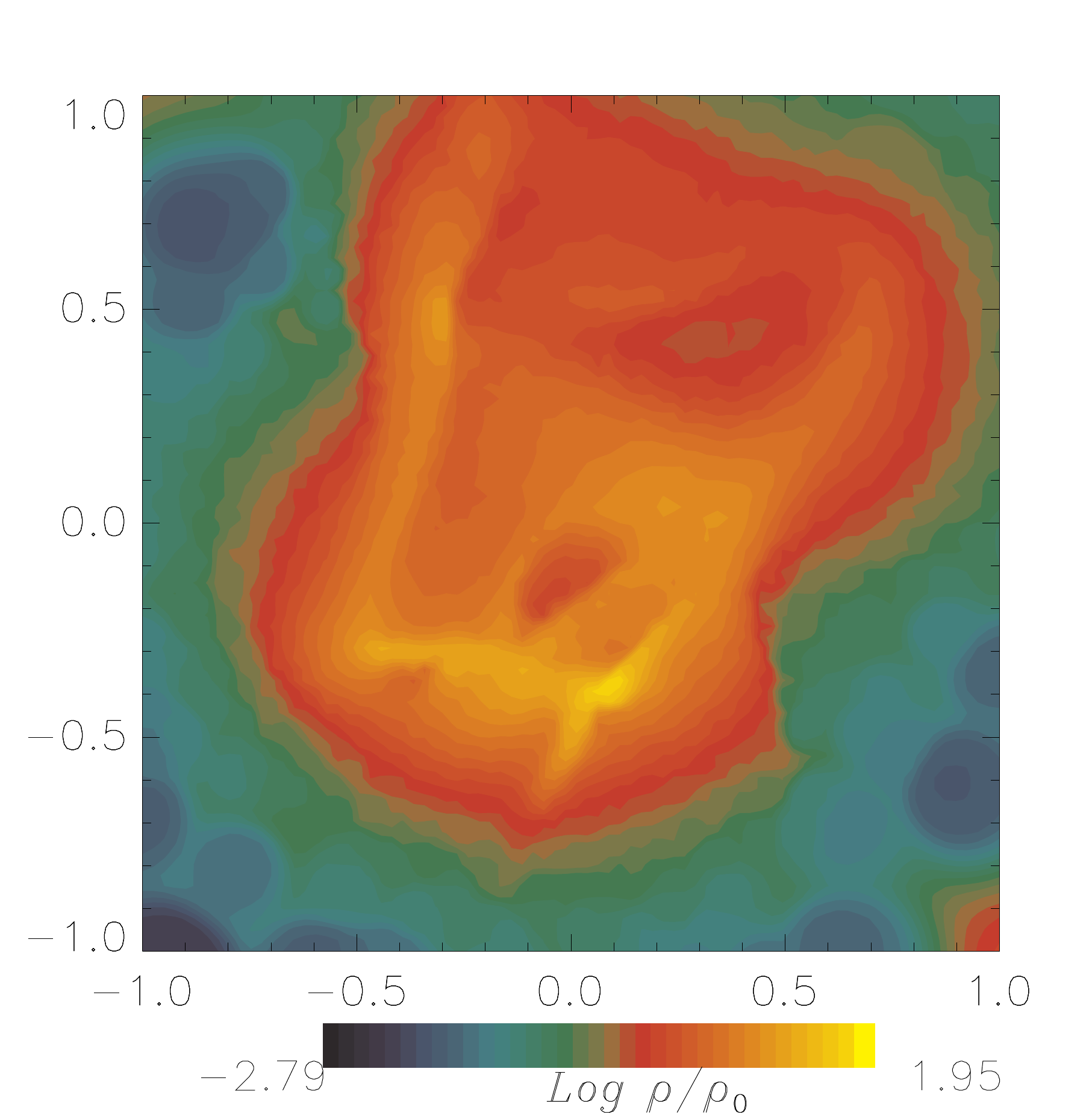} &
\includegraphics[width=2 in]{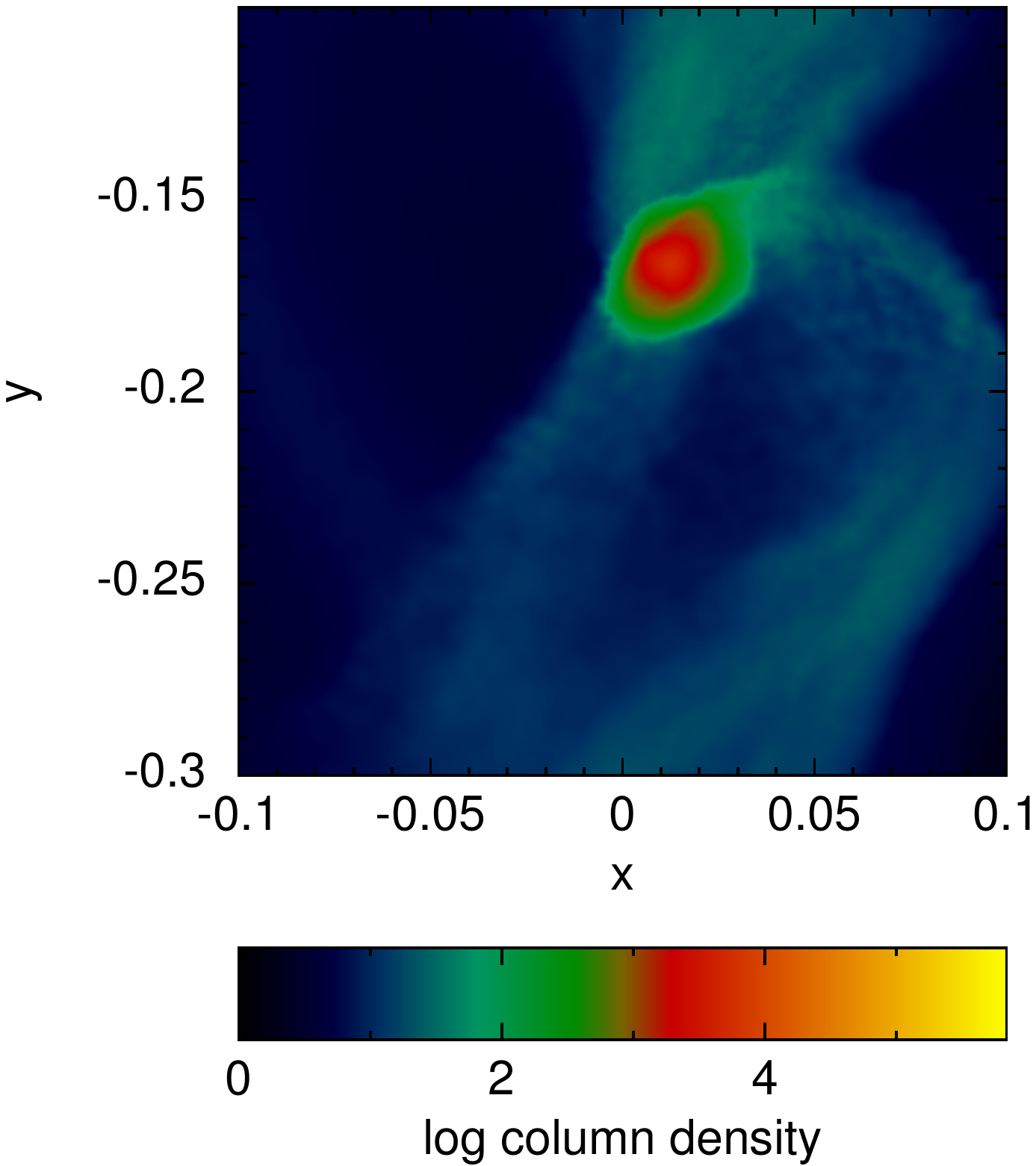}
\end{tabular}
\caption{\label{Seg9} Iso-density  plots for model
23.}
\end{figure}

\begin{figure}
\begin{tabular}{ccc}
\includegraphics[width=2 in]{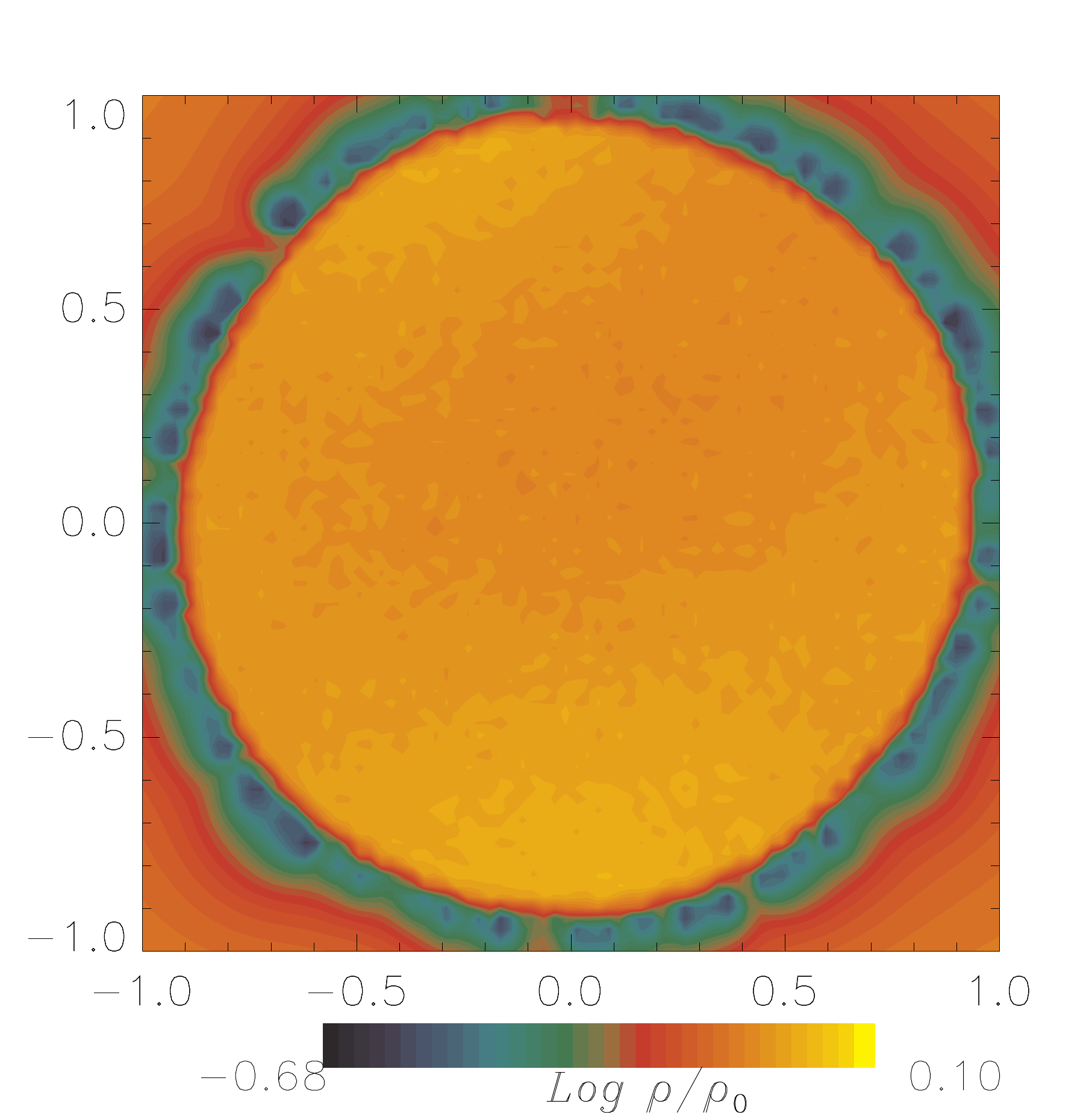} & \includegraphics[width=2 in]{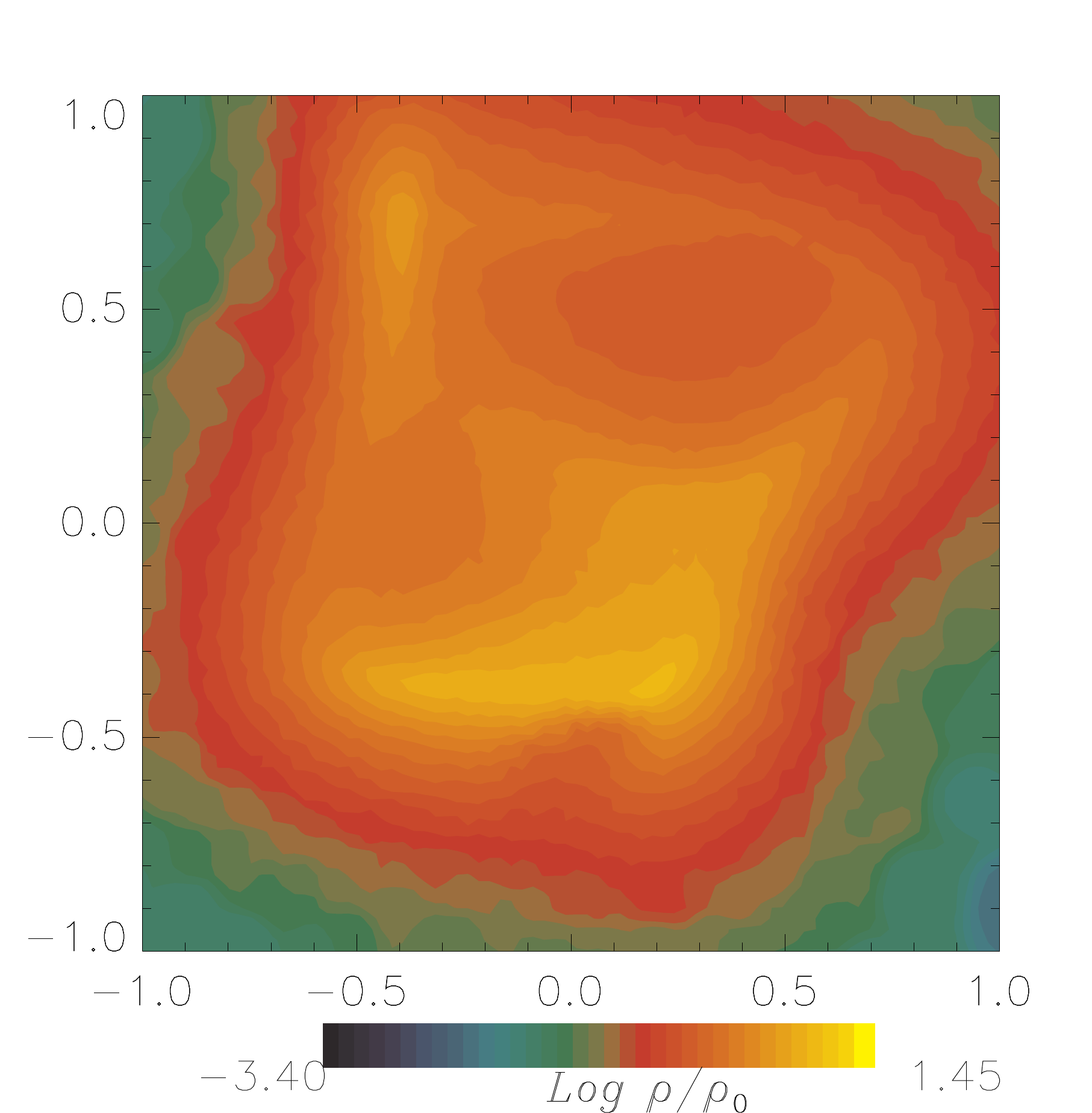} &
\includegraphics[width=2 in]{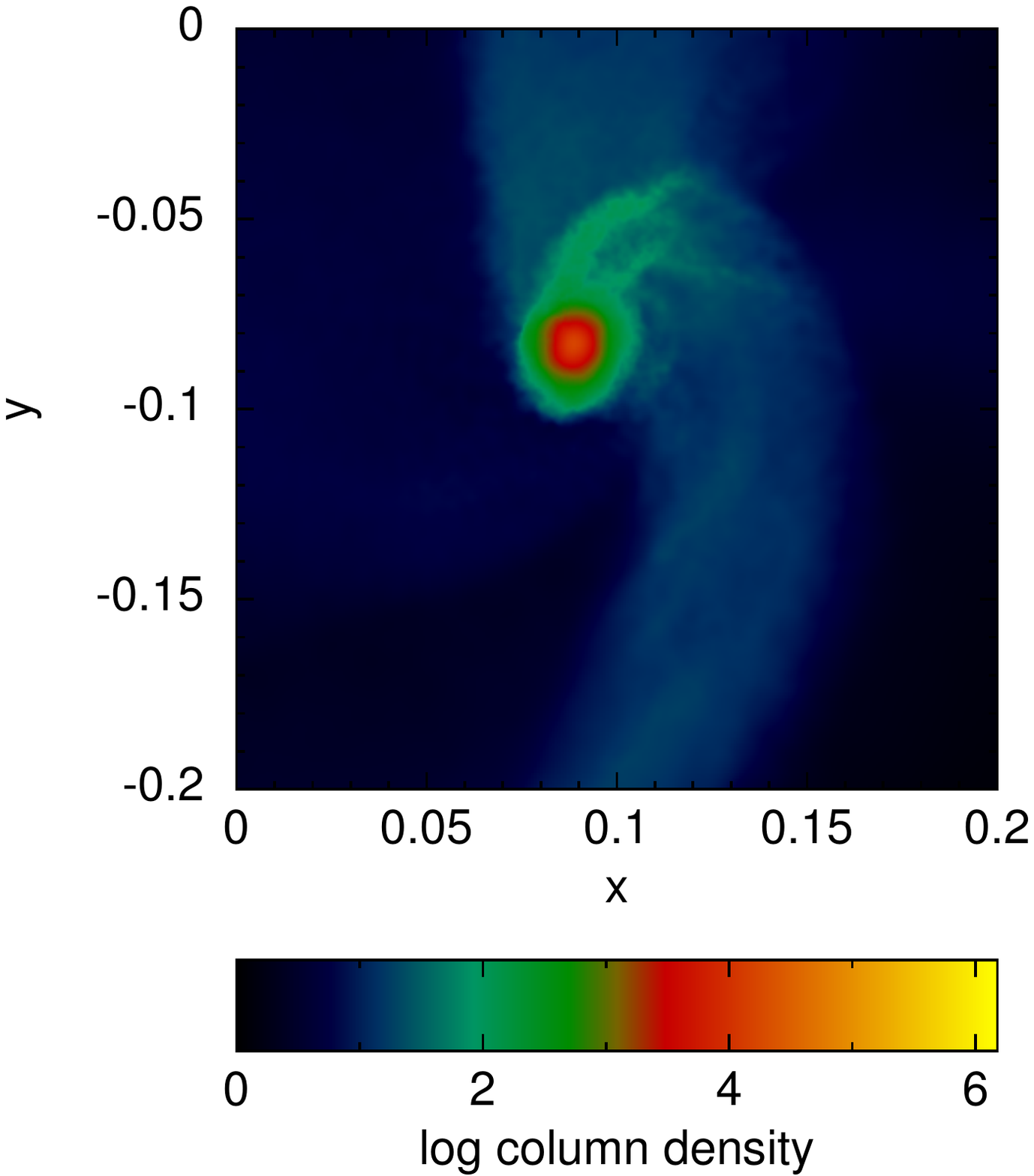}
\end{tabular}
\caption{\label{Seg10} Iso-density plots for model
24.}
\end{figure}

\begin{figure}
\begin{tabular}{ccc}
\includegraphics[width=2 in]{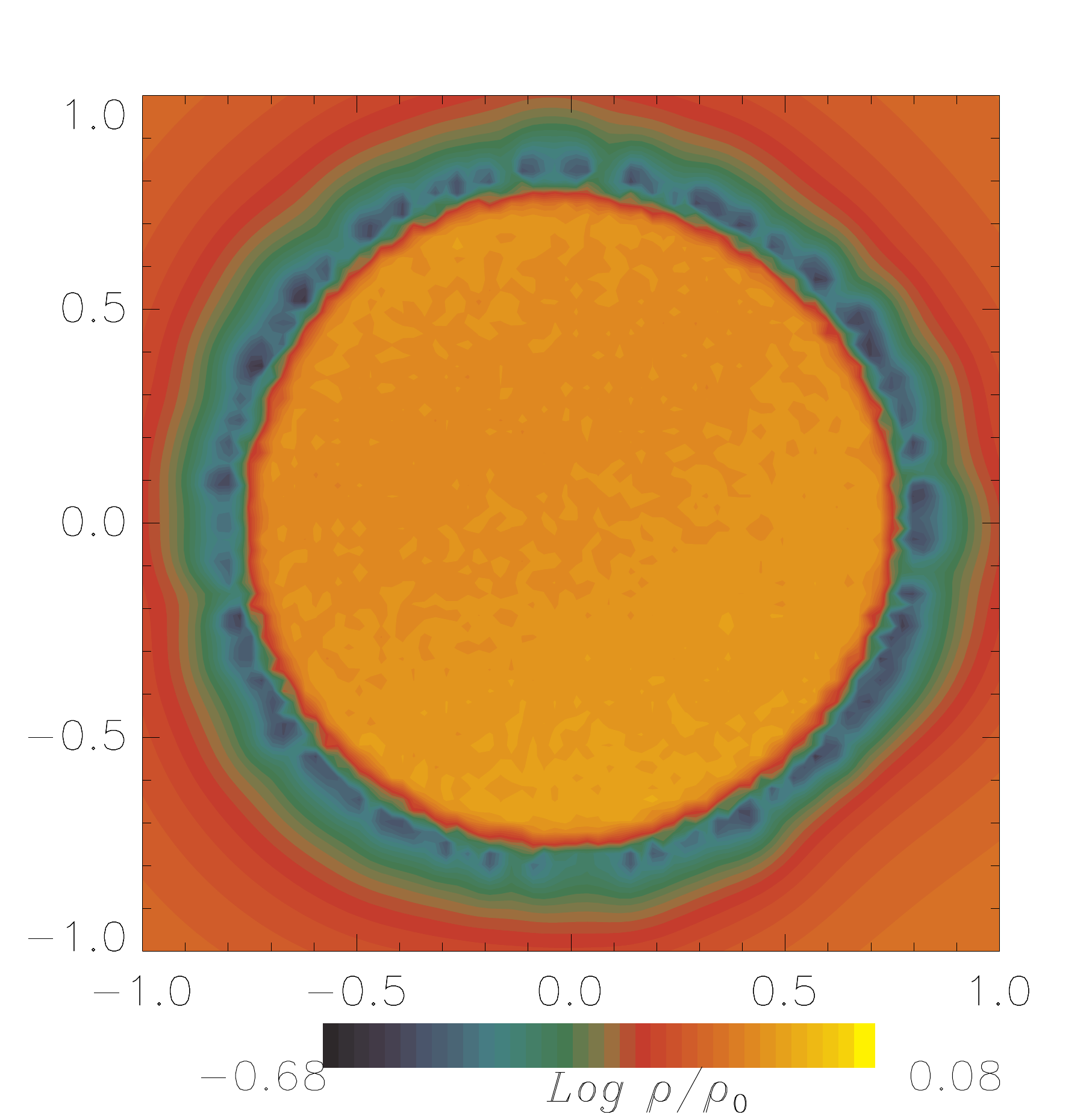} & \includegraphics[width=2 in]{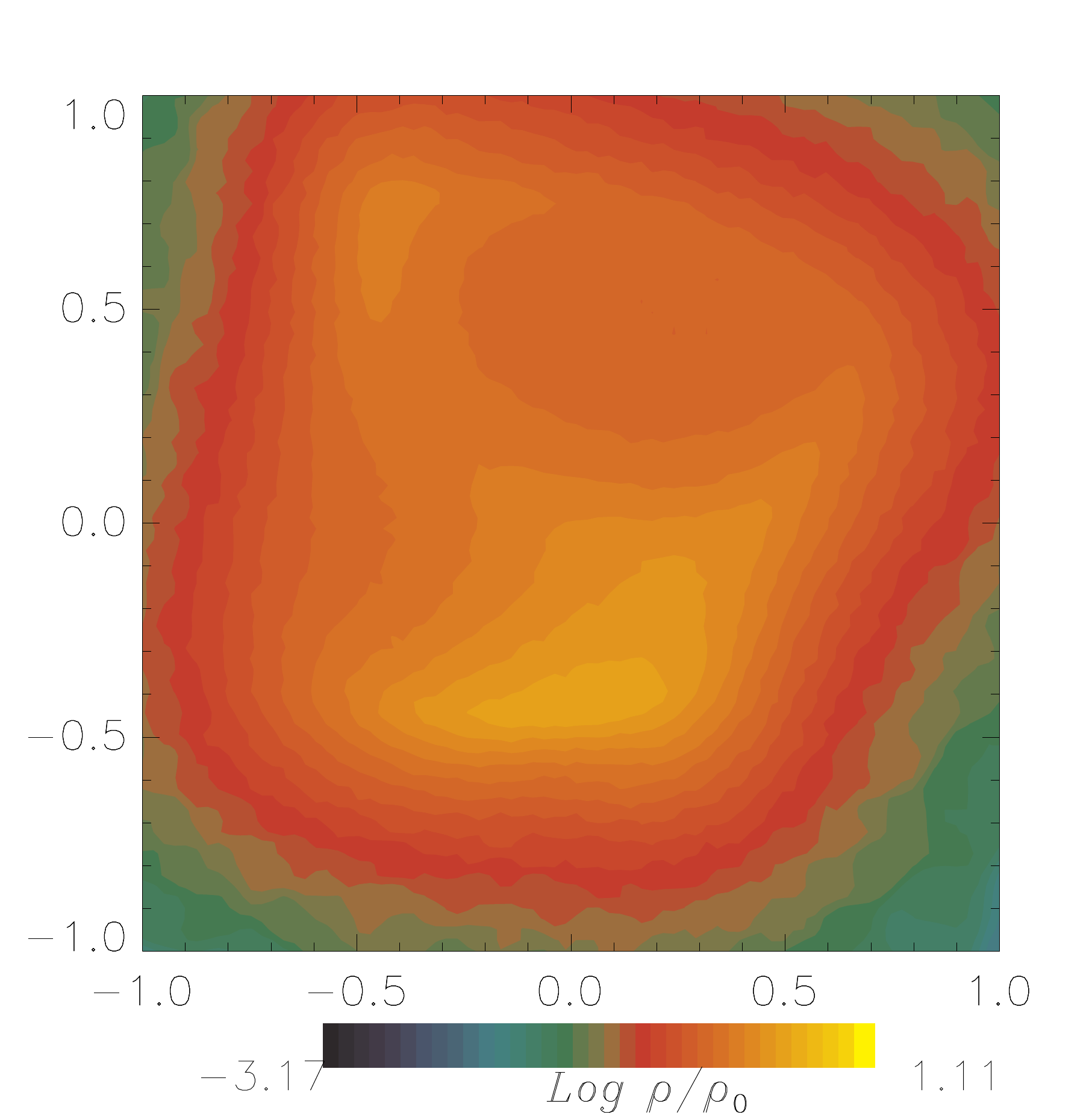} &
\includegraphics[width=2 in]{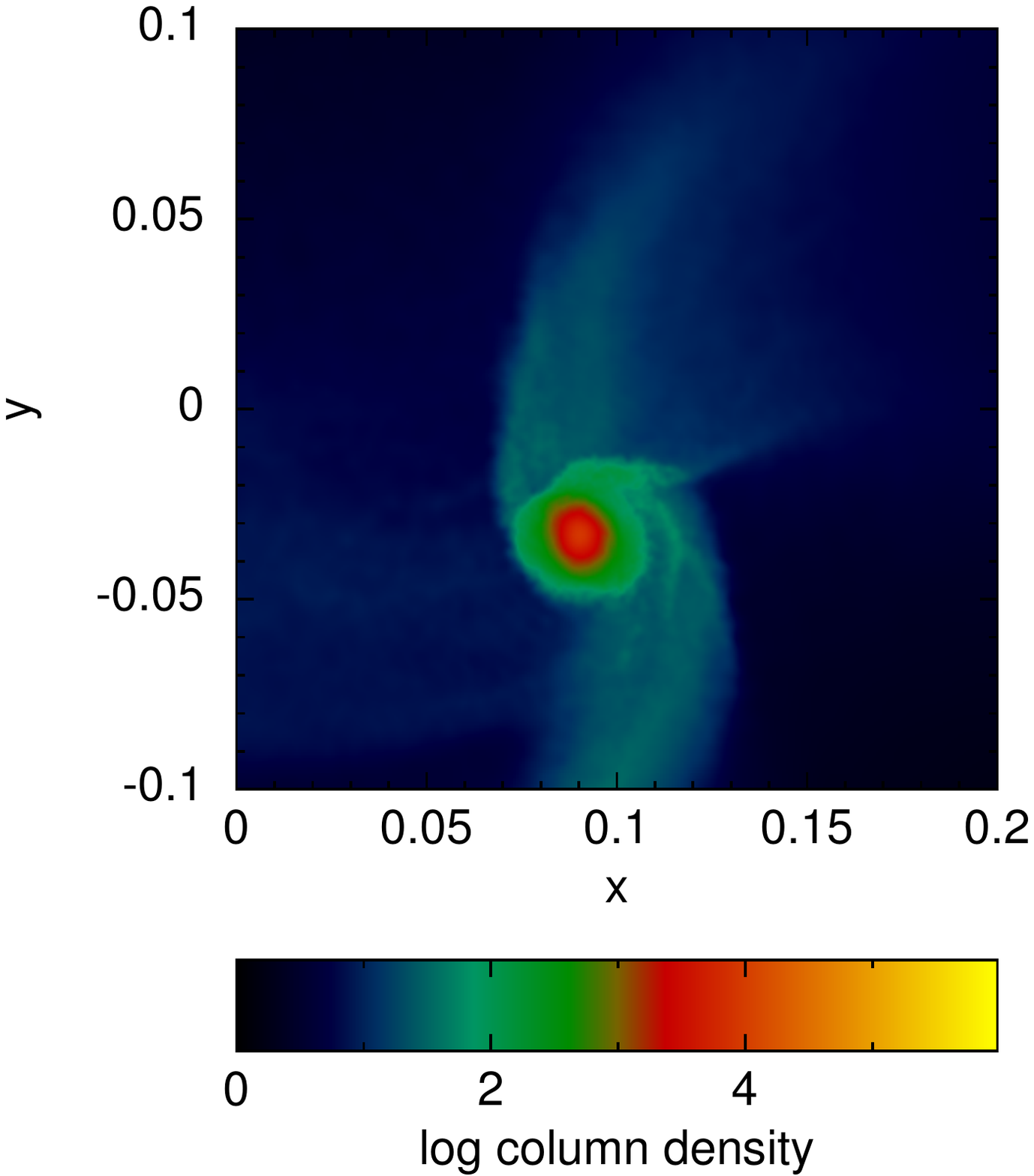}
\end{tabular}
\caption{\label{Seg11} Iso-density  plots for model
25.}
\end{figure}

When the size of the perturbation mode increases further, as for models 23--25,
then it is evident the there are only two main directions along which
the initial over-density clump distributions grow: see the left panels of
Fig.~\ref{Seg9}, \ref{Seg10} and \ref{Seg11}. For this reason, the deformation in the
core shape is very significant; see the right panels of Fig.~\ref{Seg9}, \ref{Seg10}
and \ref{Seg11}. Despite this, the final outcome of these models is still a single
protostar, which is displaced away from the central region, as a consequence of the highly
anisotropic accretion of gas.

By comparing the initial clump distributions of models 23--25 with
those of models 9--11, shown in Figs.~\ref{MosB-Osrm}, \ref{MosB-O2srm}
and \ref{MosB-Os3rm}, we note that the over-density grow in orthogonal directions
for these two set of models. This observation is expected according to Eqs.~\ref{velcasea}
and \ref{velPhi}, as the velocities are orthogonal vectors.

\subsection{Models 26--28}
\label{subsec:results26-28}

\begin{figure}
\begin{tabular}{ccc}
\includegraphics[width=2 in]{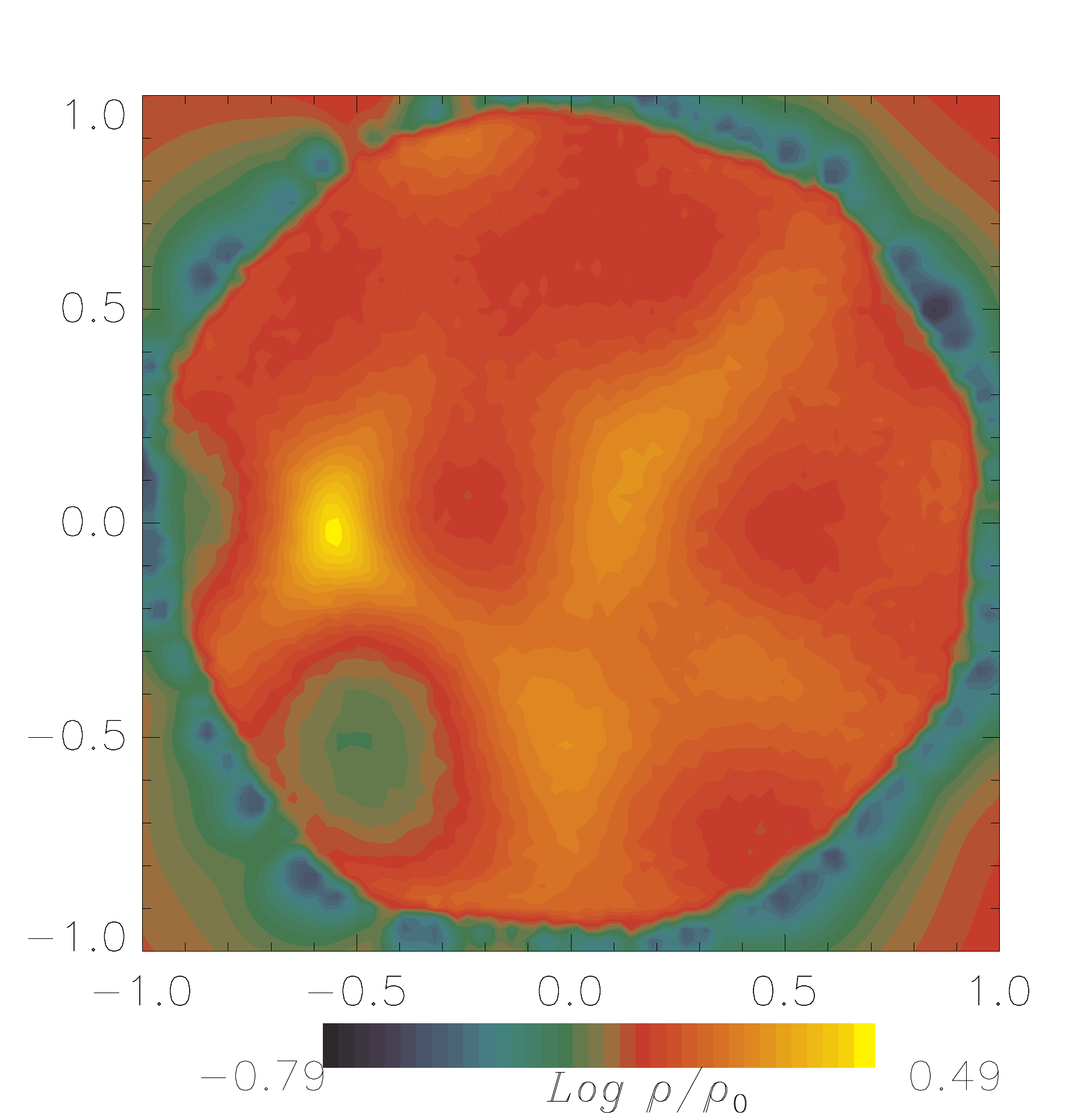} & \includegraphics[width=2 in]{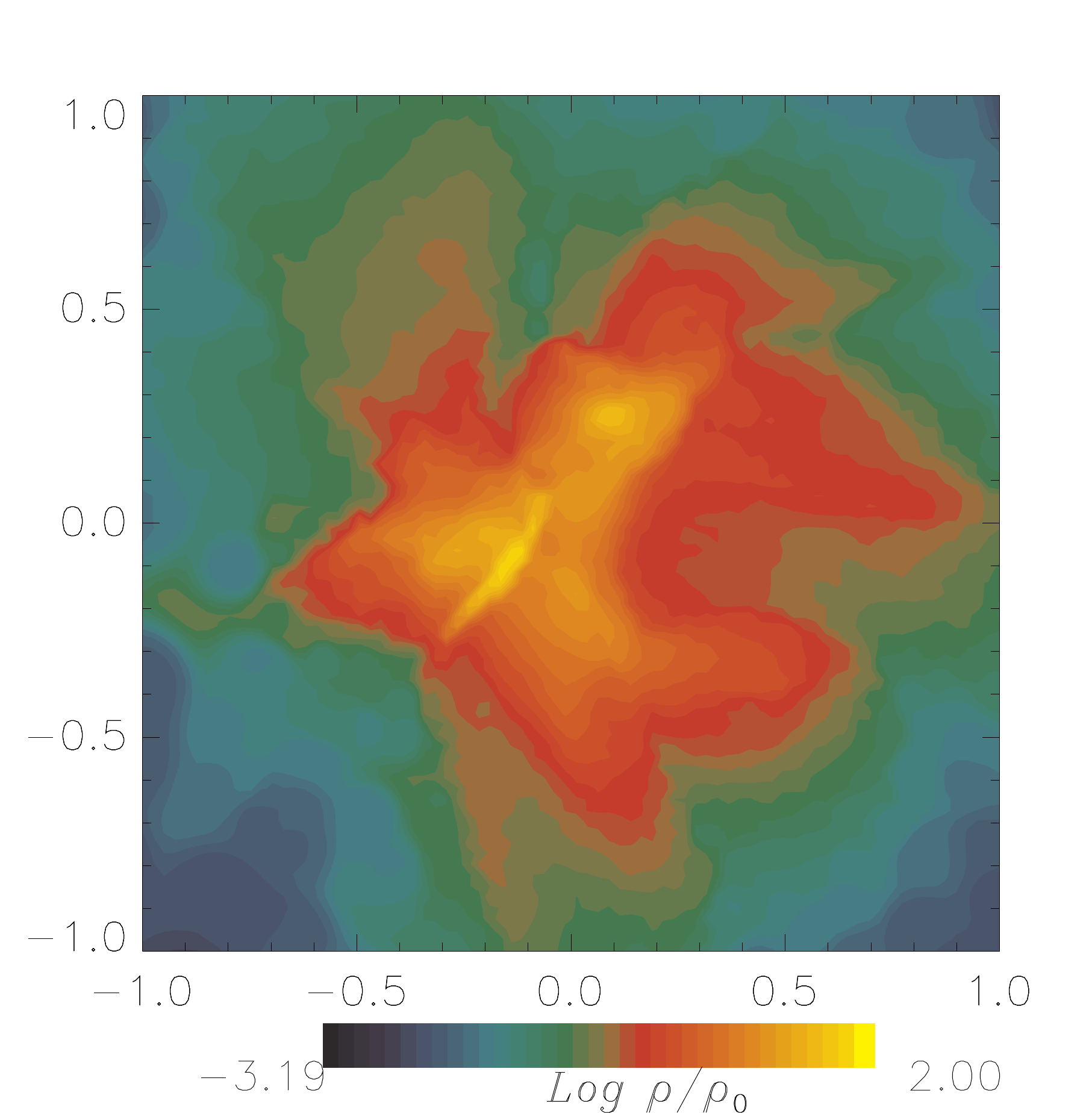} &
\includegraphics[width=2 in]{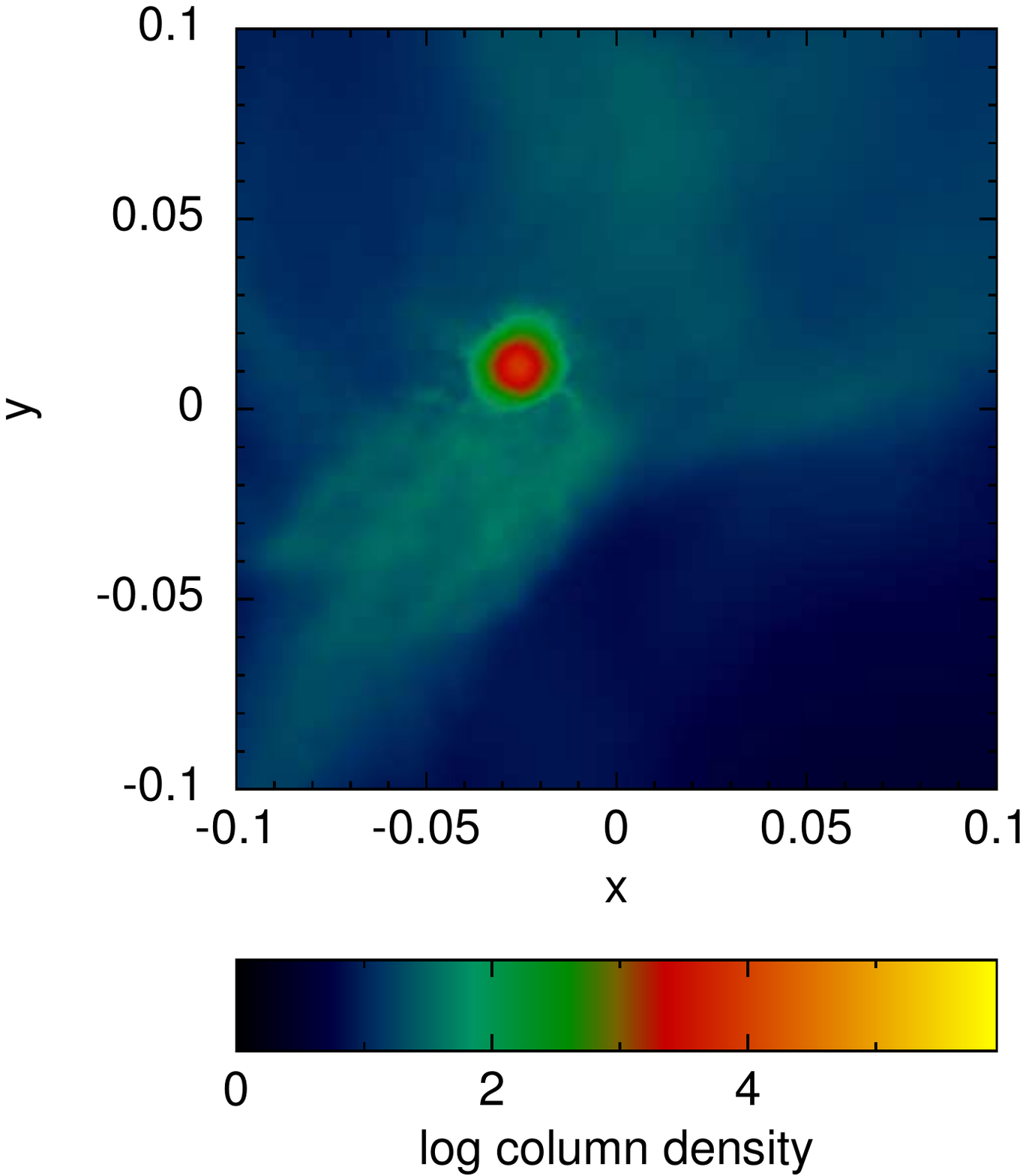}
\end{tabular}
\caption{\label{Seg12} Iso-density and plots for model
26.}
\end{figure}

\begin{figure}
\begin{tabular}{ccc}
\includegraphics[width=2 in]{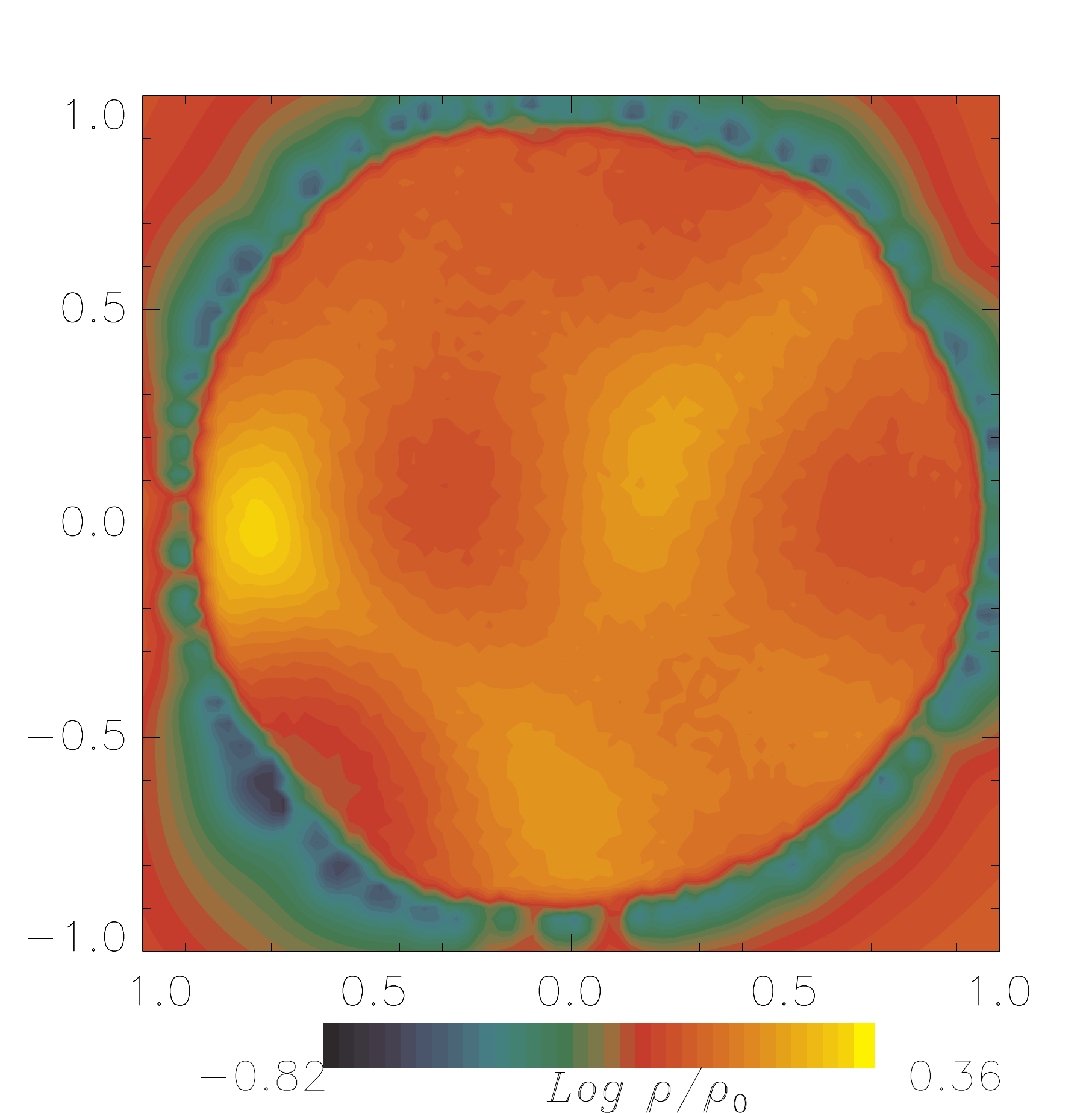} & \includegraphics[width=2 in]{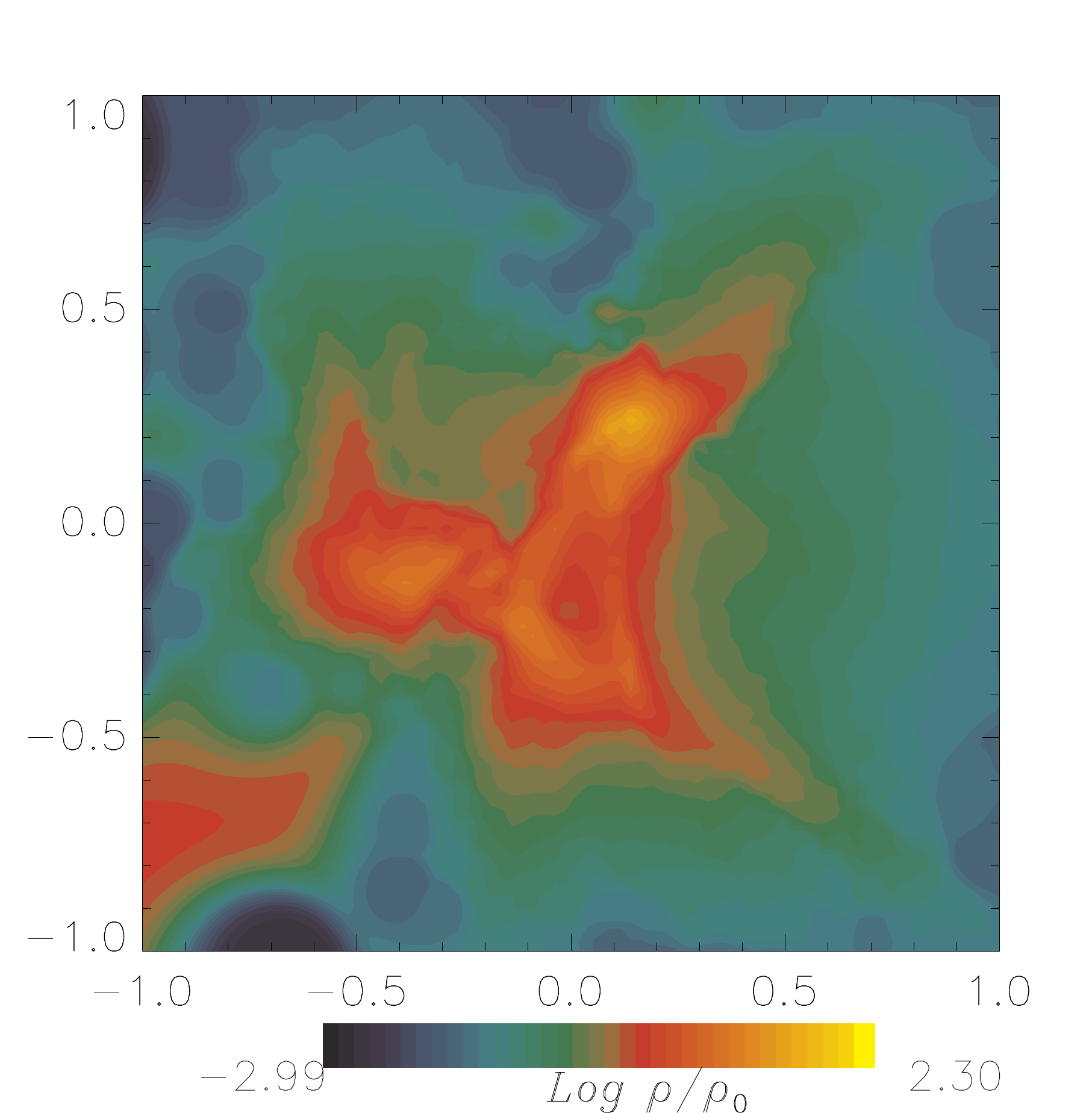} &
\includegraphics[width=2 in]{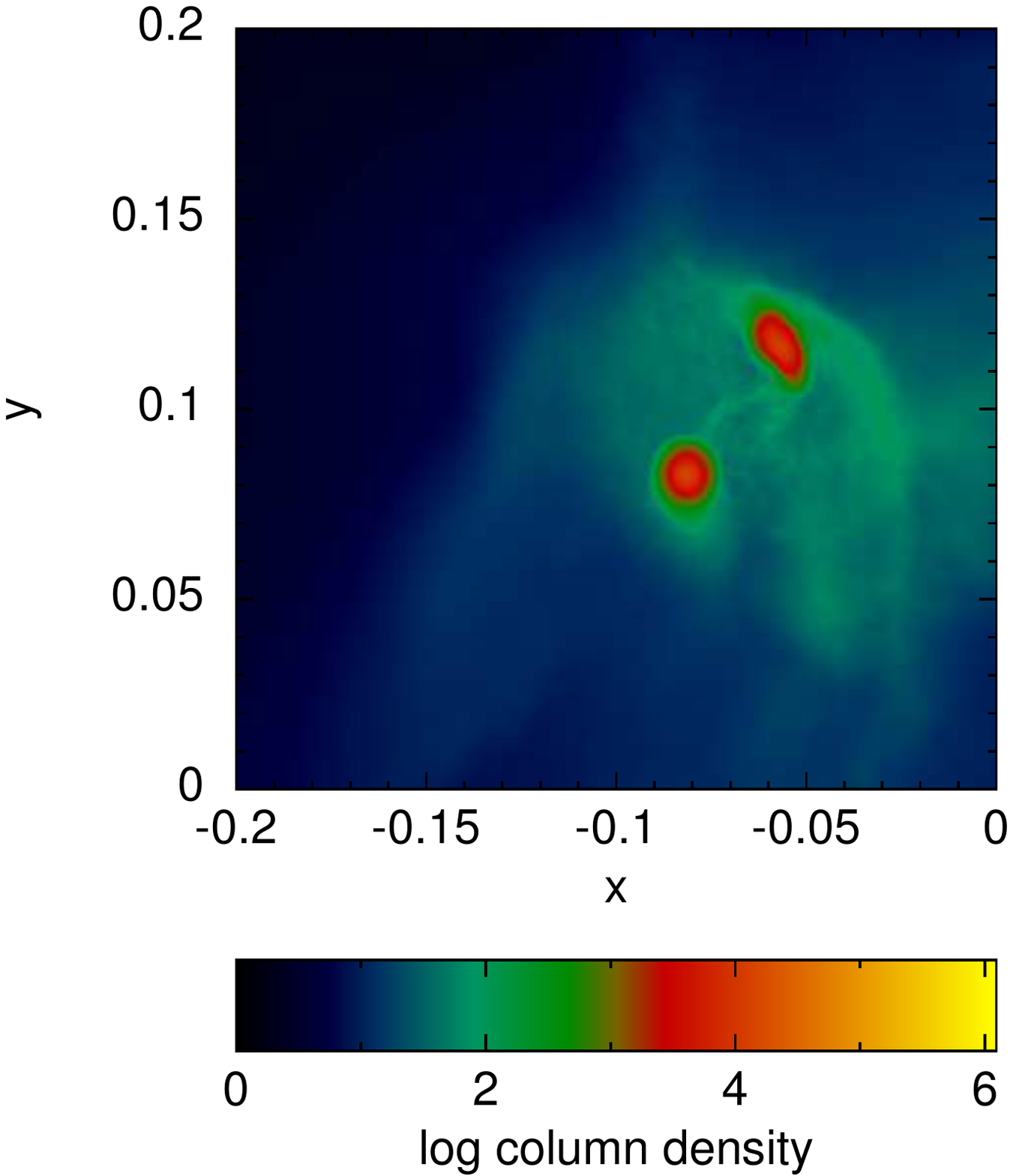}
\end{tabular}
\caption{\label{Seg13} Iso-density plots for model
27.}
\end{figure}

\begin{figure}
\begin{tabular}{ccc}
\includegraphics[width=2 in]{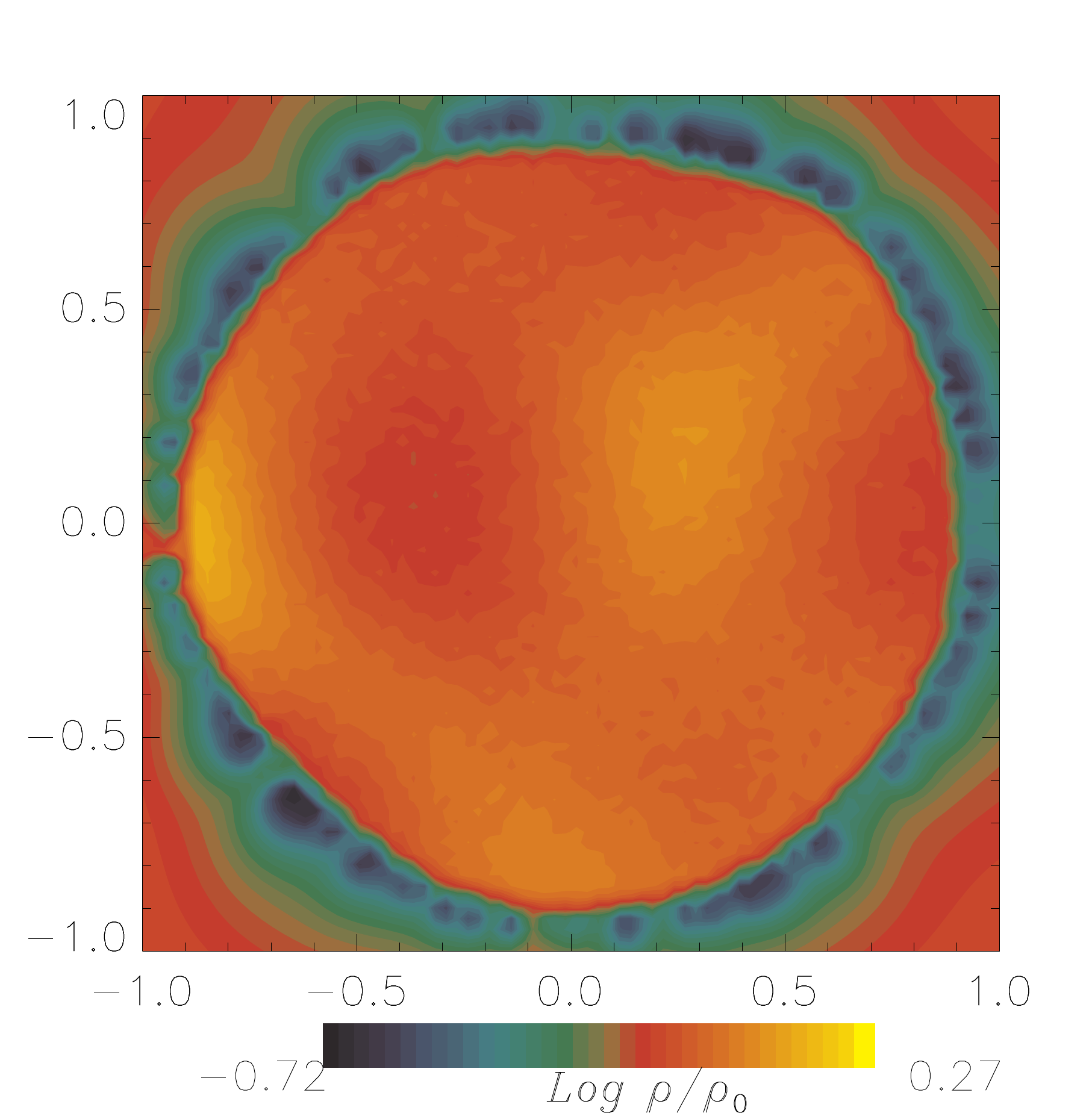} & \includegraphics[width=2 in]{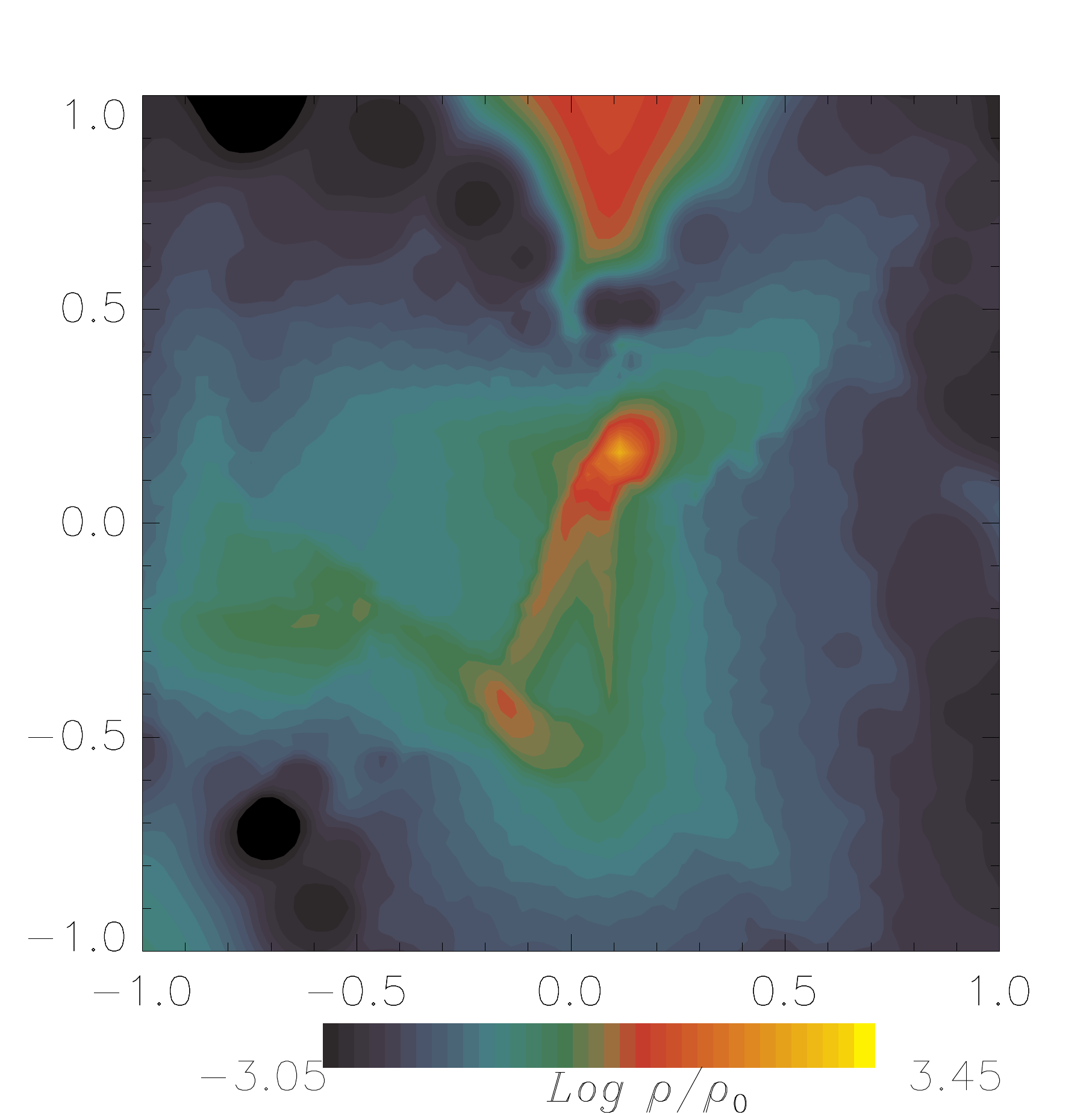} &
\includegraphics[width=2 in]{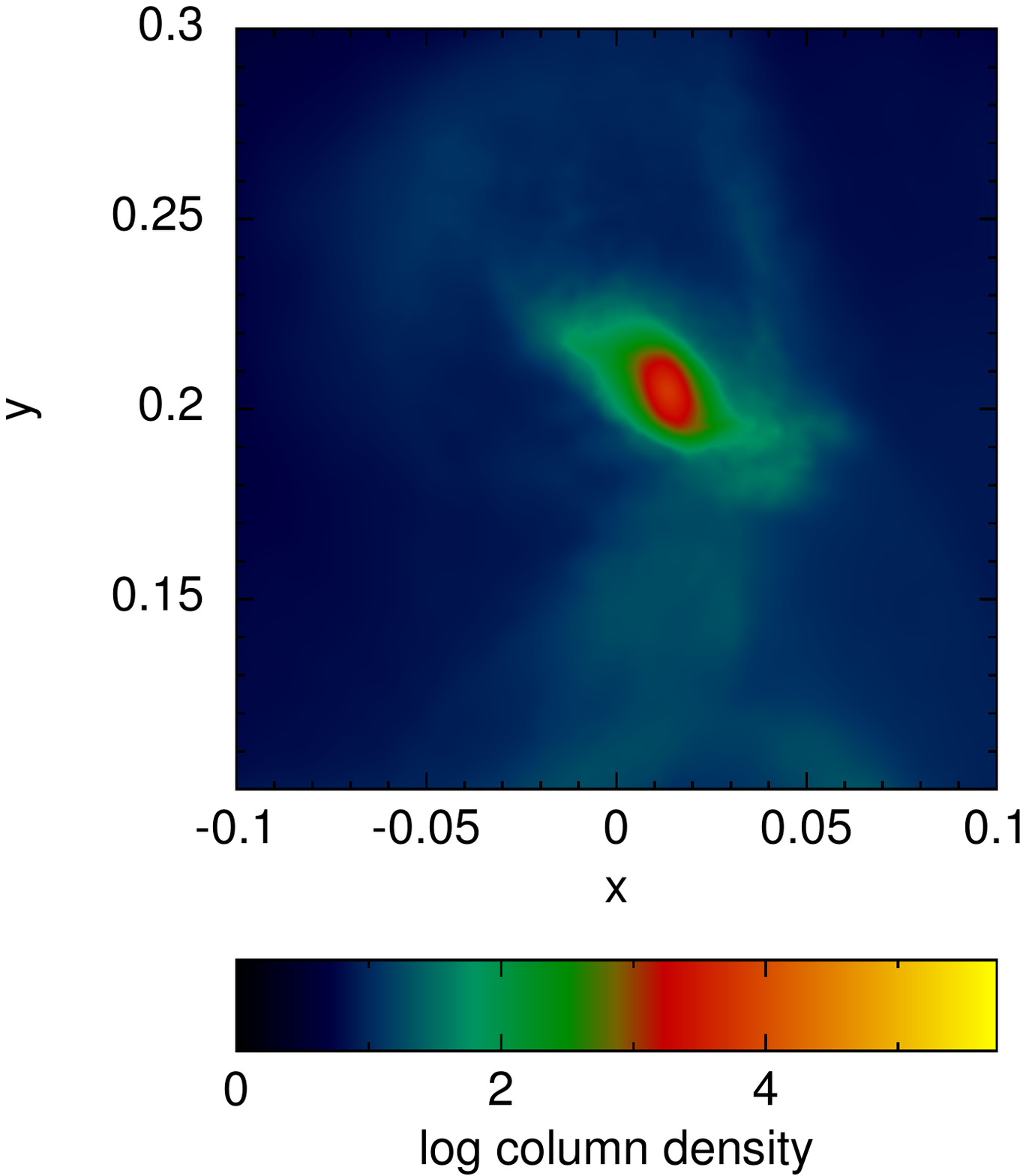}
\end{tabular}
\caption{ \label{Seg14} Iso-density  plots for model
28.}
\end{figure}

When the number of Fourier modes increases, as in models 26--27,
which are shown in Figs.~\ref{Seg12}, \ref{Seg13} and \ref{Seg14}, in
a set of models with a large number of modes, as in models 23--25 shown in
Figs.~\ref{Seg9}, \ref{Seg10} and \ref{Seg11}, then we observe that
the core is more deformed, so that even core disruption can be
achieved, as was the case with model 27, in which a binary protostars
system is formed.

\begin{figure}
\begin{tabular}{ccc}
\includegraphics[width=2 in]{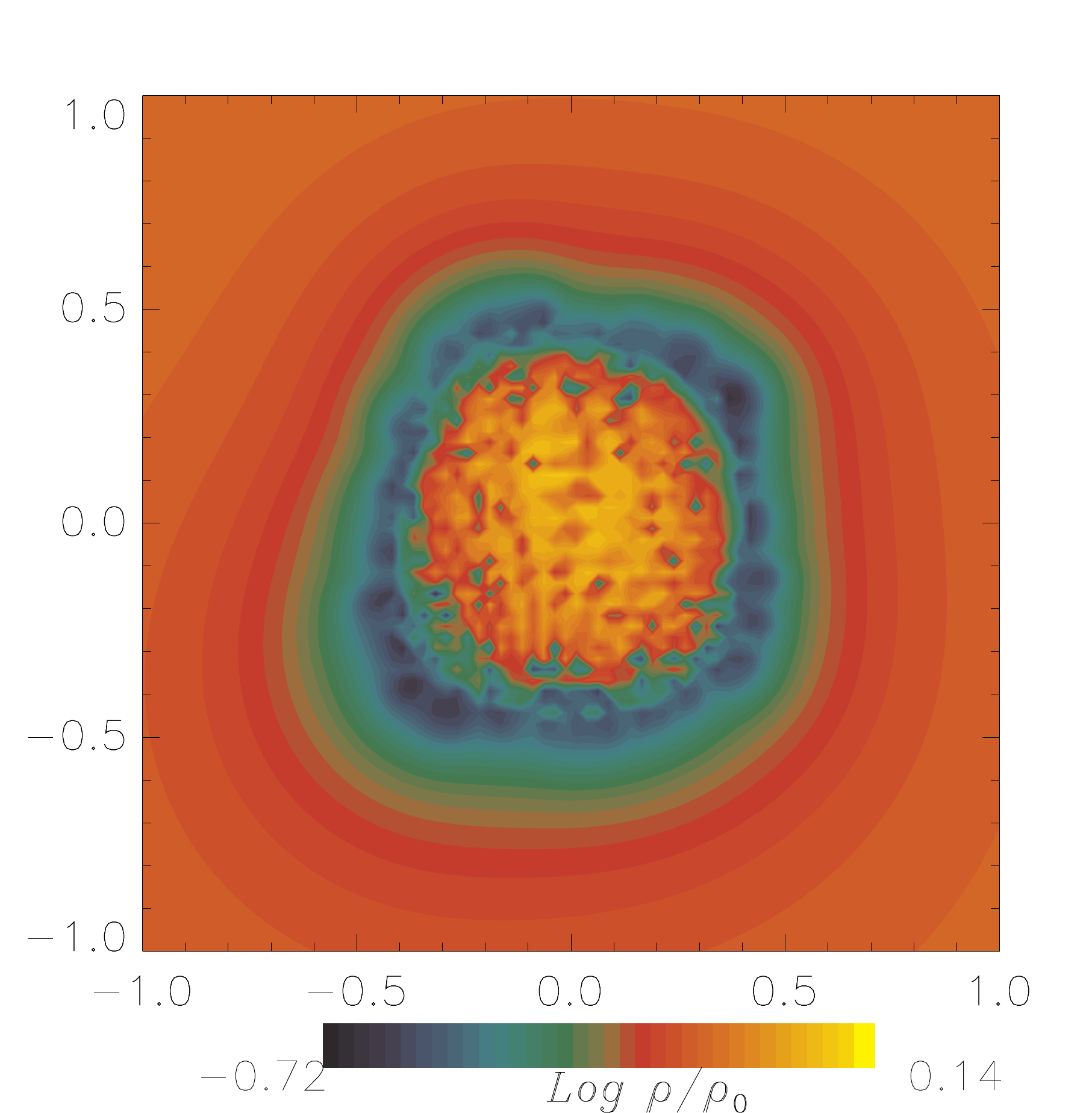} & \includegraphics[width=2 in]{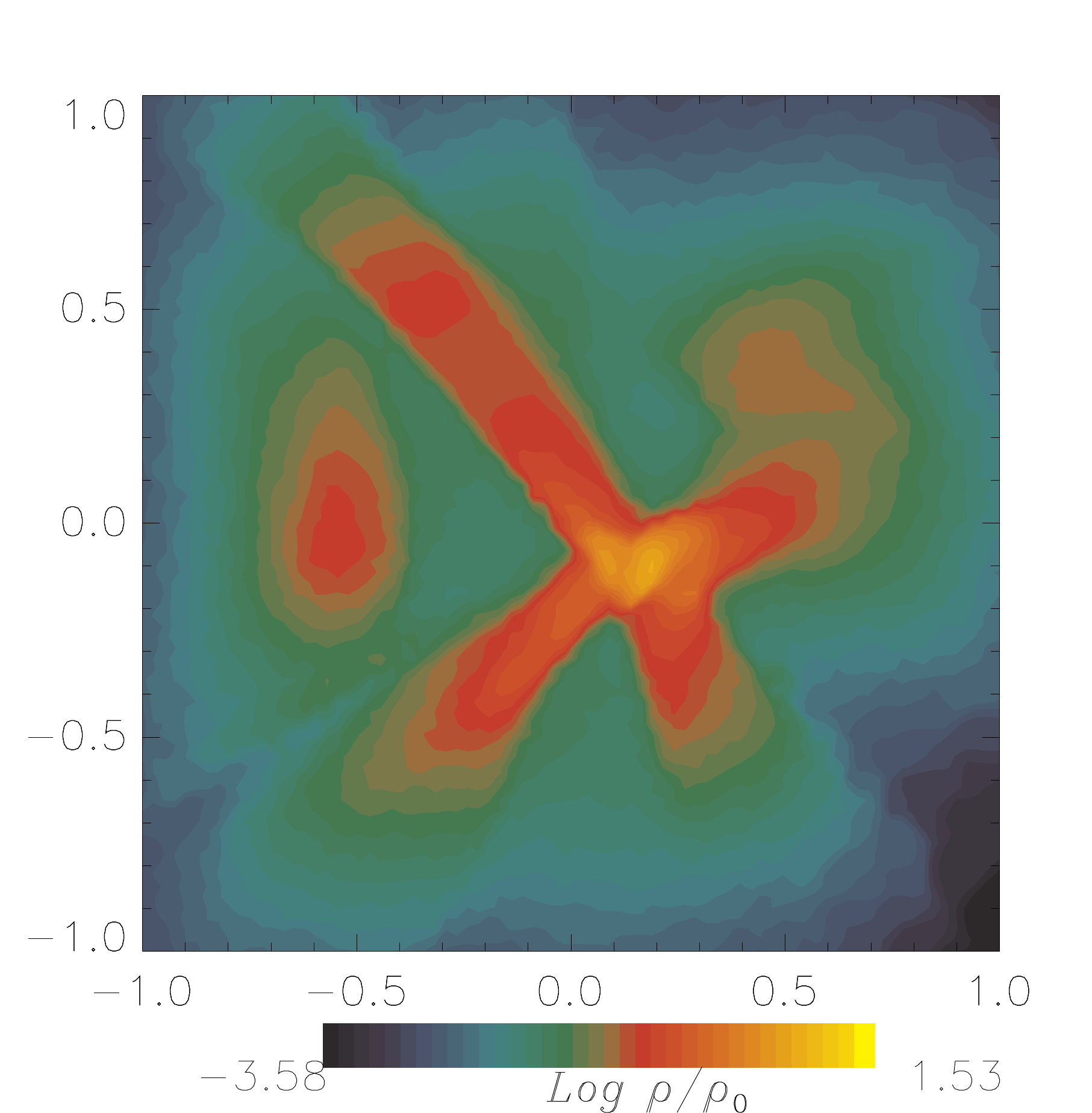} &
\includegraphics[width=2 in]{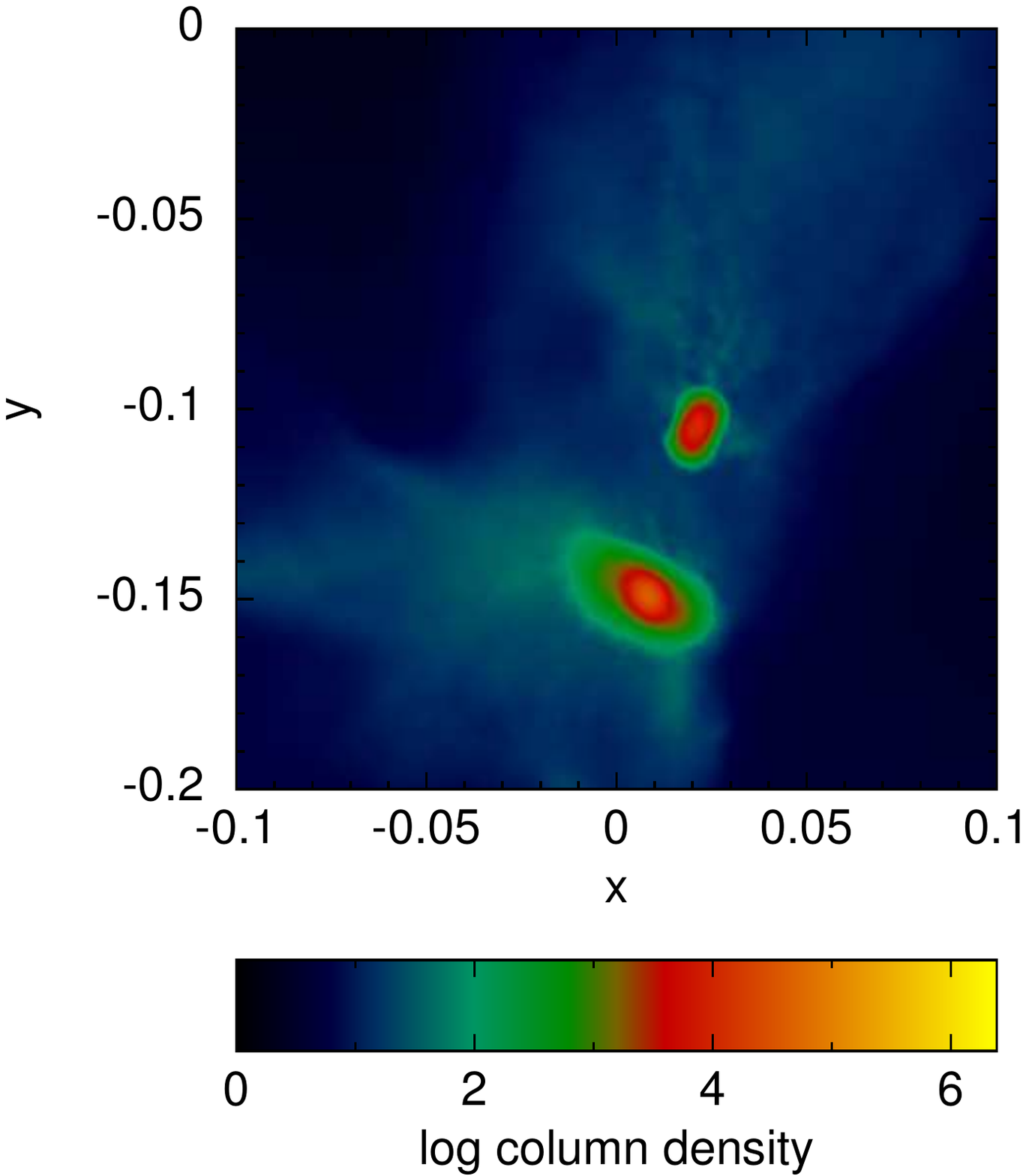}
\end{tabular}
\caption{\label{MosBOscT3rp} Iso-density  plots for
model 14.}
\end{figure}

The same behavior was observed in models 12--14, shown in
Figs.~\ref{MosBOscTrp}, \ref{MosBOscT2rp} and \ref{MosBOscT3rp}, in
which we see again the orthogonal directions of the growth  of the over-density
clumps when compared with those observed in models 26--28.
\subsection{Integral properties}
\label{subsec:intprop}

We present here some integral properties of
the resulting protostars, such as the mass and the values of the energy ratios
$\alpha_f$ and $\beta_f$. These properties are calculated by using a subset
of the simulation particles, which are determined by means of the
following procedure. We first locate the highest density
particle in the collapsed central region of the last available snapshot for each model; this
particle will be the center of the protostar. We then find all the particles which
(i) have density above some minimum value, given by
$\log_{10} \left( \rho_{min}/\rho_0 \right)=1.0$ for all
the turbulent models; (ii) are also located within a given maximum
radius r$_{max}$ from the protostar's center.

All the calculated integral properties are reported in Table \ref{tab:propint}, whose
entries are as follows. The first column shows the number of the model; the second
column shows the parameter r$_{max}$ in terms of the initial core
radius R$_0$; the third column shows the mass of the protostar given in terms
of M$_\odot$; the fourth and fifth columns give the values
of $\alpha_f$ and $\beta_f$, respectively.

There are two lines in Table \ref{tab:propint} for
models 14 and 27, as their outcomes are binary protostar systems, so that each line
indicates the properties of each binary protostar.
For these resulting binary systems, we simply define the binary separation
as the distance between the centers associated with
each protostar. We find the the separation is 163 AU for each binary system.

We find that the average mass of the protostars for models 1--13, excluding
model 14, whose outcome is a binary system, is 0.94 M$_{\odot}$, with
a standard deviation of 0.47 M$_{\odot}$. Analogously, the protostar
average mass for models 15--26 and 28, excluding the mass of the
binary of model 27, is 0.67 M$_{\odot}$, with a standard deviation
of 0.34 M$_{\odot}$. It must be emphasized that these mass relations
do not change much if we include the mass of the protostars in
the binaries.

\begin{figure}
\begin{center}
\includegraphics[width=4.2in]{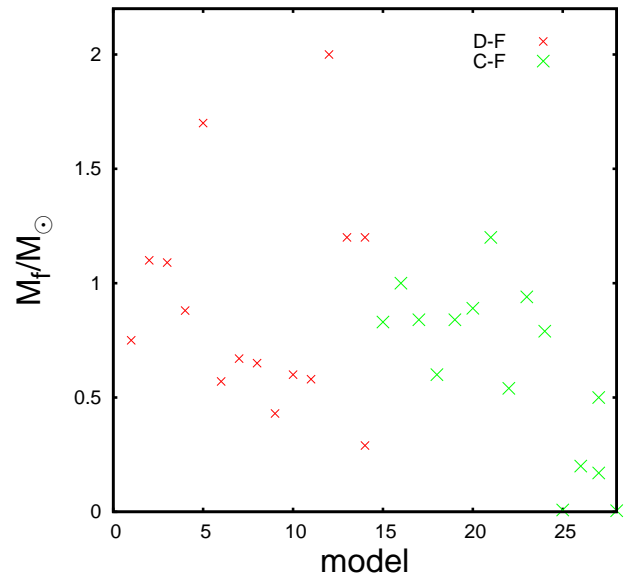}
\caption{\label{MassFrag} Proto-stellar mass in terms of the
simulation model number.}
\end{center}
\end{figure}

In Fig.~\ref{MassFrag} we show the mass of the protostars in terms of
the simulation model number. It seems that there is nothing that allows us
to clearly distinguish the turbulent spectrum used initially; neither
is there any trend to assess the effect that the increase in
the number of Fourier modes can have on the resulting protostar
mass; see also Table~\ref{tab:propint}. However, there seems to be a
tendency to low protostar masses for  higher model numbers in both
turbulence spectra. If this is true, then it would imply that the
larger the perturbation mode, the lower the protostar mass.

\begin{figure}
\begin{center}
\includegraphics[width=4.2in]{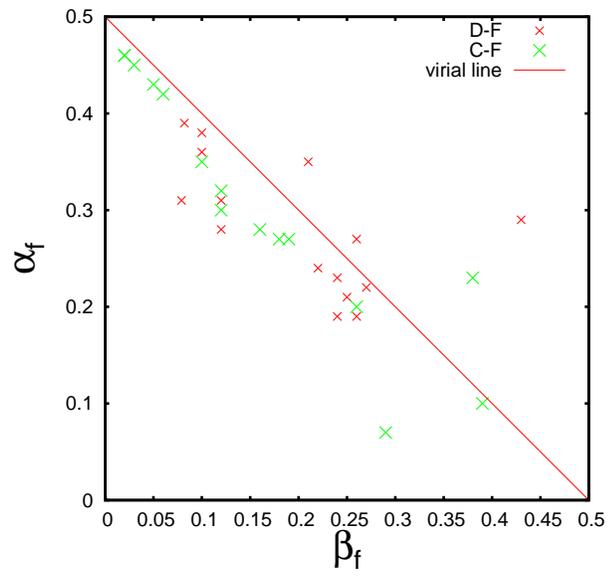}
\caption{\label{AlphavsBetaFrags} Dimensionless energy ratios of the resulting protostars.}
\end{center}
\end{figure}

In Fig.~\ref{AlphavsBetaFrags}, we show that most protostars are
near or on their way towards virialization, as they are close to the
virial line; see Section \ref{subs:energies}.It should be noticed
that this tendency of collapsing cores to evolve toward the virial
line was first pointed out by \citet{boss80} and \citet{boss81}. We
also see that higher values of $\alpha_f$ and $\beta_f$ are obtained
for the curl-free turbulence type.

\section{Discussion}
\label{sec:dis}

Much similarity of the core collapses is already noticeable by
comparing Figs.~\ref{evoluciondensidadDF} and
\ref{evoluciondensidadRF}. In these figures, there is a local
density peak in the early evolution stage, which is a consequence of
the collisions of particles whose velocities have been assigned
randomly by means of a turbulent spectrum with a given number of
Fourier modes in a Fourier mesh of a given size. By looking at the
third panel of each of these figures, we note that the 64$^3$
modes used in models 9--11 of the divergence-free turbulent type and
models 23--25 of the curl-free turbulent type are equally
insufficient to capture these local density peak.

The fact that these local peaks are of the same order irrespective
of the turbulence type indicates that the initial density
fluctuations are also of the same order. Despite this, we find
that the average protostar mass of the divergence-free turbulent
models is larger than of the curl-free turbulent models; see Section
\ref{subsec:intprop}.

With regard to the peak velocity of the collapsing core's central
region, we find a significant similarity between the two
turbulence types: the average collapsing velocity is 7.75 km/s for
the divergence-free type and 7.6 km/s for the curl-free type.
However, the interval of velocities of the former type of turbulence is
in general wider than that of the latter type of turbulence.

Another very important issue to discuss is the low level of
fragmentation observed in the suite of turbulent models. A first
consideration is that we do not use the sink technique introduced by
\citet{batebonnellprice95}, so that the simulations presented here
do not evolve further in time than t$_{ff}$. If we were able
to follow the simulations longer, we would possibly see the
fragmentation of the spiral arms seen surrounding some protostars or
the fragmentation of the highly deformed over-density clumps seen in
models with a large number of perturbation modes.

A second consideration was already mentioned in Section~\ref{subsec:core}, that the
core considered in this paper was very
dense and with a significant thermal support; see also Section~\ref{subs:energies}. For
these two reasons, we expect that the
turbulence spectra induced initially find it more difficult to compress
the gas randomly across the core. The density of the core in our
simulations is ten times larger than the central core considered by
\citet{goodwin2004a}, \citet{goodwin2004b}, \citet{goodwin2006}; see
also \citet{attwood}. We suppose that fragmentation can be prevented
in a central core of higher density.

A third consideration is the stochastic nature of the turbulent
spectra. In this paper we fixed the seed to generate random numbers, so that all the
simulations ran using the same seed. In other papers, it has been shown
that different random realizations of the same simulation can have
significant differences  in their outcomes; see for instance
\citet{walch}, who used four random seeds in their simulations.

A fourth consideration is that the turbulent velocity
fluctuations considered in this paper provide the core with a net
angular momentum; see Section~\ref{subs:mod} and
Fig.~\ref{SpecAngMom}. We calculated the specific angular momentum
with respect to the origin of coordinates (shown in
Fig.~\ref{SpecAngMom} ) and with respect to each of the X, Y and Z
axes, as well. We did not observe any significant difference in
these angular momentum calculations,  but only small changes due to
random velocity fluctuations in a frame with spherical symmetry.

Let us now mention that for rotating cores, $\beta$, defined in
Section \ref{subs:energies}, would correspond to the ratio of
rotational energy to gravitational energy, and a value of 0.21 (see
Eq.\ref{alphaybeta}) can be considered high with respect to
observational values, because their specific angular momentum varies
within a range 1 $\times 10^{20}$-2 $\times 10^{22}$ ; see
\citet{goodman93,boden95}. For this reason, fragmentation can be
easily obtained from the collapse of this kind of rotating cores;
see \citet{RMAA2012} and \citet{RMAA2016}. As we mentioned in
Section~\ref{subs:mod}, the level of angular momentum initially
given to our models is low compared to typical values of rotating
cores and this is the key to understand why we observed only two
models with binary formation.

The protostar masses in the divergence-free turbulence models are in the
range 0.29--2 M$_\odot$ while for the curl-free models the masses vary within
0.2--1.2 M$_\odot$. These masses are in any case much too much larger
than those obtained from the collapse of a rotating uniform core of similar
total mass and initial energy ratios; see for instance \citet{RMAA2016}
and \citet{miAAS}, where the masses generally obtained by numerical
simulations of binary formation are around $0.01 \, M_{\odot}$.
\citet{goodwin2004a} obtained a wide
distribution of protostar masses, with a peak around 1 M$_\odot$. Large masses,
like the ones obtained from turbulent models, are therefore in better
agreement with recent VLA and CARMA observations, which
showed that the proto-stellar masses of systems such as CB230 IRS1 and
L1165-SMM1 have been detected in the range of $0.1-0.25 \, M_{\odot}$; see \citet{tobin}.

The binary systems obtained in this paper have a mass ratio of
q$_{14}$=0.24 and q$_{27}$=0.34 for models 14 and 27, respectively.
\citet{goodwin2004a} obtained a $q$ distribution for wide binaries in
the range 0.1--0.7 with a peak in the range 0.6--0.7; however, as
the binaries evolved, the peak of the $q$ distributions moved to smaller
values. In any case, our $q$ values are in agreement with
\citet{goodwin2004a}.

\section{Concluding Remarks}
\label{sec:conclu}

In this paper we considered the gravitational collapse of a uniform
core, in which only two types of turbulent velocity spectra have been
initially implemented.

It must be emphasized to all the simulations satisfy the
same energy requirements, contained in the ratios
$\alpha$ and $\beta$; see Eq.~\ref{alphaybeta} of Section
\ref{subs:energies}. Therefore, it is because of this last statement that
we observe a lot similarity in the outcomes of the models, irrespective
of the turbulence type considered.

Besides, as we mentioned at the end of
Section \ref{subs:energies}, the models do
not have different levels of turbulence, as the ratio of the turbulent energy
to the gravitational energy was fixed at $\gamma_{turb}=0.11$ for all the simulations.
It should also be noted that the differences observed between the models can not
be attributed to a different random realization of a given simulation, as the random
seed used to generate each model was fixed at the same value for all the simulations.

For these reasons, we consider that we achieved the main objective of this
paper, so that the different outcomes in the models are due to the change in the
number and size of the Fourier modes of the two types of velocity spectra considered. By comparing
the plots shown in Section \ref{sec:results},  we are able to summarize the following
observations:

\begin{enumerate}

\item The larger the wave length of the perturbation mode, the longer
the collapse; in fact, for the models with the largest wavelength modes, we
observed a collapse delay up to 0.2 $t_{ff}$ with respect to the
models with the shortest wavelength modes.

\item We observe that a larger wave length of the initial perturbation lengthens
the initial over-dense clumps, softens the density contrast, and decreases the velocity of the
particles in the region of collapse.

\item When the Fourier mesh size begins to increase in terms of the core radius, then the divergence-free
turbulence type loses the homogeneity and isotropy of the initial
over-density clump distribution earlier than does the curl-free
turbulence.

\item The initial appearance of elongated clumps is delayed by increasing the number of
Fourier modes.

\item Fragmentation can be induced by increasing the number of perturbation modes.

\item The resulting protostars of all the models with very large perturbation
modes have a net angular momentum, which is also a consequence of the highly anisotropic
accretion of gas.

\item The protostars obtained from divergence-free turbulence are more massive, in general, than
those from models with a curl-free turbulence.

\item In particular, the binary protostar system formed from divergence-free turbulence is also
more massive than
that formed from the curl-free turbulence.

\item The larger the perturbation mode, the lower the resulting protostar mass.

\item The protostar masses in turbulent models are larger than those obtained from the
collapse of a rotating uniform core of similar total mass and initial energy.

However, due to the stochastic nature of the turbulent spectra, these
observations must be validated by more simulations; therefore it is necessary
to perform a larger ensemble of simulations to have a statistically
representative sample of results, which hopefully can help to validate our observations.

\end{enumerate}
\section*{Acknowledgements}
The author gratefully acknowledges the computer resources, technical expertise, and support provided by the
Laboratorio Nacional de Superc\'omputo del Sureste de M\'exico through grant number O-2016/047.

\clearpage

\begin{table}[ph]
\caption{ Turbulent models and their resulting configurations}
{ \begin{tabular}{|c|c|c|c|c|c|c|c|}
\hline \hline
Model  & $C_R$  & $N_x=N_y=N_z$   & ${\cal M}_{max}$  & v$_{max}$ [km/s] & Figure & Turbulence Type & Configuration \\
1  & 1  &  64  & 4.5  & 8.1 & \ref{MosBsrm}    &  D-F   &   S\\
2  & 2  &  64  &  4.48  & 7.8 & \ref{MosBs2rm}   &  D-F   &  S \\
3  & 3  &  64  &   4.49 & 7.2 & \ref{MosBs3rm}   &  D-F   &  S \\
4  & 4  &  64  &   4.48 & 6.3 & \ref{MosBs4rm}   &  D-F    &  S \\
\hline
5  & 1  &  128 &  5.09 & 7.9 & \ref{MosB2srmp}  &  D-F  & S\\
6  & 2  &  128 &  4.84 & 7.1 & \ref{MosB2s2rmp} &  D-F  &  S\\
7  & 3  &  128 &   4.78 & 6.8 & \ref{MosB2s3rmp} &  D-F  & S\\
8  & 4  &  128 &   4.6   & 5.9 & \ref{MosB2s4rmp} &  D-F  & S\\
\hline
9  & 6  &  64  &    3.5   & 5.7 & \ref{MosB-Osrm}  &  D-F   & S\\
10 & 8  &  64  &   3.5   & 7.0 & \ref{MosB-O2srm} &  D-F & S \\
11 & 10 &  64  &  3.25 & 7.0 & \ref{MosB-Os3rm} &  D-F & S \\
\hline
12 & 6  &  128 &  3.9    & 11.1 & \ref{MosBOscTrp} &  D-F  & S \\
13 & 8  &  128 &   3.97  & 10.9 & \ref{MosBOscT2rp}&  D-F  & S\\
14 & 10 &  128 &  4.04  & 9.7 & \ref{MosBOscT3rp}&  D-F  & B\\
\hline
\hline
15 & 1  &  64  & 5.09   & 8.6 & \ref{Seg1} &  C-F  & S \\ 
16 & 2  &  64  & 4.19  & 9.5 & \ref{Seg2} &  C-F  & S \\ 
17 & 3  &  64  &  3.99 & 7.4 & \ref{Seg3} &  C-F & S \\ 
18 & 4  &  64  &  3.82 & 6.6 & \ref{Seg4} &  C-F & S \\   
\hline
19 & 1  &  128 & 4.73  & 9.5 & \ref{Seg5} &  C-F  & S \\
20 & 2  &  128 &  4.9 & 9.2 & \ref{Seg6} &  C-F & S \\ 
21 & 3  &  128 &  5.02 & 10.5 & \ref{Seg7} &  C-F  & S \\
22 & 4  &  128 &  4.3 & 5.7 & \ref{Seg8} &  C-F & S \\ 
\hline
23 & 6  &  64  &  3.5  & 6.6  & \ref{Seg9} &  C-F  & S \\
24 & 8  &  64  &  3.0 & 7.4  & \ref{Seg10}&  C-F  & S \\
25 & 10 &  64 &  2.7   & 6.4  & \ref{Seg11}&  C-F  & S \\
\hline
26 & 6  &  128 &  4.11  & 5.9 & \ref{Seg12}&  C-F & S  \\
27 & 8  &  128 &   4.03 & 7.1 & \ref{Seg13}&  C-F  & B \\
28 & 10 &  128 &  3.43 & 6.0 & \ref{Seg14}&  C-F  & S \\
\hline \hline
\end{tabular} }
 \label{tab:mod}
\end{table}

\begin{table}[ph]
\caption{Physical properties of protostars}
{\begin{tabular}{|c|c|c|c|c|}
\hline\hline
Model  & r$_{max}$/R$_0$ & M$_f$/M$_\odot$   & $\left| \alpha_{f} \right| $ & $ \left| \beta_{f} \right| $\\
1    &    0.008  &  0.75   &  0.36  & 0.10 \\
2    &    0.01   &  1.1     &  0.21  & 0.25 \\
3    &    0.017  &   1.09  &  0.19  & 0.26 \\
4    &    0.01   &   0.88  &  0.19  & 0.24 \\
5   &     0.005  &   1.7    &  0.35  & 0.21\\
6   &     0.005  &   0.57  &  0.39 & 0.082\\
7   &     0.008  &  0.67   &  0.27 & 0.26\\
8   &     0.01   &  0.65    &  0.23 & 0.24 \\
9   &      0.007   &  0.43    &  0.31 & 0.079\\
10   &    0.008   &  0.6      &  0.28 & 0.12 \\
11   &   0.007    &   0.58   &  0.31 & 0.12 \\
12   &   0.013   &    2.0      &  0.22 & 0.27 \\
13   &   0.004    &   1.2      &   0.38 & 0.1 \\
\hline
14   &   0.0095   &   1.2     &   0.24 & 0.22\\
14    &   0.005    &    0.29   &  0.29 & 0.43\footnote{The binary separation is 163.8 AU and
its mass ratio q=0.24}\\
\hline
15    &   0.005    &   0.83    &   0.42 &0.06\\
16   &   0.005     &    1.0     &   0.46  & 0.02\\
17   &    0.01      &     0.84   &   0.27 & 0.18\\
18   &    0.007    &     0.6      &   0.35 & 0.1 \\
19   &    0.004     &    0.84    &   0.46 & 0.02\\
20   &    0.005     &     0.89   &   0.43 & 0.05 \\
21   &    0.005       &     1.2     &    0.45 & 0.03 \\
22   &   0.0095      &     0.54   &    0.28 & 0.16\\
23   &  0.016        &      0.94    &    0.2 & 0.26\\
24   &  0.0095     &      0.79    &   0.27 & 0.19 \\
25   &   0.01         &       0.008   & 0.07 & 0.29\footnote{Only 1758 particles entered in the
subset of particles with the criteria described in Sect.\ref{subsec:intprop}; this lack of particles
explains the anomalous values
obtained for integral properties.}\\
26   &    0.0095    &      0.2       &   0.32 &  0.12\\
\hline
27   &    0.008       &       0.5       &  0.3  & 0.12\\
27    &   0.007       &        0.17    &   0.23 & 0.38\footnote{The binary separation is 163.3 AU and its mass ratio q=0.34}\\
\hline
28    &   0.015       &        0.005  & 0.10 &  0.39\\
\hline
\hline
\end{tabular}}
\label{tab:propint}
\end{table}
\newpage
\clearpage
\clearpage
\clearpage
\clearpage
\clearpage
\clearpage
\clearpage
\end{document}